
%
\def\unredoffs{}
\tolerance=1000\hfuzz=2pt
\catcode`\@=11 
\ifx\hyperdef\UNd@FiNeD\def\hyperdef#1#2#3#4{#4}\def\hyperref#1#2#3#4{#4}\def\href#1#2{#2}\fi
\magnification=1200\unredoffs\baselineskip=16pt plus 2pt minus 1pt
\def\Date#1{\vfill\leftline{#1}\tenpoint\supereject%
\footline={\hss\tenrm\hyperdef\hypernoname{page}\folio\folio\hss}}%

{\count255=\time\divide\count255 by 60 \xdef\hourmin{\number\count255}
 \multiply\count255 by-60\advance\count255 by\time
 \xdef\hourmin{\hourmin:\ifnum\count255<10 0\fi\the\count255}
}
\def\date{\number\day.\number\month.\number\year\ at \hourmin}


\def\nolabels{\def\wrlabeL##1{}\def\eqlabeL##1{}\def\reflabeL##1{}}
\def\writelabels{\def\wrlabeL##1{\leavevmode\vadjust{\rlap{\smash%
{\line{{\escapechar=` \hfill\rlap{\sevenrm\hskip.03in\string##1}}}}}}}%
\def\eqlabeL##1{{\escapechar-1\rlap{\sevenrm\hskip.05in\string##1}}}%
\def\reflabeL##1{\noexpand\llap{\noexpand\sevenrm\string\string\string##1}}}
\nolabels

\global\newcount\secno \global\secno=0
\global\newcount\meqno \global\meqno=1
\def\s@csym{}

\def\newsec#1\par{\global\advance\secno by1%
{\toks0{#1}\message{(\the\secno. \the\toks0)}}%
\global\subsecno=0\eqnres@t\let\s@csym\secsym\xdef\secn@m{\the\secno}\noindent
{\bf\hyperdef\hypernoname{section}{\the\secno}{\the\secno.} #1}%
\writetoca{{\string\hyperref{}{section}{\the\secno}{\bf \the\secno\quad}} {\bf #1}}\par%
\nobreak\medskip\nobreak\noindent\ignorespaces}
\def\eqnres@t{\xdef\secsym{\the\secno.}\global\meqno=1\bigbreak\bigskip}
\def\sequentialequations{\def\eqnres@t{\bigbreak}}\xdef\secsym{}

\global\newcount\subsecno \global\subsecno=0
\def\subsec#1\par{\global\advance\subsecno by1%
{\toks0{#1}\message{(\s@csym\the\subsecno. \the\toks0)}}%
\global\subsubsecno=0%
\ifnum\lastpenalty>9000\else\bigbreak\fi
\noindent{\it\hyperdef\hypernoname{subsection}{\secn@m.\the\subsecno}%
{\secn@m.\the\subsecno.} #1}\writetoca{\string\hskip1.45cm
{\string\hyperref{}{subsection}{\secn@m.\the\subsecno}{\secn@m.\the\subsecno.}}
{#1}}\par\nobreak\medskip\nobreak\noindent\ignorespaces}

\def\appendix#1#2{\global\meqno=1\global\subsecno=0\xdef\secsym{\hbox{#1.}}%
\bigbreak\bigskip\noindent{\bf Appendix \hyperdef\hypernoname{appendix}{#1}%
{#1.} #2}{\toks0{(#1. #2)}\message{\the\toks0}}%
\xdef\s@csym{#1.}\xdef\secn@m{#1}%
\writetoca{{\string\hyperref{}{appendix}{#1}{\bf {#1}\quad}} {\bf #2}}%
\par\nobreak\medskip\nobreak}

%
\def\checkm@de#1#2{\ifmmode{\def\f@rst##1{##1}\hyperdef\hypernoname{equation}%
{#1}{#2}}\else\hyperref{}{equation}{#1}{#2}\fi}
\def\eqnn#1{\DefWarn#1\xdef #1{(\noexpand\relax\noexpand\checkm@de%
{\s@csym\the\meqno}{\secsym\the\meqno})}%
\wrlabeL#1\writedef{#1\leftbracket#1}\global\advance\meqno by1}
\def\f@rst#1{\c@t#1a\em@ark}\def\c@t#1#2\em@ark{#1}
\def\eqna#1{\DefWarn#1\wrlabeL{#1$\{\}$}%
\xdef #1##1{(\noexpand\relax\noexpand\checkm@de%
{\s@csym\the\meqno\noexpand\f@rst{##1}1}{\hbox{$\secsym\the\meqno##1$}})}
\writedef{#1\numbersign1\leftbracket#1{\numbersign1}}\global\advance\meqno by1}
\def\eqn#1#2{\DefWarn#1%
\xdef #1{(\noexpand\hyperref{}{equation}{\s@csym\the\meqno}%
{\secsym\the\meqno})}$$#2\eqno(\hyperdef\hypernoname{equation}%
{\s@csym\the\meqno}{\secsym\the\meqno})\eqlabeL#1$$%
\writedef{#1\leftbracket#1}\global\advance\meqno by1}
\def\xeqn{\expandafter\xe@n}\def\xe@n(#1){#1}
\def\xeqna#1{\expandafter\xe@n#1}
\def\eqns#1{(\e@ns #1{\hbox{}})}
\def\e@ns#1{\ifx\UNd@FiNeD#1\message{eqnlabel \string#1 is undefined.}%
\xdef#1{(?.?)}\fi{\let\hyperref=\relax\xdef\next{#1}}%
\ifx\next\em@rk\def\next{}\else%
\ifx\next#1\xeqn#1\else\def\n@xt{#1}\ifx\n@xt\next#1\else\xeqna#1\fi
\fi\let\next=\e@ns\fi\next}

\def\DefWarn#1{\ifx\UNd@FiNeD#1\else
\immediate\write16{*** WARNING: the label \string#1 is already defined ***}\fi}
%
\newskip\footskip\footskip14pt plus 1pt minus 1pt 
\def\footnotefont{\ninepoint}\def\f@t#1{\footnotefont #1\@foot}
\def\f@@t{\baselineskip\footskip\bgroup\footnotefont\aftergroup\@foot\let\next}
\setbox\strutbox=\hbox{\vrule height9.5pt depth4.5pt width0pt}
\global\newcount\ftno \global\ftno=0
\def\foot{\global\advance\ftno by1\def\foot@rg{\hyperref{}{footnote}%
{\the\ftno}{\the\ftno}\xdef\foot@rg{\noexpand\hyperdef\noexpand\hypernoname%
{footnote}{\the\ftno}{\the\ftno}}}\footnote{$^{\foot@rg}$}}
%
%
%
\global\newcount\refno \global\refno=1
\newwrite\rfile
\def\ref{[\hyperref{}{reference}{\the\refno}{\the\refno}]\nref}
\def\nref#1{\DefWarn#1%
\xdef#1{[\noexpand\hyperref{}{reference}{\the\refno}{\the\refno}]}%
\writedef{#1\leftbracket#1}%
\ifnum\refno=1\immediate\openout\rfile=\jobname.refs\fi
\chardef\wfile=\rfile\immediate\write\rfile{\noexpand\item{[\noexpand\hyperdef%
\noexpand\hypernoname{reference}{\the\refno}{\the\refno}]\ }%
\reflabeL{#1\hskip.31in}\pctsign}\global\advance\refno by1\findarg}
\def\findarg#1#{\begingroup\obeylines\newlinechar=`\^^M\pass@rg}
{\obeylines\gdef\pass@rg#1{\writ@line\relax #1^^M\hbox{}^^M}%
\gdef\writ@line#1^^M{\expandafter\toks0\expandafter{\striprel@x #1}%
\edef\next{\the\toks0}\ifx\next\em@rk\let\next=\endgroup\else\ifx\next\empty%
\else\immediate\write\wfile{\the\toks0}\fi\let\next=\writ@line\fi\next\relax}}
\def\striprel@x#1{} \def\em@rk{\hbox{}}
\def\lref{\begingroup\obeylines\lr@f}
\def\lr@f#1#2{\DefWarn#1\gdef#1{\let#1=\UNd@FiNeD\ref#1{#2}}\endgroup\unskip}
\def\semi{;\hfil\break}
\def\addref#1{\immediate\write\rfile{\noexpand\item{}#1}} 
\def\listrefs{\vfill\supereject\immediate\closeout\rfile\writestoppt
\baselineskip=\footskip\centerline{{\bf References}}\bigskip{\parindent=20pt%
\frenchspacing\escapechar=` \input \jobname.refs\vfill\eject}\nonfrenchspacing}
\def\startrefs#1{\immediate\openout\rfile=\jobname.refs\refno=#1}
\def\xref{\expandafter\xr@f}\def\xr@f[#1]{#1}
\def\refs#1{\count255=1[\r@fs #1{\hbox{}}]}
\def\r@fs#1{\ifx\UNd@FiNeD#1\message{reflabel \string#1 is undefined.}%
\nref#1{need to supply reference \string#1.}\fi%
\vphantom{\hphantom{#1}}{\let\hyperref=\relax\xdef\next{#1}}%
\ifx\next\em@rk\def\next{}%
\else\ifx\next#1\ifodd\count255\relax\xref#1\count255=0\fi%
\else#1\count255=1\fi\let\next=\r@fs\fi\next}
%

%
\newwrite\ffile\global\newcount\figno \global\figno=1
\def\fig{fig.~\hyperref{}{figure}{\the\figno}{\the\figno}\nfig}
\def\nfig#1{\DefWarn#1%
\xdef#1{fig.~\noexpand\hyperref{}{figure}{\the\figno}{\the\figno}}%
\writedef{#1\leftbracket fig.\noexpand~\xfig#1}%
\ifnum\figno=1\immediate\openout\ffile=\jobname.figs\fi\chardef\wfile=\ffile%
{\let\hyperref=\relax
\immediate\write\ffile{\noexpand\medskip\noexpand\item{Fig.\ %
\noexpand\hyperdef\noexpand\hypernoname{figure}{\the\figno}{\the\figno}. }
\reflabeL{#1\hskip.55in}\pctsign}}\global\advance\figno by1\findarg}
\def\xfig{\expandafter\xf@g}\def\xf@g fig.\penalty\@M\ {}
\def\figs#1{figs.~\f@gs #1{\hbox{}}}
\def\f@gs#1{{\let\hyperref=\relax\xdef\next{#1}}\ifx\next\em@rk\def\next{}\else
\ifx\next#1\xfig #1\else#1\fi\let\next=\f@gs\fi\next}
%
\def\figin{\epsfcheck\figin}\def\figins{\epsfcheck\figins}
\def\epsfcheck{\ifx\epsfbox\UnDeFiNeD
\message{(NO epsf.tex, FIGURES WILL BE IGNORED)}
\gdef\figin##1{\vskip2in}\gdef\figins##1{\hskip.5in}
\else\message{(FIGURES WILL BE INCLUDED)}%
\gdef\figin##1{##1}\gdef\figins##1{##1}\fi}
\def\DefWarn#1{}
\def\figinsert{\goodbreak\topinsert}
\def\ifig#1#2#3{\DefWarn#1\xdef#1{fig.~\the\figno}
\writedef{#1\leftbracket fig.\noexpand~\the\figno}%
\figinsert\figin{\centerline{#3}}
\smallskip
\leftskip=20pt \rightskip=20pt
\baselineskip12pt\noindent
{{\bf Fig.~\the\figno}\ \ninepoint #2}
\medskip
\global\advance\figno by1\par\endinsert}
\newwrite\lfile
{\escapechar-1\xdef\pctsign{\string\%}\xdef\leftbracket{\string\{}
\xdef\rightbracket{\string\}}\xdef\numbersign{\string\#}}
\def\writedefs{\immediate\openout\lfile=label.defs \def\writedef##1{%
{\let\hyperref=\relax\let\hyperdef=\relax\let\hypernoname=\relax
 \immediate\write\lfile{\string\def\string##1\rightbracket}}}}%
\def\writestop{\def\writestoppt{\immediate\write\lfile{\string\pageno
 \the\pageno\string\startrefs\leftbracket\the\refno\rightbracket
 \string\def\string\secsym\leftbracket\secsym\rightbracket
 \string\secno\the\secno\string\meqno\the\meqno}\immediate\closeout\lfile}}
\def\writestoppt{}\def\writedef#1{}

\def\seclab#1{\DefWarn#1%
\xdef #1{\noexpand\hyperref{}{section}{\the\secno}{\the\secno}}%
\writedef{#1\leftbracket#1}\wrlabeL{#1=#1}}
\def\subseclab#1{\DefWarn#1%
\xdef #1{\noexpand\hyperref{}{subsection}{\the\secno.\the\subsecno}%
{\the\secno.\the\subsecno}}\writedef{#1\leftbracket#1}\wrlabeL{#1=#1}}
\def\applab#1{\DefWarn#1%
\xdef #1{\noexpand\hyperref{}{appendix}{\secn@m}{\secn@m}}%
\writedef{#1\leftbracket#1}\wrlabeL{#1=#1}}
\newwrite\tfile \def\writetoca#1{}
\def\leaderfill{\leaders\hbox to 1em{\hss.\hss}\hfill}
\def\writetoc{\immediate\openout\tfile=\jobname.toc
   \def\writetoca##1{{\edef\next{\write\tfile{\noindent ##1
   \string\leaderfill{
   \string\hyperref{}{page}{\noexpand\number\pageno}%
   {\noexpand\number\pageno}} \par}}\next}}
}
\newread\ch@ckfile
\def\listtoc{\immediate\closeout\tfile\immediate\openin\ch@ckfile=\jobname.toc
\ifeof\ch@ckfile\message{no file \jobname.toc, no table of contents this pass}%
\else\closein\ch@ckfile\centerline{\bf Contents}\nobreak\medskip%
{\baselineskip=16pt\footnotefont\parskip=0pt\catcode`\@=11\input\jobname.toc
\catcode`\@=12\bigbreak\bigskip}\fi}
\catcode`\@=12 
\def\tenpoint{\def\rm{\fam0\tenrm}
\textfont0=\tenrm \scriptfont0=\sevenrm \scriptscriptfont0=\fiverm
\textfont1=\teni  \scriptfont1=\seveni  \scriptscriptfont1=\fivei
\textfont2=\tensy \scriptfont2=\sevensy \scriptscriptfont2=\fivesy
\textfont\itfam=\tenit \def\it{\fam\itfam\tenit}\def\footnotefont{\ninepoint}%
\textfont\bffam=\tenbf \def\bf{\fam\bffam\tenbf}\def\sl{\fam\slfam\tensl}\rm}
\font\ninerm=cmr9 \font\sixrm=cmr6 \font\ninei=cmmi9 \font\sixi=cmmi6
\font\ninesy=cmsy9 \font\sixsy=cmsy6 \font\ninebf=cmbx9
\font\nineit=cmti9 \font\ninesl=cmsl9 \skewchar\ninei='177
\skewchar\sixi='177 \skewchar\ninesy='60 \skewchar\sixsy='60
\def\ninepoint{\def\rm{\fam0\ninerm}
\textfont0=\ninerm \scriptfont0=\sixrm \scriptscriptfont0=\fiverm
\textfont1=\ninei \scriptfont1=\sixi \scriptscriptfont1=\fivei
\textfont2=\ninesy \scriptfont2=\sixsy \scriptscriptfont2=\fivesy
\textfont\itfam=\ninei \def\it{\fam\itfam\nineit}\def\sl{\fam\slfam\ninesl}%
\textfont\bffam=\ninebf \def\bf{\fam\bffam\ninebf}\rm}
%
\hyphenation{anom-aly anom-alies coun-ter-term coun-ter-terms}

\global\newcount\subsubsecno \global\subsubsecno=0
\def\subsubsec#1\par{\global\advance\subsubsecno by1%
{\toks0{#1}\message{(\the\secno\the\subsecno\the\subsubsecno. \the\toks0)}}%
\ifnum\lastpenalty>9000\else\bigbreak\fi
\noindent{\it\hyperdef\hypernoname{subsubsection}{\the\secno.\the\subsecno\the\subsubsecno}%
{\the\secno.\the\subsecno.\the\subsubsecno.} #1}
\par\nobreak\medskip\nobreak\noindent\ignorespaces}

\def\DefWarn#1{}
\def\tikzcaption#1#2{\DefWarn#1\xdef#1{Fig.~\the\figno}
\writedef{#1\leftbracket Fig.\noexpand~\the\figno}%
{
\smallskip
\leftskip=20pt \rightskip=20pt \baselineskip12pt\noindent
{{\bf Fig.~\the\figno}\ \ninepoint #2}
\bigskip
\global\advance\figno by1 \par}}

\def\ntoalpha#1{%
\ifcase#1%
@%
\or A\or B\or C\or D\or E\or F\or G\or H\or I
\fi
}

\global\newcount\appno \global\appno=1
\def\applab#1{\xdef #1{\ntoalpha\appno}\writedef{#1\leftbracket#1}\wrlabeL{#1=#1}
\global\advance\appno by1}

\def\preprint#1 #2\par{\rightline{\vbox{\baselineskip12pt\hbox{#1}\hbox{#2}}}\vskip2cm}
%
\def\title#1\par{\centerline{\bf #1}\nopagenumbers\pageno=0}
\def\author#1\par{\bigskip\bigskip\centerline{#1}}

\newcount\addressno

\def\email#1#2{\unskip$^#1$\footnote{\null}{\kern-\parindent \llap{$^#1$\hskip1pt}email: #2}}

\def\startcenter{%
  \par
  \begingroup
  \leftskip=0pt plus 1fil
  \rightskip=\leftskip
  \parindent=0pt
  \parfillskip=0pt
}
\def\stopcenter{\endgroup}

\def\address{\bigskip%
  \ifnum\the\addressno=0\else\stopcenter\endgroup\fi
  \advance\addressno by 1%
  \begingroup
  \startcenter
  \it
  \obeylines
  \addressAux
}
\def\addressAux#1{#1}

\def\abstract{\stopcenter\endgroup\bigskip\bigskip\noindent}

\def\Dsl{\,\raise.15ex\hbox{/}\mkern-13.5mu D} 
\def\dsl{\raise.15ex\hbox{/}\kern-.57em\partial}
 
\def\boxeqn#1{\vcenter{\vbox{\hrule\hbox{\vrule\kern3pt\vbox{\kern3pt
	\hbox{${\displaystyle #1}$}\kern3pt}\kern3pt\vrule}\hrule}}}


\def\ap{{\alpha^{\prime}}}

\def\a{\alpha}

\def\g{{\gamma}}

\def\e{{\epsilon}}
\def\l{\lambda}

\def\t{{\theta}}

\def\half{{1\over 2}}
\def\p{{\partial}}

\def\bar{\overline}
\def\({\left(}
\def\){\right)}

\def\Im{\mathop{{\rm Im}}} 


\def\qed{\hbox{\hskip 3pt
\vbox{\hrule\hbox to 7pt{\vrule height 7pt\hfill\vrule}
\hrule}}\hskip3pt}

\overfullrule=0pt\relax

\frenchspacing

\newread\instream \openin\instream= label.defs
\ifeof\instream \message{No labels in advance yet. Wait till next pass.}
\else \closein\instream \input label.defs
\fi
\writedefs

\def\arXiv:#1].{\hepthStrip#1 \nil}
\def\hepthStrip#1 #2\nil{\href{http://arxiv.org/abs/#1}{arXiv:#1 #2\unskip}].}

\input epsf

\newdimen\pageremains\newdimen\pdepth
\newdimen\figwidth
\newdimen\figheight
\newcount\figlines
\newcount\flevel

\def\figflow#1#2#3{
\ifnum\flevel>0
\message{******Figure collision. Ignoring second figure.******}
\else
\figwidth=#1
\figheight=#2
\def\contents{#3}
\def\figure{\let\temp=\par \let\par=\plainpar
  \line{\overfullrule=0pt
   \ifdim \figwidth<0pt \hsize=-\figwidth \hss\else \hsize=\figwidth\fi
   \advance \hsize by -10pt
   \vbox to \figheight{\vfil\noindent\contents\vfill}
   \ifdim \figwidth>0pt \hss\fi
  } \vskip-\figheight \vskip-5pt
  \let\par=\temp%
}
\advance\figheight by \baselineskip
\divide\figheight by \baselineskip
\figlines=\figheight \multiply\figheight by \baselineskip
\begingroup\overfullrule=0pt
\tolerance=1000
\flevel=1
\let\plainpar=\par
\def\par{
  \ifnum\flevel=1
   \plainpar
   \pageremains=\pagegoal \advance\pageremains by -\pagetotal
   \ifdim\pageremains<\figheight \message{Moving figure...}
   \else
      \pdepth=\prevdepth
      \nointerlineskip
      \figure
      \hangindent \figwidth \hangafter -\figlines \hfuzz 5 pt
      \flevel=2
      \prevgraf=0
      \figheight=\baselineskip
   \fi
  \else
   \ifnum\flevel=2
    \ifdim\figheight<\parskip
       \advance\figlines -1 \advance\hangafter 1
       \advance\figheight\baselineskip
    \else
       \advance\figheight -\parskip
    \fi
    \hangcarrypar\relax
   \fi
  \fi
}
\par
\vskip-\pdepth
\fi
}
\def\endflow{\global\let\par=\plainpar\endgroup}
\def\hangcarrypar{
\edef\next{\hangafter=\the\hangafter\hangindent=\the\hangindent}
\plainpar\next
\edef\next{\prevgraf=\the\prevgraf}
\ifnum\prevgraf>0
   \ifnum\prevgraf>\figlines \endflow \flevel=0
   \else
     \message{FIGFLOW: line \the\prevgraf, of \the\figlines.}
     \leavevmode
     \next
   \fi
\fi
}

\preprint AEI--2014--053 DAMTP--2014--59

\title Towards one-loop SYM amplitudes from the pure spinor BRST cohomology

\author Carlos R. Mafra\email{\dagger}{c.r.mafra@damtp.cam.ac.uk} and
	Oliver Schlotterer\email{\ddagger}{olivers@aei.mpg.de}

\address
$^\dagger$DAMTP, University of Cambridge
Wilberforce Road, Cambridge, CB3 0WA, UK

\address
$^\ddagger$Max--Planck--Institut f\"ur Gravitationsphysik
Albert--Einstein--Institut, 14476 Potsdam, Germany

\abstract
In this paper, we outline a method to compute supersymmetric one-loop integrands in ten-dimensional SYM
theory. It relies on the constructive interplay between their cubic-graph organization and BRST
invariance of the underlying pure spinor superstring description. The five- and six-point amplitudes
are presented in a manifestly local form where the kinematic dependence is furnished
by BRST-covariant expressions in pure spinor superspace. At five points, the local kinematic numerators
are shown to satisfy the BCJ duality between color and kinematics leading to
type IIA/B supergravity amplitudes as a byproduct. At six points, the sources of the hexagon anomaly are
identified in superspace as systematic obstructions to BRST invariance. Our results are expected to
reproduce any integrated SYM amplitude in dimensions $D< 8$.

\Date {October 2014}

\newif\iffig
\figfalse

\def\Box#1,#2,#3,#4,{{\cal N}^{(4)}_{#1|#2,#3,#4}\, I^{(4)}_{#1,#2,#3,#4}}
\def\Pentagon#1,#2,#3,#4,#5,{{\cal N}^{(5)}_{#1|#2,#3,#4,#5}(\ell) I^{(5)}_{#1,#2,#3,#4,#5}}


\lref\MasonSVA{
  L.~Mason and D.~Skinner,
  ``Ambitwistor strings and the scattering equations,''
JHEP {\bf 1407}, 048 (2014).
[arXiv:1311.2564 [hep-th]].
\semi
 T.~Adamo, E.~Casali and D.~Skinner,
  ``Ambitwistor strings and the scattering equations at one loop,''
JHEP {\bf 1404}, 104 (2014).
[arXiv:1312.3828 [hep-th]].
}

\lref\GomezWZA{
  H.~Gomez and E.~Y.~Yuan,
  ``N-point tree-level scattering amplitude in the new Berkovits` string,''
JHEP {\bf 1404}, 046 (2014).
[arXiv:1312.5485 [hep-th]].
}

\lref\TsuchiyaVA{
  A.~Tsuchiya,
  ``More on One Loop Massless Amplitudes of Superstring Theories,''
Phys.\ Rev.\ D {\bf 39}, 1626 (1989).
}

\lref\twoloop{
  N.~Berkovits,
  ``Super-Poincare covariant two-loop superstring amplitudes,''
JHEP {\bf 0601}, 005 (2006).
[hep-th/0503197].
}

\lref\BernUF{
  Z.~Bern, J.~J.~M.~Carrasco, L.~J.~Dixon, H.~Johansson and R.~Roiban,
  ``Simplifying Multiloop Integrands and Ultraviolet Divergences of Gauge Theory and Gravity Amplitudes,''
Phys.\ Rev.\ D {\bf 85}, 105014 (2012).
[arXiv:1201.5366 [hep-th]].
\semi
Z.~Bern, S.~Davies, T.~Dennen and Y.~t.~Huang,
  ``Absence of Three-Loop Four-Point Divergences in N=4 Supergravity,''
Phys.\ Rev.\ Lett.\  {\bf 108}, 201301 (2012).
[arXiv:1202.3423 [hep-th]].
\semi
Z.~Bern, S.~Davies, T.~Dennen and Y.~t.~Huang,
  ``Ultraviolet Cancellations in Half-Maximal Supergravity as a Consequence of the Double-Copy Structure,''
Phys.\ Rev.\ D {\bf 86}, 105014 (2012).
[arXiv:1209.2472 [hep-th]].
\semi
 Z.~Bern, S.~Davies, T.~Dennen, A.~V.~Smirnov and V.~A.~Smirnov,
  ``Ultraviolet Properties of N=4 Supergravity at Four Loops,''
Phys.\ Rev.\ Lett.\  {\bf 111}, no. 23, 231302 (2013).
[arXiv:1309.2498 [hep-th]].
\semi
 Z.~Bern, S.~Davies and T.~Dennen,
  ``Enhanced Ultraviolet Cancellations in N = 5 Supergravity at Four Loop,''
[arXiv:1409.3089 [hep-th]].
}

\lref\OchirovXBA{
J.J.M.~Carrasco, M.~Chiodaroli, M.~G\"unaydin and R.~Roiban,
  ``One-loop four-point amplitudes in pure and matter-coupled $N \le 4$ supergravity,''
JHEP {\bf 1303}, 056 (2013).
[arXiv:1212.1146 [hep-th]].
\semi
  Z.~Bern, S.~Davies, T.~Dennen, Y.t.~Huang and J.~Nohle,
  ``Color-Kinematics Duality for Pure Yang-Mills and Gravity at One and Two Loops,''
[arXiv:1303.6605 [hep-th]]
\semi
 J.~Nohle,
  ``Color-Kinematics Duality in One-Loop Four-Gluon Amplitudes with Matter,''
Phys.\ Rev.\ D {\bf 90}, 025020 (2014).
[arXiv:1309.7416 [hep-th]].
\semi
M.~Chiodaroli, Q.~Jin and R.~Roiban,
  ``Color/kinematics duality for general abelian orbifolds of N=4 super Yang-Mills theory,''
JHEP {\bf 1401}, 152 (2014).
[arXiv:1311.3600 [hep-th]].
\semi
  A.~Ochirov and P.~Tourkine,
  ``BCJ duality and double copy in the closed string sector,''
JHEP {\bf 1405}, 136 (2014).
[arXiv:1312.1326 [hep-th]].
\semi
H.~Johansson and A.~Ochirov,
  ``Pure Gravities via Color-Kinematics Duality for Fundamental Matter,''
[arXiv:1407.4772 [hep-th]].
\semi
M.~Chiodaroli, M.~Gunaydin, H.~Johansson and R.~Roiban,
  ``Scattering amplitudes in N=2 Maxwell-Einstein and Yang-Mills/Einstein supergravity,''
  arXiv:1408.0764 [hep-th].
}

\lref\BjerrumBohrVC{
  N.E.J.~Bjerrum-Bohr and P.~Vanhove,
  ``Explicit Cancellation of Triangles in One-loop Gravity Amplitudes,''
JHEP {\bf 0804}, 065 (2008).
[arXiv:0802.0868 [hep-th]].
}

\lref\ChenEVA{
  W.M.~Chen, Y.~t.~Huang and D.~A.~McGady,
  ``Anomalies without an action,''
[arXiv:1402.7062 [hep-th]].
}

\lref\worldline{
	Z.~Bern and D.~A.~Kosower,
  ``Efficient calculation of one loop QCD amplitudes,''
Phys.\ Rev.\ Lett.\  {\bf 66}, 1669 (1991)..
\semi
  Z.~Bern and D.~A.~Kosower,
  ``The Computation of loop amplitudes in gauge theories,''
Nucl.\ Phys.\ B {\bf 379}, 451 (1992)..
\semi
M.~J.~Strassler,
  ``Field theory without Feynman diagrams: One loop effective actions,''
Nucl.\ Phys.\ B {\bf 385}, 145 (1992).
[hep-ph/9205205].
\semi
 Z.~Bern, D.~C.~Dunbar and T.~Shimada,
  ``String based methods in perturbative gravity,''
Phys.\ Lett.\ B {\bf 312}, 277 (1993).
[hep-th/9307001].
\semi
D.~C.~Dunbar and P.~S.~Norridge,
  ``Calculation of graviton scattering amplitudes using string based methods,''
Nucl.\ Phys.\ B {\bf 433}, 181 (1995).
[hep-th/9408014].
\semi
C.~Schubert,
  ``Perturbative quantum field theory in the string inspired formalism,''
Phys.\ Rept.\  {\bf 355}, 73 (2001).
[hep-th/0101036].
\semi
  N.~E.~J.~Bjerrum-Bohr and P.~Vanhove,
  ``Absence of Triangles in Maximal Supergravity Amplitudes,''
JHEP {\bf 0810}, 006 (2008).
[arXiv:0805.3682 [hep-th]].
}

\lref\InfiniteTension{
  N.~Berkovits,
  ``Infinite Tension Limit of the Pure Spinor Superstring,''
JHEP {\bf 1403}, 017 (2014).
[arXiv:1311.4156 [hep-th], arXiv:1311.4156].
}

\lref\GomezSLA{
	H.~Gomez and C.R.~Mafra,
  	``The closed-string 3-loop amplitude and S-duality,''
  	JHEP {\bf 1310}, 217 (2013).
	[arXiv:1308.6567 [hep-th]].
}

\lref\oneloopbb{
	C.R.~Mafra and O.~Schlotterer,
	``The Structure of n-Point One-Loop Open Superstring Amplitudes,''
JHEP {\bf 1408}, 099 (2014).
[arXiv:1203.6215 [hep-th]].
}
\lref\fivetree{
C.R.~Mafra,
  	``Pure Spinor Superspace Identities for Massless Four-point Kinematic Factors,''
	JHEP {\bf 0804}, 093 (2008).
	[arXiv:0801.0580 [hep-th]].
	\semi
	C.R.~Mafra,
	``Simplifying the Tree-level Superstring Massless Five-point Amplitude,''
	JHEP {\bf 1001}, 007 (2010).
	[arXiv:0909.5206 [hep-th]].
\semi
	C.R.~Mafra, O.~Schlotterer, S.~Stieberger and D.~Tsimpis,
  	``Six Open String Disk Amplitude in Pure Spinor Superspace,''
	Nucl.\ Phys.\ B {\bf 846}, 359 (2011).
	[arXiv:1011.0994 [hep-th]].
}
\lref\towards{
	C.R.~Mafra,
	``Towards Field Theory Amplitudes From the Cohomology of Pure Spinor Superspace,''
	JHEP {\bf 1011}, 096 (2010).
	[arXiv:1007.3639 [hep-th]].
}
\lref\nptMethod{
	C.R.~Mafra, O.~Schlotterer, S.~Stieberger and D.~Tsimpis,
	``A recursive method for SYM n-point tree amplitudes,''
	Phys.\ Rev.\ D {\bf 83}, 126012 (2011).
	[arXiv:1012.3981 [hep-th]].
}
\lref\nptTree{
	C.R.~Mafra, O.~Schlotterer and S.~Stieberger,
	``Complete N-Point Superstring Disk Amplitude I. Pure Spinor Computation,''
	Nucl.\ Phys.\ B {\bf 873}, 419 (2013).
	[arXiv:1106.2645 [hep-th]].
\semi
  	C.R.~Mafra, O.~Schlotterer and S.~Stieberger,
	``Complete N-Point Superstring Disk Amplitude II. Amplitude and Hypergeometric Function Structure,''
	Nucl.\ Phys.\ B {\bf 873}, 461 (2013).
	[arXiv:1106.2646 [hep-th]].
}
\lref\eombbs{
  	C.R.~Mafra and O.~Schlotterer,
  	``Multiparticle SYM equations of motion and pure spinor BRST blocks,''
	JHEP {\bf 1407}, 153 (2014).
	[arXiv:1404.4986 [hep-th]].
}

\lref\siegel{
	W.~Siegel,
	``Classical Superstring Mechanics,''
	Nucl.\ Phys.\  {\bf B263}, 93 (1986).
}

\lref\GSWII{
  M.B.~Green, J.H.~Schwarz and E.~Witten,
  ``Superstring Theory. Vol. 2: Loop Amplitudes, Anomalies And Phenomenology,''
Cambridge, UK: Univ.~Pr.~(1987) 596 P. (Cambridge Monographs On Mathematical Physics).
}

\lref\wittentwistor{
	E.Witten,
        ``Twistor-Like Transform In Ten-Dimensions''
        Nucl.Phys. B {\bf 266}, 245~(1986)
}
\lref\psf{
 	N.~Berkovits,
	``Super-Poincare covariant quantization of the superstring,''
	JHEP {\bf 0004}, 018 (2000)
	[arXiv:hep-th/0001035].
}
\lref\multiloop{
        N.~Berkovits,
	``Multiloop amplitudes and vanishing theorems using the pure spinor formalism for the superstring,''
	JHEP {\bf 0409}, 047 (2004).
	[hep-th/0406055].
}
\lref\oneloopMichael{
	M.B.~Green, C.R.~Mafra and O.~Schlotterer,
	``Multiparticle one-loop amplitudes and S-duality in closed superstring theory,''
	JHEP {\bf 1310}, 188 (2013).
	[arXiv:1307.3534].
}
\lref\wipNpt{
C.R.~Mafra and O.~Schlotterer, work in progress
}

\lref\anomaly{
	N.~Berkovits and C.R.~Mafra,
	``Some Superstring Amplitude Computations with the Non-Minimal Pure Spinor Formalism,''
	JHEP {\bf 0611}, 079 (2006).
	[hep-th/0607187].
}
\lref\PSBCJ{
	C.R.~Mafra, O.~Schlotterer and S.~Stieberger,
	``Explicit BCJ Numerators from Pure Spinors,''
	JHEP {\bf 1107}, 092 (2011).
	[arXiv:1104.5224 [hep-th]].
}

\lref\BCJ{
	Z.~Bern, J.J.M.~Carrasco and H.~Johansson,
	``New Relations for Gauge-Theory Amplitudes,''
	Phys.\ Rev.\ D {\bf 78}, 085011 (2008).
	[arXiv:0805.3993 [hep-ph]].
}
\lref\PSS{
	C.R.~Mafra,
	``PSS: A FORM Program to Evaluate Pure Spinor Superspace Expressions,''
	[arXiv:1007.4999 [hep-th]].
\semi
	J.A.M.~Vermaseren,
	``New features of FORM,''
	arXiv:math-ph/0010025.
\semi
	M.~Tentyukov and J.A.M.~Vermaseren,
	``The multithreaded version of FORM,''
	arXiv:hep-ph/0702279.
}

\lref\GSanomaly{
	M.B.~Green and J.H.~Schwarz,
	``Anomaly Cancellation in Supersymmetric D=10 Gauge Theory and Superstring Theory,''
	Phys.\ Lett.\ B {\bf 149}, 117 (1984).
\semi
	M.B.~Green and J.H.~Schwarz,
	``The Hexagon Gauge Anomaly in Type I Superstring Theory,''
	Nucl.\ Phys.\ B {\bf 255}, 93 (1985).
}

\lref\YuanRG{
  E.~Y.~Yuan,
  ``Virtual Color-Kinematics Duality: 6-pt 1-Loop MHV Amplitudes,''
JHEP {\bf 1305}, 070 (2013).
[arXiv:1210.1816 [hep-th]].
}

\lref\bigHowe{
  P.S.~Howe,
  ``Pure Spinors Lines In Superspace And Ten-Dimensional Supersymmetric
  Theories,''
  Phys.\ Lett.\  B {\bf 258}, 141 (1991)
  [Addendum-ibid.\  B {\bf 259}, 511 (1991)].
\semi
  P.S.~Howe,
  ``Pure Spinors, Function Superspaces And Supergravity Theories In
  Ten-Dimensions And Eleven-Dimensions,''
  Phys.\ Lett.\  B {\bf 273}, 90 (1991).
}

\lref\WWW{
	C.R.~Mafra, O.~Schlotterer,
http://www.damtp.cam.ac.uk/user/crm66/SYM/pss.html
}

\lref\piotr{
  A.~Ochirov and P.~Tourkine,
  ``BCJ duality and double copy in the closed string sector,''
[arXiv:1312.1326 [hep-th]].
}

\lref\GreenFT{
  M.B.~Green, J.H.~Schwarz and L.~Brink,
  ``N=4 Yang-Mills and N=8 Supergravity as Limits of String Theories,''
Nucl.\ Phys.\ B {\bf 198}, 474 (1982).
}
\lref\BCJloop{
  Z.~Bern, J.J.M.~Carrasco and H.~Johansson,
  ``Perturbative Quantum Gravity as a Double Copy of Gauge Theory,''
Phys.\ Rev.\ Lett.\  {\bf 105}, 061602 (2010).
[arXiv:1004.0476 [hep-th]].
}

\lref\BrinkBC{
  L.~Brink, J.H.~Schwarz and J.~Scherk,
  ``Supersymmetric Yang-Mills Theories,''
Nucl.\ Phys.\ B {\bf 121}, 77 (1977)..
}

\lref\ICTP{
	N.~Berkovits,
  	``ICTP lectures on covariant quantization of the superstring,''
	[hep-th/0209059].
}

\lref\Oscar{
  O.A.~Bedoya and N.~Berkovits,
  ``GGI Lectures on the Pure Spinor Formalism of the Superstring,''
[arXiv:0910.2254 [hep-th]].
}
\lref\partI{
	C.R.~Mafra and O.~Schlotterer,
  	``Cohomology foundations of one-loop amplitudes in pure spinor superspace,''
	[arXiv:1408.3605 [hep-th]].
}

\lref\WWW{
	C.R.~Mafra and O.~Schlotterer,
	{\tt http://www.damtp.cam.ac.uk/user/crm66/SYM/pss.html}
}
\lref\CJfive{
  J.J.~Carrasco and H.~Johansson,
  ``Five-Point Amplitudes in N=4 Super-Yang-Mills Theory and N=8 Supergravity,''
Phys.\ Rev.\ D {\bf 85}, 025006 (2012).
[arXiv:1106.4711 [hep-th]].
}

\lref\BernUE{
  Z.~Bern, J.~J.~M.~Carrasco and H.~Johansson,
  ``Perturbative Quantum Gravity as a Double Copy of Gauge Theory,''
Phys.\ Rev.\ Lett.\  {\bf 105}, 061602 (2010).
[arXiv:1004.0476 [hep-th]].
}
\lref\BernZX{
  Z.~Bern, L.~J.~Dixon, D.~C.~Dunbar and D.~A.~Kosower,
  ``One loop n point gauge theory amplitudes, unitarity and collinear limits,''
Nucl.\ Phys.\ B {\bf 425}, 217 (1994).
[hep-ph/9403226].
}
\lref\MonteiroOx{
  N.E.J.~Bjerrum-Bohr, T.~Dennen, R.~Monteiro and D.~O'Connell,
  ``Integrand Oxidation and One-Loop Colour-Dual Numerators in N=4 Gauge Theory,''
  JHEP {\bf 1307}, 092 (2013).
[arXiv:1303.2913 [hep-th]].
}

\lref\BerkovitsAIA{
  N.~Berkovits,
  ``Twistor Origin of the Superstring,''
[arXiv:1409.2510 [hep-th]].
}
\lref\thetaSYM{
        J.P.~Harnad and S.~Shnider,
        ``Constraints And Field Equations For Ten-Dimensional Superyang-Mills
        Theory,''
        Commun.\ Math.\ Phys.\  {\bf 106}, 183 (1986)
\semi
        P.A.~Grassi and L.~Tamassia,
        ``Vertex operators for closed superstrings,''
        JHEP {\bf 0407}, 071 (2004)
        [arXiv:hep-th/0405072].
\semi
	G.~Policastro and D.~Tsimpis,
        ``$R^4$, purified,''
        Class.\ Quant.\ Grav.\  {\bf 23}, 4753 (2006)
        [arXiv:hep-th/0603165].
}

\lref\ElvangCUA{
  H.~Elvang and Y.~t.~Huang,
  ``Scattering Amplitudes,''
[arXiv:1308.1697 [hep-th]].
}
\lref\FramptonAnomaly{
  P.~H.~Frampton and T.~W.~Kephart,
  ``Explicit Evaluation of Anomalies in Higher Dimensions,''
  Phys.\ Rev.\ Lett.\  {\bf 50}, 1343 (1983), [Erratum-ibid.\  {\bf 51}, 232 (1983)].
\semi
  P.H.~Frampton and T.~W.~Kephart,
  ``The Analysis of Anomalies in Higher Space-time Dimensions,''
Phys.\ Rev.\ D {\bf 28}, 1010 (1983).
}
\lref\NMPS{
	N.~Berkovits,
	``Pure spinor formalism as an N=2 topological string,''
	JHEP {\bf 0510}, 089 (2005).
	[hep-th/0509120].
}
\lref\yutin{
  Z.~Bern, T.~Dennen, Y.~t.~Huang and M.~Kiermaier,
  ``Gravity as the Square of Gauge Theory,''
Phys.\ Rev.\ D {\bf 82}, 065003 (2010).
[arXiv:1004.0693 [hep-th]].
}
\lref\GreenED{
  M.B.~Green and J.H.~Schwarz,
  ``Infinity Cancellations in SO(32) Superstring Theory,''
Phys.\ Lett.\ B {\bf 151}, 21 (1985)..
}

\listtoc
\writetoc
\filbreak

\newsec Introduction

Recent developments have shown that scattering amplitudes take a much simpler form than the multitude
of Feynman diagrams would seem to suggest \ElvangCUA. In a variety of theories and spacetime dimensions, the
laborious computations based on textbook methods were successfully sidestepped by new approaches to
determine amplitudes from first principles manifesting their hidden simplicity. Along these lines, this
work describes a method to obtain the integrands of one-loop amplitudes in ten-dimensional ${\cal N}=1$
super-Yang--Mills theory (SYM) \BrinkBC\ on the basis of two fundamental principles: locality and BRST
symmetry.

Locality refers to the expansion of amplitudes in terms of cubic graphs whose propagators encode the
structure of poles and branch cuts in the scattering data \BCJ. BRST invariance is embedded into pure
spinor superspace where it guarantees supersymmetry and gauge invariance as originally described in the
context of the pure spinor superstring\foot{See \refs{\ICTP,\Oscar} for reviews of the pure spinor
formalism and \BerkovitsAIA\ for a recent derivation of the BRST operator from gauge fixing a reparametrization invariant worldsheet action.} \psf.
This underpins the observation of Howe \bigHowe\ that pure spinor variables simplify the description
of ten-dimensional ${\cal N}=1$ SYM.

\medskip
\figflow{-1.7 truein}{1.8 truein}{{\epsfxsize=1.00\hsize\epsfbox{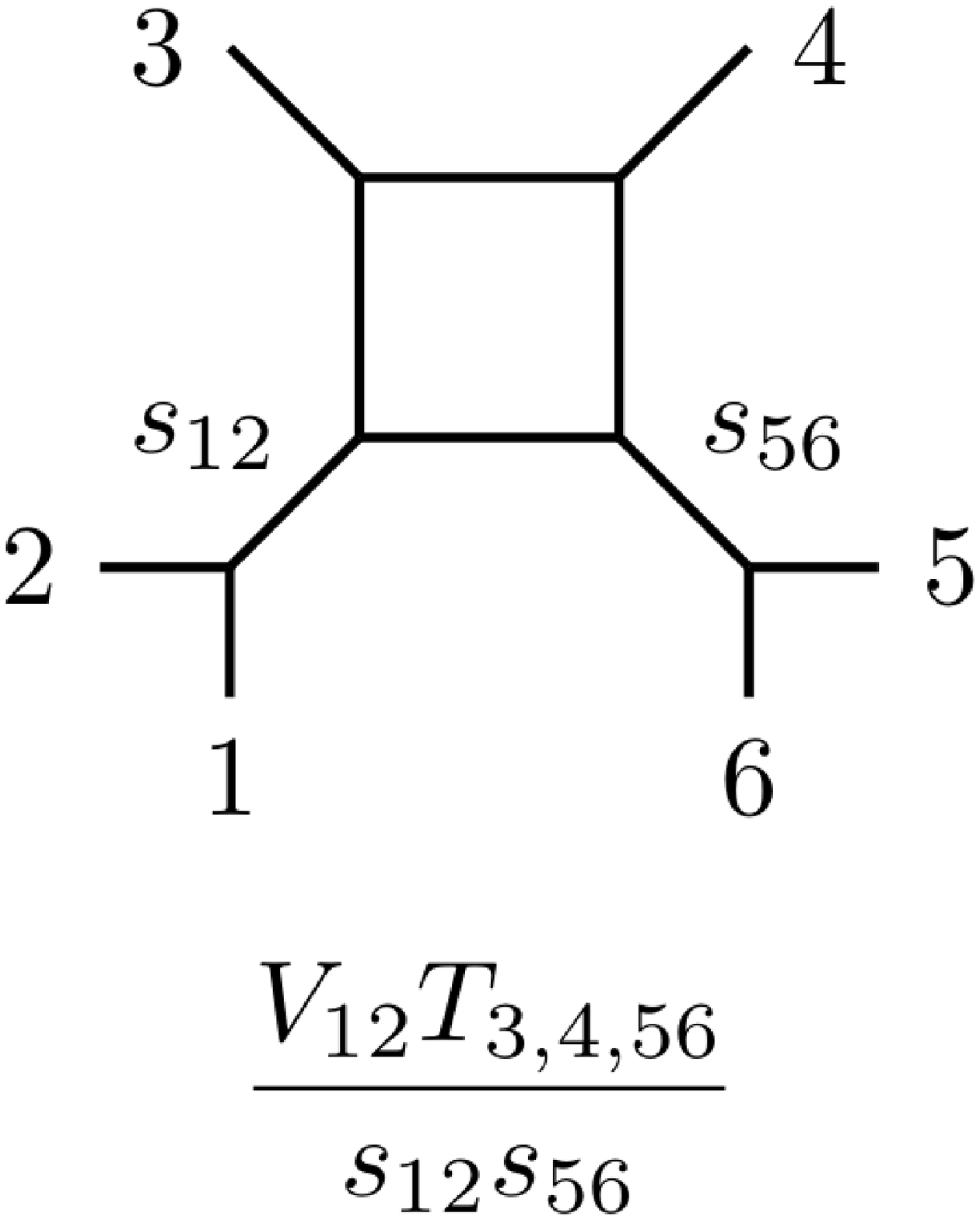}}\vfill}
In the subsequent, we will explain how to combine locality and BRST invariance to a constructive and
intuitive prescription to assemble
one-loop integrands of SYM from their cubic-graph expansion. The long-term goal of this approach is to
find a general and intuitive mapping from cubic graphs at any loop order to superfields such as the objects $V_{12}$ and
$T_{3,4,56}$ as seen on the right. The tree-level mappings have already been worked out
in \refs{\nptMethod,\nptTree} and the main result of this paper is a one-loop implementation of this
dictionary. Using this method we obtain a local form of the five- and six-point one-loop integrands of
ten-dimensional SYM.


The superspace expressions encoding the integrands are in the cohomology of the pure spinor BRST
charge, whose action on kinematic factors follows from simple equations of motion
\refs{\wittentwistor,\eombbs}. The cohomology requirement on scattering amplitudes tightly constrains
the admissible combinations of superfields which, when supplemented by the required propagator
structure set by cubic diagrams, empirically leads to unique answers.

The cohomology approach was successfully applied to assemble tree amplitudes from its cubic graphs
\refs{\towards, \nptMethod}. The kinematic objects which describe individual subdiagrams share the
symmetries of the associated color factors, leading to a manifestation of the BCJ duality \BCJ\ in tree
amplitudes \PSBCJ. Their generalization to multiparticle superfields of SYM in \eombbs\ provides a
superspace representation for any tree-level subdiagram in a one-loop amplitude. For example, the
four-point box numerator can be written as $V_1 T_{2,3,4}$ while the six-point box in the above figure
is represented by $V_{12}T_{3,4,56}$, where two of the superfields are exchanged by their multiparticle
representatives. Hence, the leftover challenge boils down to fixing the irreducible $n$-gon diagram in
the $n$-point one-loop amplitude using BRST invariance. Multiparticle superfields then allow to infer
the structure of massive $n$-gons at higher multiplicity.

This paper is structured as follows. Section~2 reviews the diagrammatic construction of tree-level
amplitudes based on BRST properties of the kinematic numerators in superspace. In section~3, we
introduce a set of one-loop specific superfields as selected by the zero-mode saturation of the open
superstring. They furnish the alphabet of BRST-covariant building blocks for any kinematic numerator in
one-loop integrands. Their precise matching with box, pentagon and hexagon diagrams is dictated by the
BRST algebra and explained in section~4. This leads to a superspace description of the hexagon anomaly
\FramptonAnomaly\ inherent to the chiral fermions of ten-dimensional SYM.

Even though BRST invariance serves as a driving force to determine the kinematic numerators, it is not
manifest at the level of individual diagrams of the local integrands. Hence, we provide an alternative
representation in section~5 where the propagators are reorganized such that any kinematic factor
is manifestly BRST {\it pseudo-invariant\/} -- meaning BRST closed up to anomaly effects \partI.
These cohomology objects have been classified in
\partI\ and shown to obey a rich network of relations under permutations of external legs and
contractions with momenta.
Representations with manifest BRST properties obscure locality but allow to check cyclic symmetry of the
integrated amplitudes in superspace -- again up to the hexagon anomaly.

Finally, section~6 is devoted to the BCJ duality. The five-point integrand is shown to obey all
kinematic Jacobi relations, and the corresponding type IIA/B supergravity amplitudes are presented as a
corollary -- in lines with the field theory limit of the closed superstring. However, the six-point
amplitude suffers from obstructions to satisfy the BCJ duality whose precise form might signal a subtle
relation to the anomaly.

The gluon and gluino components of any superspace numerator presented in this work can be obtained
by combining the known $\theta$ expansions of the superfields \thetaSYM\ with the
prescription $\langle \lambda^3 \theta^5 \rangle=1$ of pure spinor superspace, see for example \PSS.
The gluon components of any kinematic factor in the amplitude representations
of section~5 are available on the website \WWW.

\newsec Tree-level cohomology construction of SYM amplitudes

\seclab\sectwo

\noindent
In this section, we review the pure spinor cohomology derivation of the tree-level amplitudes of SYM
theory \refs{\towards,\nptMethod} using the BRST block techniques of \eombbs. This will prove useful to
undertake the analogous construction of one-loop amplitudes.

\subsec BRST-covariant building blocks from the pure spinor string

\subseclab\sectwoone

\noindent The tree-level amplitude among $n$ massless open superstring states is encoded
in iterated integrals along the boundary of a worldsheet of disk topology parametrized by real $z_i$.
The prescription in the pure spinor formalism is given by \psf
\eqn\treepresc{
{\cal A}^{{\rm tree}}_n = \int \prod_{j=2}^{n-2} dz_j\langle V_1(z_1)\, U_2(z_2) \, \ldots \, U_{n-2}(z_{n-2}) \, V_{n-1}(z_{n-1}) \, V_n(z_{n})  \rangle
}
where $V(z)$ and $U(z)$ are the vertex operators for the gluon super-multiplet
\eqn\PSvertices{
V_i = \l^\a A^i_\a, \qquad
U_i = \p\t^\a A^i_\a + \Pi_m A_i^m + d_\a W^\a_i + \half N_{mn} F^{mn}_i  \ ,
}
and $[A_\a, A^m, W^\a, F_{mn}]$ are the ten-dimensional superfields of ${\cal N}=1$ SYM \wittentwistor.
They are contracted into the spinorial ghost $\lambda^\alpha$ subject to the pure spinor constraint
$\lambda \gamma^m \lambda =0$, and $[\partial \theta^\alpha , \Pi_m , d_\alpha , N_{mn}]$ are conformal
primaries of weight $h=1$, see \siegel\ for their OPEs.

The correlation function in \treepresc\ is determined by the OPEs among the vertices, $V(z)U(w)$ and
$U(z)U(w)$. Based on the experience from four- to six-point computations \fivetree, the general
solution to this problem was argued in \eombbs\ to be captured by multiparticle superfields $[A^B_\a,
A^m_B, W^\a_B, F^B_{mn}]$. As shown in \eombbs, they generalize the standard (single-particle) SYM
superfields of \wittentwistor\ and can be interpreted as representing a multiperipheral tree level
subdiagram with an off-shell leg, see \figone. The on-shell legs $b_j$ are collected in multiparticle
labels $B=b_1b_2 \ldots b_p$, usually denoted by capital Latin letters.

\ifig\figone{The correspondence of cubic graphs and the BRST blocks with
multiparticle label $B=b_1b_2 \ldots b_p$. For the unintegrated vertex $V_B =
\lambda^\alpha A_\alpha^B$, this mapping implies that the BRST variation of cubic graph numerators
cancels propagators and allows the construction of BRST invariant tree-level amplitudes. }
{\epsfxsize=0.70\hsize\epsfbox{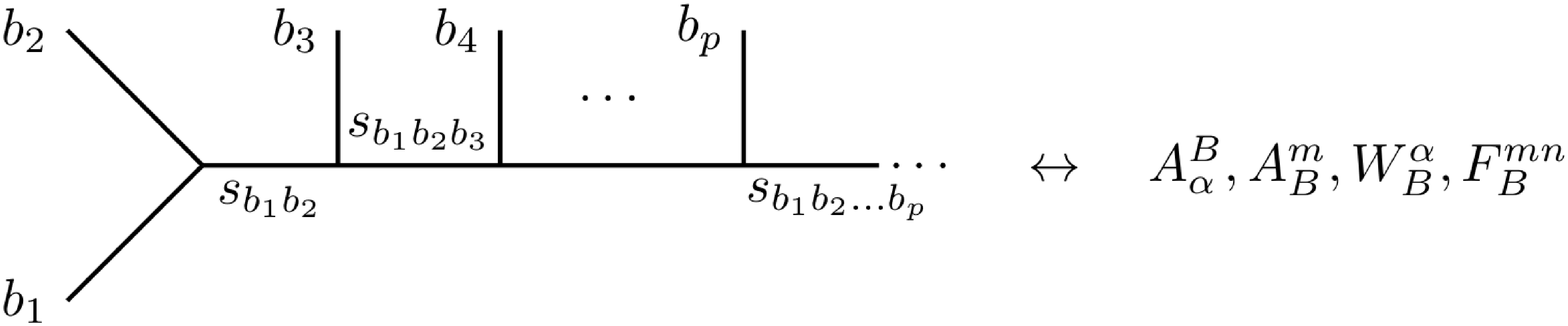}}


More precisely, kinematic factors of the $n$-point
string amplitude at tree level can always be written in terms of multiparticle
vertex operators,
\eqn\MultV{
V_B \equiv \l^\a A^B_\a\ , \ \ \ U_B \equiv \partial \theta^\alpha A_{\alpha}^B + \Pi_m A^m_B + d_\alpha W^\alpha_B + \half N_{mn} F^{mn}_B \ .
}
The simplest two-particle instance reads
\eqn\MultVtwo{
V_{12} \equiv \half\bigl[ V_2 (k^2\cdot A^1) + A^2_m (\g^m W^1)_\a - (1\leftrightarrow 2)\bigr] \,,
}
and a recursive prescription for cases with three and more particles is given in \eombbs. The resulting multiparticle fields $[A^B_\alpha,A_B^m,W_B^\alpha,F^{mn}_B]$ satisfy Lie-symmetries such as
\eqn\Lie{
V_{12} = - V_{21}, \quad V_{123} = - V_{213}, \quad V_{123} + V_{231} + V_{312} = 0 \,,
}
in lines with the dual color tensors $f^{12a}$ and $f^{12a} f^{a3b}$ \BCJ.

Multiparticle vertices  build  up by iterating OPEs of schematic form $U_A U_B \rightarrow U_C$ and $V_A
U_B \rightarrow V_C$. Once  the  conformal fields $[\partial \theta^\alpha , \Pi_m , d_\alpha , N_{mn}]$
in  \treepresc\ are integrated out along these lines, the most general kinematic pattern in superstring
tree amplitudes is furnished by $V_A V_B  V_C$.  Their  ghost number  three is compatible with the
component prescription \psf
\eqn\lambdapresc{
\langle (\l\g^m\t)(\l\g^n\t)(\l\g^p\t)(\t\g_{mnp}\t)\rangle = 2880 \ .
}
Since the open superstring reduces to ten-dimensional ${\cal N}=1$ SYM
in the field-theory limit $\ap\to0$ \GreenFT, these same ingredients of the form  $V_A V_B V_C$ suffice to write
down SYM amplitudes. Furthermore, supersymmetry of
the string amplitudes in the pure spinor formalism is a consequence of BRST
invariance independently of the $\ap$ order, so the
SYM amplitudes must also be BRST invariant.

The above reasoning led to the conjecture in \towards\ that the
$n$-point tree amplitudes of SYM could be obtained by requiring BRST invariance
of linear combinations of $V_A V_B V_C$ with the
appropriate kinematic pole structure. This conjecture eventually led to a
recursive algebraic method for the $n$-point SYM tree amplitude in \nptMethod, but it
will be convenient to recall the diagrammatic construction suggested in
\towards\ since it will be generalized to one-loop below.

\subsec BRST variations and diagrammatic interpretation

\subseclab\sectwotwo

\noindent
At the level of SYM superfields, the BRST operator acts as a fermionic derivative,
\eqn\BRSTcharge{
Q = \lambda^\alpha D_\alpha \ , \ \ \ D_\alpha \equiv {\partial \over \partial \theta^\alpha} + \half k_m (\gamma^m \theta)_\alpha \ .
}
The multiparticle equations of motion for $A_\alpha^B$ \eombbs\ imply covariant BRST variations for$V_B$,
\eqnn\QBRSTV
$$\eqalignno{
QV_{12} &= s_{12} V_1  V_1 \ , \ \ \ \  Q V_{123} = (s_{123}-s_{12}) V_{12}V_3 + s_{12}(V_1 V_{23}+V_{13} V_2) \ ,
&\QBRSTV
}$$
see \eombbs\ for generalizations to higher multiplicity $V_{12\ldots p}$. The Mandelstam invariants in \QBRSTV\ are defined by
\eqn\mandconv{
s_{ij} \equiv (k_i\cdot k_j) = \half (k_i+k_j)^2 \ , \ \ \ \ s_{i_1 i_2\ldots i_p } \equiv \half (k_{i_1}+k_{i_2}+ \ldots + k_{i_p})^2
}
and guide the diagrammatic interpretation shown in \figone. For example, the variation $QV_{12}=
s_{12}V_1V_2$ suggests to associate $V_{12}$ with a propagator $s_{12}^{-1}$. The latter in turn
describes a cubic vertex with on-shell particles 1 and 2 as well as an off-shell leg carrying the
overall momentum $k^m_{12}\equiv k^m_1+k^m_2$. Higher-multiplicity $V_{12\ldots p}$ have analogous BRST
variations with Mandelstam invariants $s_{12\ldots j} , \ j=2,3,\ldots ,p$, so they are the natural
superspace representatives of cubic subdiagrams with these propagators. The resulting multiperipheral
tree level subdiagrams with a terminal off-shell leg are depicted in \figone. The trilinears $V_A V_B
V_C$ selected by the above string theory prescription allows to connect the three off-shell legs
through an additional vertex and to form an on-shell tree-level diagram, see \figonshell.

\ifig\figonshell{On-shell diagrams represented by $V_AV_BV_C$ connect three off-shell subdiagrams.
Propagators such as $s_{a_1a_2}$ and $s_{a_1a_2a_3}$ are suppressed on the left-hand side.}
{\epsfxsize=0.60\hsize\epsfbox{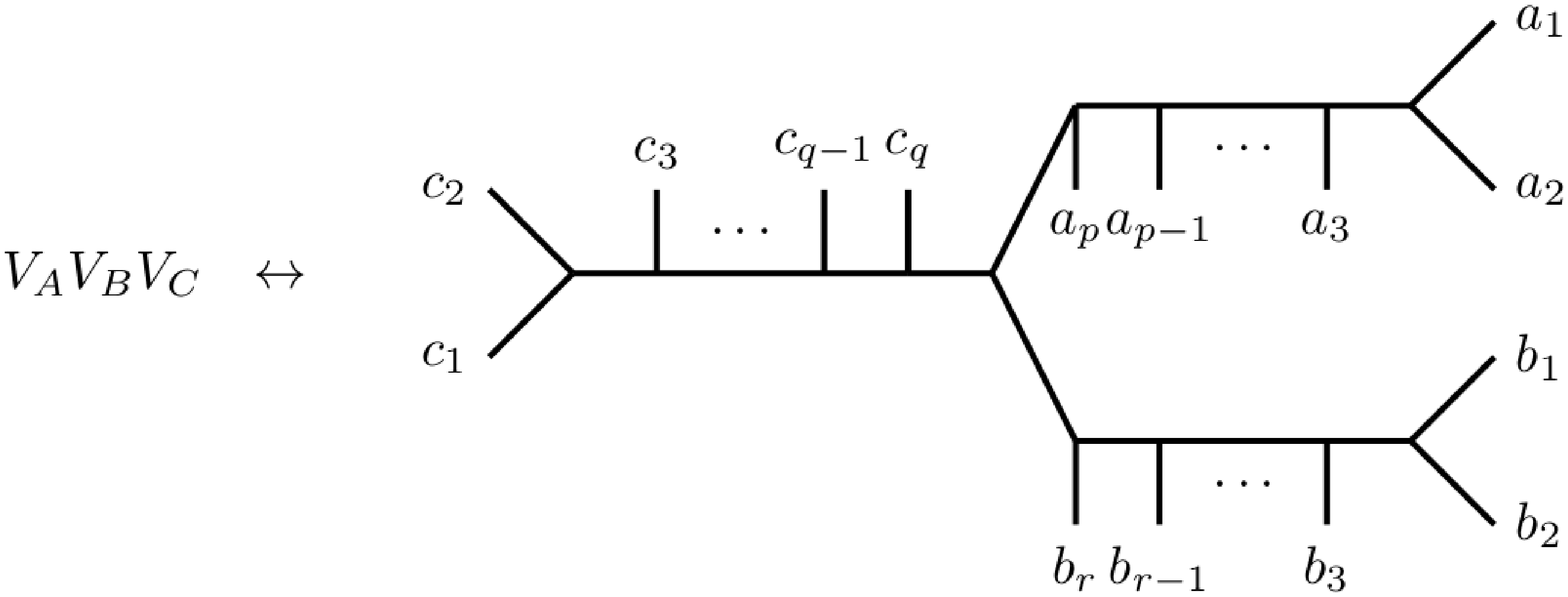}}


\subsec Tree-level SYM amplitudes from the cohomology of pure spinor superspace

\subseclab\sectwothree

\noindent The color-ordered tree-level SYM amplitudes can be organized in terms of
cubic on-shell graphs capturing their kinematic pole structure \BCJ,
\eqn\BCJorg{
A^{{\rm tree}}(1,2, \ldots,n) = \sum_{\Gamma_i}{N_i \over \prod_k P_{k,i}} \ .
}
The sum encompasses all cubic diagrams $\Gamma_i$ compatible with the color ordering, and $P_{k,i}$
denote their corresponding propagators, see \KKdiags\ for four- and five-point examples. The kinematic
numerators $N_i$ carry the polarization dependence and will be specified below.

\ifig\KKdiags{The four- and five-point tree amplitudes represented in terms of cubic graphs.}
{\epsfxsize=0.64\hsize\epsfbox{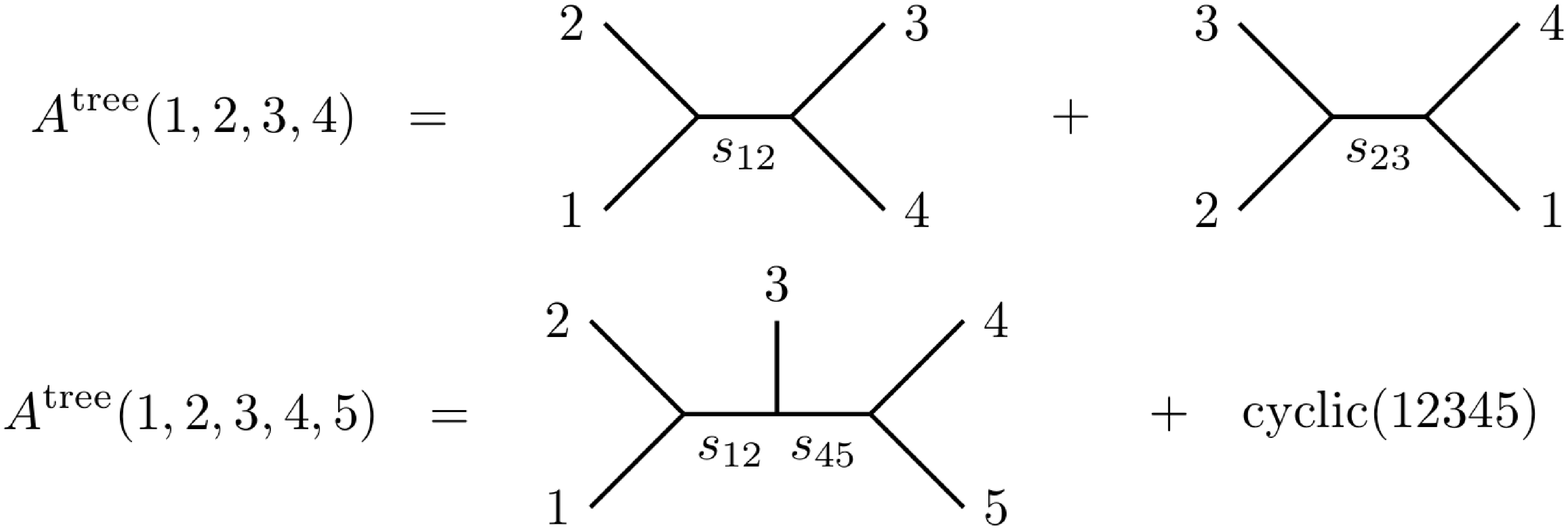}}

The pure spinor cohomology method of \nptMethod\ exploits the dictionary of \figonshell\ between cubic
diagrams and superfields to associate trilinear $V_AV_BV_C$ with each numerators $N_i$ in \BCJorg. The
BRST-covariant properties of the BRST blocks (see \QBRSTV\ and \eombbs) guarantee that these
superspace numerators satisfy the necessary condition for BRST invariance:
\eqn\principleTree{
\hbox{\it each term of  $QN_i$ must have a factor of $P_{k,i}$ with  $k=1,2,\ldots,n-3$.}
}
Otherwise, the BRST variation of the amplitude would have a nonzero residue at the simultaneous pole $\prod_{k} P_{k,i}$ and could not vanish.

One can check that the
following mapping between cubic graphs and pure spinor superspace expressions leads
to BRST-invariant four- and five-point amplitudes:
\medskip
\centerline{{\epsfxsize=0.75\hsize\epsfbox{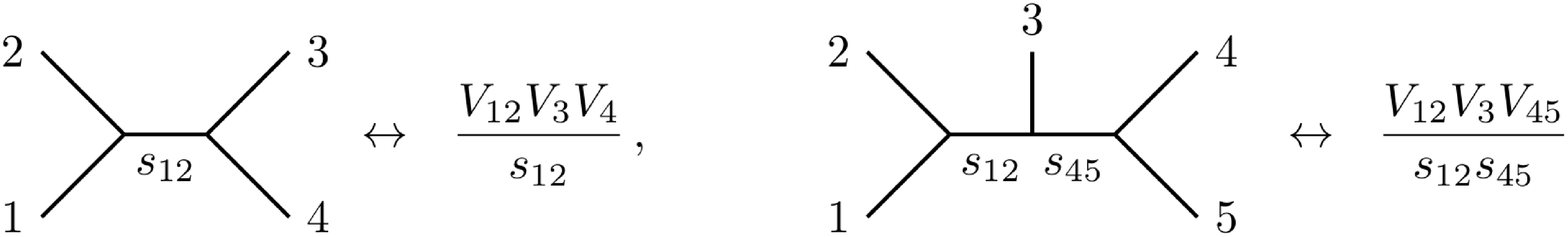}}}
\smallskip

\noindent Using these mappings the four- and five-point amplitudes displayed in \KKdiags\
become
\eqnn\AmpsEx
$$
\eqalignno{
A^{{\rm tree}}(1,2,3,4) &= {\langle V_{12}V_3V_4\rangle \over s_{12}} + {\langle
V_{23}V_4V_1\rangle\over s_{23}} &\AmpsEx\cr
A^{{\rm tree}}(1,2,3,4,5) &= {\langle V_{12}V_3V_{45}\rangle\over s_{12}s_{45}}
 + {\rm cyclic}(1,2,3,4,5) \ .\cr
}$$
BRST invariance follows from the cancellation of the propagators e.g.
\eqn\cancellation{
{ Q V_{12} V_3 V_{4} \over s_{12} } = V_1 V_2 V_3 V_4\,,\quad
{ Q V_{12} V_3 V_{45} \over s_{12} s_{45} } = { V_{1} V_2 V_3 V_{45} \over s_{45}} + { V_{12} V_3 V_4 V_5 \over s_{12}}\ ,
}
and both terms on the right-hand side of the five-point diagram can cancel against further diagrams
which share one of the propagators $s_{12},s_{45}$. Note that any pure spinor superspace numerator
$\langle V_{A}V_B V_C\rangle$ is a {\it local} expression of polarizations and momenta.

Higher-point amplitudes can be similarly obtained using vertices $V_B$ of higher multiplicity. SYM tree
amplitudes up to seven points can be found in \towards, and the $n$-point solution is presented in
\nptMethod\ based on a recursive method. Furthermore, explicit component expansions of SYM trees up to
multiplicity eight can be found in the website \WWW.

Note that the assignment of superfields to a cubic graph is not unique. By choosing different vertices
to play the role of the center of \figonshell, one can arrive at three representations for the
following five-point diagram:
\medskip
\centerline{{\epsfxsize=0.70\hsize\epsfbox{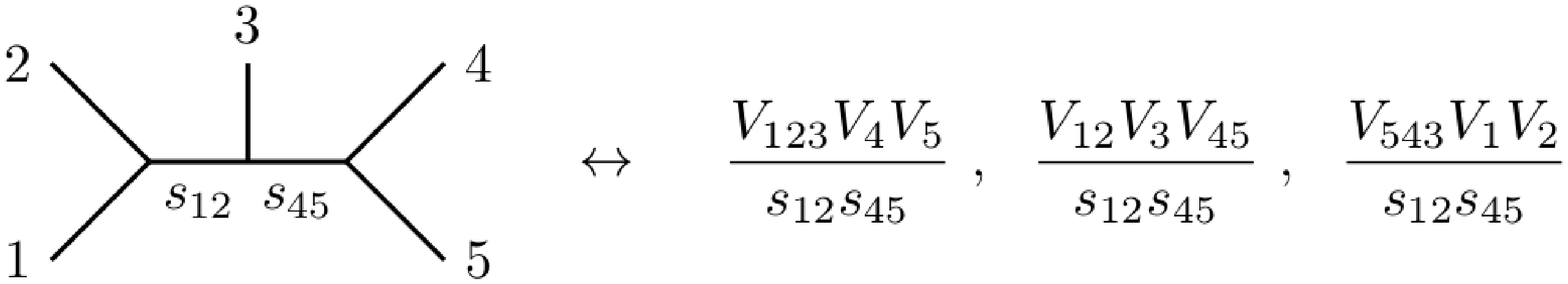}}}


\noindent They differ by contact terms and have to be chosen coherently such as to render the resulting
amplitude \BCJorg\ BRST invariant. At tree level, a consistent choice consists of trilinears of the
schematic form $V_{1\ldots} V_{n-1\ldots} V_n$. The legs of the integrated vertices $2,3,\ldots,n-2$ in
\treepresc\ are distributed along the ellipses of $V_{1\ldots}$ or $V_{n-1\ldots} $, and the single
particle nature of $V_n$ reflects the choice of worldsheet position $z_n \rightarrow \infty$. This is
also a $(n-2)!$ basis of local tree-level numerators found in \PSBCJ\ which satisfy the duality between
color and kinematics.

As we will see in the next sections, similar ambiguities arise at one-loop and can be settled through the
choice of unintegrated vertex operator $V_1$ such that the first particle one can only enter through $V_B$ with $B=1b_2 b_3\ldots$.

\newsec One-loop cohomology construction of SYM integrands

\seclab\secthree

\noindent In this section, the one-loop string amplitude prescription will be used to propose kinematic
building blocks in superspace for SYM one-loop integrands. This will be done in a
similar diagrammatic fashion as reviewed for tree-level amplitudes in the previous section.

\subsec Review of the four point amplitude

\subseclab\secthreeone

\noindent The four-point one-loop amplitudes of ten-dimensional SYM and supergravity were firstly
determined in 1982 by Brink, Green and Schwarz by taking the field-theory limit of their superstring
ancestors. Remarkably, the only contributing Feynman integral was found to be the box graph \GreenFT,
\eqnn\loopSYM
\eqnn\loopSUGRA
$$\eqalignno{
A(1,2,3,4) &= 
 \int {d^D\ell \over (2\pi)^D} \, { s_{12} s_{23} A^{{\rm tree}}(1,2,3,4) \over \ell^2 (\ell-k_1)^2 (\ell-k_{12})^2 (\ell-k_{123})^2} &\loopSYM
\cr
M_4 &=  \int {d^D\ell \over (2\pi)^D} \, { s_{12} s_{23} s_{13} M^{{\rm tree}}_4 \over \ell^2 (\ell-k_1)^2 (\ell-k_{12})^2 (\ell-k_{123})^2} + {\rm symm}(2,3,4)
   \ , &\loopSUGRA
}$$
where the ordering of particles in \loopSYM\ refers to a single color--trace. The analogous pure spinor
derivations have been performed in \multiloop\ resulting in superspace kinematic factors
\eqnn\loopSYMps
\eqnn\loopSUGRAps
$$\eqalignno{
s_{12} s_{23} A^{{\rm tree}}(1,2,3,4) &=  \langle V_1(\lambda \gamma_m W_2) (\lambda \gamma_n W_3) F^{mn}_4 \rangle&\loopSYMps
\cr
s_{12} s_{23} s_{13} M^{{\rm tree}}_4  &=    \langle | V_1(\lambda \gamma_m W_2) (\lambda \gamma_n W_3) F^{mn}_4|^2 \rangle   \ . &\loopSUGRAps 
}$$
In the remainder of this paper, we develop systematic methods to determine their generalization at
higher multiplicity, making either locality or BRST pseudo-invariance manifest. As a first step, the
superfields in \loopSYMps\ will be generalized below to BRST-covariant building blocks suitable to
represent non-trivial tree subdiagrams and $\ell$-dependent parts.

\subsec BRST-covariant building blocks from the pure spinor prescription

\subseclab\secthreetwo

\noindent The one-loop pure spinor amplitude prescription of \multiloop\ leads to a richer set of
BRST-covariant building blocks when compared to the tree-level prescription. As explained in \partI,
the zero-mode saturation patterns following from different contributions from the b-ghost suggest that
building blocks of arbitrary tensor ranks appear in the one-loop string amplitudes.  More precisely,
the superstring prescription for one-loop amplitudes is given by an integral over conformally
inequivalent cylinder\foot{For the purpose of deriving single-trace color-ordered SYM amplitudes, we
neglect the worldsheet topologies beyond the planar cylinder even though they play an important role in
string theory for the cancellation of anomalies and divergences \refs{\GSanomaly,\GreenED,\GSWII}.} diagrams with
circumference $t$ \multiloop,
\eqn\looppresc{
{\cal A}_n = \int_0^{\infty} dt   \! \! \! \! \!  \! \! \! \! \!
\int \limits _{0 \leq \Im z_{i} \leq \Im z_{i+1} \leq t} \! \! \! \! \!  \! \! \! \! \!
dz_2 \, dz_3\, \ldots \, dz_n \langle \int  \mu \, b \,{\cal Z}\, V_1(z_1)\, U_2(z_2) \, \ldots \, U_{n}(z_{n})   \rangle \ .
}
In contrast to the tree-level prescription \treepresc, only one vertex operator $V_1$ appears in the
unintegrated picture whereas $n-1$ vertices $U_j$ are integrated along the cylinder boundary
parametrized by purely imaginary $z_j$. The b-ghost, the various picture changing operators
collectively denoted by ${\cal Z}$ and the Beltrami differencial $\mu$
are explained in \multiloop, and the subsequent discussion only
requires the schematic form of their zero-mode structure: Among the worldsheet fields of conformal
weight one, zero-modes of $d_\alpha d_\beta N^{mn}$ must necessarily by saturated by the integrated
vertices, regardless of the contribution from $b$ and ${\cal Z}$. Given the single particle integrated
vertex \PSvertices, this mechanism gives rise to the superfields $W_2^\alpha W_3^\beta F_4^{mn}$ in the
four point amplitude \loopSYMps.

At higher multiplicity, the correlator in \looppresc\ is determined by a cascade of OPEs among multiparticle vertices of schematic
form $U_A U_B \rightarrow U_C$ and $V_A U_B \rightarrow V_C$. The
expression \MultV\ for their integrated version identifies the multiparticle fields $W_A^\alpha
W_B^\beta F_C^{mn}$ along with the zero-modes $d_\alpha d_\beta N^{mn}$. The two-particle
superfields beyond $V_{12}$ in \MultVtwo\ are given by
\eqnn\Wtwo
$$\eqalignno{
A^{12}_m &=  \half\Bigl[ A^1_p F^2_{pm} - A^1_m(k^1\cdot A^2) + (W^1\g_m W^2) - (1\leftrightarrow 2)\Bigr] \cr
W_{12}^\a &= {1\over 4}(\g^{mn}W^2)^\a F^1_{mn} + W_2^\a (k^2\cdot A^1) - (1\leftrightarrow 2) &\Wtwo \cr
F^{12}_{mn} &= F^2_{mn}(k^2\cdot A^1) +  F^2_{[m}{}^{p}F_{n]p}^1 + k^{12}_{[m}(W_1\g_{n]}W_2) - (1\leftrightarrow 2) 
}$$
with $k_{12}^m \equiv k_1^m +k_2^m$, and generalizations to higher multiplicity can be found in
\eombbs. The unique tensor structure combining the
superfields $W_A^\alpha W_B^\beta F_C^{mn}$ to a ghost-number-two expression as required by the
$\langle \lambda^3 \theta^5 \rangle=1$ prescription \psf\ is given by the scalar
\eqnn\Tsc
$$\eqalignno{
T_{A,B,C} &\equiv {1\over 3} (\lambda \gamma_m W_A) (\lambda \gamma_n W_B) F^{mn}_C + (C \leftrightarrow B,A) \ . &\Tsc 
}$$
It is symmetric under exchange of the slots $A,B,C$ and generalizes the four--point kinematic factor in
\loopSYMps\ to incorporate multiparticle tree-level subdiagrams.

The five- and six-point amplitudes presented in the following also involve vector and
symmetric tensor building blocks
\eqnn\Tvec
\eqnn\Ttens
$$\eqalignno{
T_{A,B,C,D}^m &\equiv \big[ T_{A,B,C} A_D^m + (D \leftrightarrow C,B,A) \big] + W^m_{A,B,C,D} \ &\Tvec 
\cr
T_{A,B,C,D,E}^{mn} &\equiv T^m_{A,B,C,D} A^n_E  +  W^n_{A,B,C,D} A^m_E + (E \leftrightarrow D,C,B,A)  \ . &\Ttens
}$$
They stem from additional saturations of $\Pi^m$ zero-modes from the integrated vertices \MultV\
leaving behind the multiparticle superfield $A^m_B$. Moreover, absorption of $\Pi^m$ also involves
another b-ghost sector which is represented through the shorthand
\eqnn\wwww
$$\eqalignno{
W^m_{A,B,C,D} &\equiv {1\over 12} (\lambda \gamma_n W_A) (\lambda \gamma_p W_B) (W_C \gamma^{mnp} W_D) + (A,B|A,B,C,D) \ . &\wwww
}$$
The notation $(A_1,{\ldots } , A_p \,|\, A_1,{\ldots} ,A_n)$ instructs to sum over all possible ways to choose $p$
elements $A_1,A_2,\ldots ,A_p$ out of the set $\{A_1,{\ldots} ,A_n\}$, for a total of ${n\choose p}$ terms.

Similar to the BRST variation \QBRSTV\ of tree-level constituents $V_B$, the one-loop building blocks \Tsc\ to \Ttens\ transform covariantly under $Q$ \eombbs, e.g.
\eqnn\BRSTTs
$$\eqalignno{
QT_{1,2,3} &= 0 \ , \ \ \ QT_{12,3,4} = s_{12} (V_1 T_{2,3,4} - V_2 T_{1,3,4})  \cr
QT_{12,34,5} &= s_{12} (V_1 T_{2,34,5} - V_2 T_{1,34,5})+ s_{34} (V_3 T_{12,4,5} - V_4 T_{12,3,5}) &\BRSTTs \cr 
QT_{123,4,5} &= (s_{123}-s_{12}) (V_{12} T_{3,4,5} - V_3 T_{12,4,5}) \cr
& \ \ \ \ \ + s_{12} (V_1 T_{23,4,5} + V_{13} T_{2,4,5} - V_{23} T_{1,4,5} - V_2 T_{13,4,5})  \ .
}$$
The variation of vectors and tensors additionally involves terms proportional to $k_{i}^m$ where the vector index is carried by a momentum:
\eqnn\BRSTTv
$$\eqalignno{
QT^m_{1,2,3,4} &= k_1^m V_1 T_{2,3,4} + (1\leftrightarrow 2,3,4) &\BRSTTv \cr
QT^m_{12,3,4,5} &= s_{12} (V_1 T^m_{2,3,4,5} - V_2 T^m_{1,3,4,5}) + k_{12}^m V_{12} T_{3,4,5}  + \big[ k_3^m V_3 T_{12,4,5} + (3\leftrightarrow 4,5) \big]  \cr
Q T^{mn}_{1,2,3,4,5} &= \big[ 2 k_1^{(m} V_1 T^{n)}_{2,3,4,5}  + (1\leftrightarrow 2,3,4,5)\big]+ \delta^{mn} Y_{1,2,3,4,5}  \ .
}$$
The last term in the tensor variation was firstly considered in the pure spinor description of
the would-be hexagon anomaly of the superstring\foot{As shown by Green and Schwarz in 1984, the hexagon
anomaly cancels in the superstring for the gauge group SO$(32)$ \GSanomaly.} in \anomaly\
\eqnn\BRSTTw
$$\eqalignno{
Y_{A,B,C,D,E} &\equiv \half (\l\g^m W_A)(\l\g^n W_B)(\l\g^p W_C)(W_D\g_{mnp}W_E) \ , &\BRSTTw 
}$$
it is totally symmetric in $A,B,\ldots,E$ by the pure spinor constraint. Generalizations to higher rank
were introduced in \partI. Another building block $J_{1|2,3,4,5}$ \partI\ capturing subtleties of the six-point
anomaly in pure spinor superspace will be discussed in section \secfourfour.

\subsec The diagrammatic structure of one-loop amplitudes

\subseclab\secthreethree

\noindent
Since string theory reduces to field theory in the $\ap\to0$ limit, the above building blocks together
with $V_A$ should suffice to describe SYM one-loop amplitudes up to multiplicity six. In the subsequent, we will focus on the
superspace integrand $A(1,2,3, \ldots,n| \ell)$ governing the integrated color ordered
single-trace\foot{At one-loop, SYM subamplitudes associated with double-trace color factors can be
recovered through linear combinations of single-trace subamplitudes \BernZX.} amplitude
$A(1,2,3, \ldots,n)$ via
\eqn\defintegrand{
A(1,2,3, \ldots,n) = \int {d^D\ell \over (2\pi)^D} \langle A(1,2,3, \ldots,n| \ell) \rangle \ .
}
Similar to the cubic graph organization of the tree-level amplitude \BCJorg,
also one-loop SYM amplitudes can be described in terms
cubic graphs $\Gamma_i$,
\BCJloop,
\eqn\LoopGraphs{
A(1,2,3, \ldots,n| \ell) = \sum_{\Gamma_i}
 {N_i(\ell)\over \prod_k P_{k,i}(\ell)} \ .
}
The sum over cubic diagrams $\Gamma_i$ ranges from boxes to $n$-gons whose external tree-level
subdiagrams respect the color ordering on the left-hand side. The no-triangle property \BernZX\
of maximally supersymmetric SYM excludes triangles, bubbles and tadpoles. The superspace
numerators $N_i(\ell)$ and the propagators $P_{k,i}(\ell)$ now depend on the loop momentum~$\ell$, in
addition to the external kinematics. Moreover, supersymmetry bounds the powers of
loop momenta $\ell$ in the numerators $N_i(\ell)$ of a $p$-gon diagram to be $\leq p-4$.

The observations above will be exploited to propose a novel supersymmetric description of one-loop SYM
integrands following two steps. Firstly, we propose mappings between $n$-gon numerators and pure spinor
superspace expressions such as $V_A T_{B,C,D}$ and generalizations to higher rank. This mapping is naturally suggested by the propagators cancelled by the
BRST-covariant variation of the building blocks,
\eqn\principle{
\hbox{\it each term of $QN_i(\ell)$ must have a factor of $P_{k,i}(\ell)$ with  $k=1,2,\ldots,n$,}
}
generalizing the tree-level counterpart \principleTree\ to $\ell$-dependent $P_{k,i}$. And secondly, an overall BRST-invariant superspace expression 
compatible with the cubic graphs in \LoopGraphs\ must
be assembled with the help of the mappings proposed in step one.

Even though the construction is carried out in a ten-dimensional setup, the momenta and external
polarizations can still be restricted to lower dimensions \BrinkBC. The one-loop integrals in \BCJloop\ are
UV-finite if $D<8$, the dimensional reduction \BrinkBC\ of our results is then expected to integrate to
SYM amplitudes in this dimension. In $D=4$, for instance, the subsequent integrands are checked\foot{We
thank Song He for checking it.} to reproduce MHV amplitudes with the right unitarity cuts.

\newsec Local SYM superspace integrands at four, five and six points

\seclab\secfour

\noindent In this section the cubic-graph organization
of the SYM integrands \defintegrand\ will be exploited in connection with their BRST
pseudo-invariance inherited from the pure spinor superstring description. Using the building blocks reviewed
in the previous section, manifestly local integrands will be constructed following BRST cohomology
arguments.

\subsec Local form of the one-loop four-point SYM integrand

\subseclab\secfourone

\noindent
Let us rewrite the four-point SYM integrand
using the above superfield definitions in order to appreciate the natural structure of its
higher-point generalizations. The integrand of the color-ordered amplitude defined by \defintegrand\ contains
only one box:
\medskip
\centerline{{\epsfxsize=0.30\hsize\epsfbox{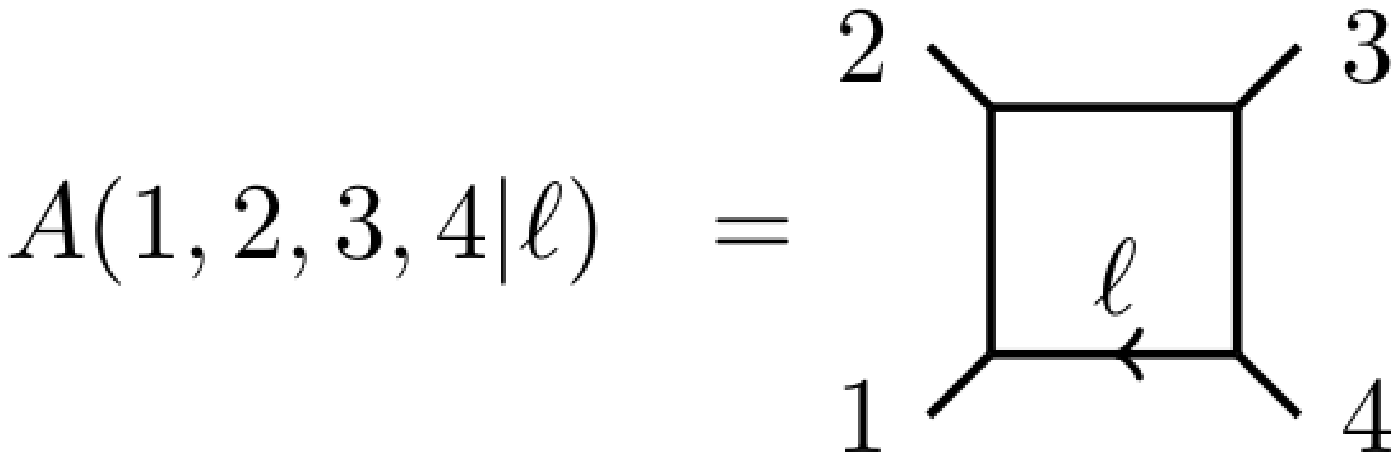}}}
\smallskip
\noindent
and its pure spinor superspace expression is given by
\eqn\fourptS{
A(1,2,3,4|\ell) = { V_1T_{2,3,4} \over \ell^2 (\ell-k_1)^2 (\ell-k_{12})^2 (\ell-k_{123})^2}\,.
}
This simple example provides the essential intuition on setting up a mapping
between boxes and superspace expressions. It will be seen below
that a general higher-point box with corners encoded by multiparticle labels $A$, $B$, $C$ and $D$ (having
the structure of cubic-graph tree subamplitudes) is mapped to $V_A T_{B,C,D}$.

\subsec Local form of the one-loop five-point SYM integrand

\subseclab\secfourtwo

\noindent In the color-ordered SYM five-point one-loop integrand the only cubic graphs compatible with
the no-triangle property are boxes and pentagons:
\smallskip
\centerline{{\epsfxsize=0.70\hsize\epsfbox{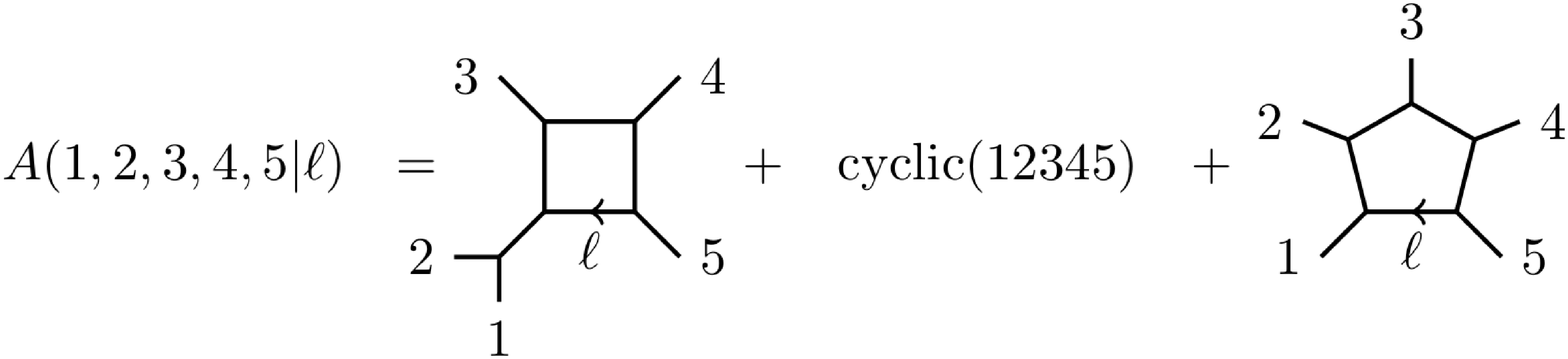}}}


\ifig\FigBoxOneTwo{The explicit superspace representation of the five-point box numerators.
The form of the numerators depends on the location of leg $1$, and the origin of this difference
is due to the string one-loop amplitude prescription \looppresc\ fixing the position of its first vertex operator $V_1$.}
{\epsfxsize=0.70\hsize\epsfbox{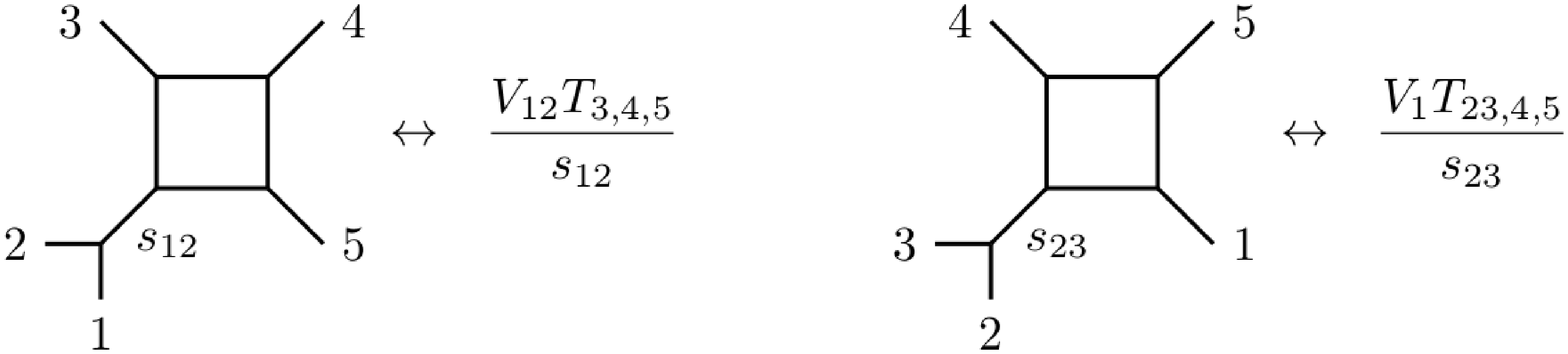}}

\noindent Hence, we split its integrand according to box and pentagon contributions
\eqn\OneLoopFive{
A(1,2,3,4,5|\ell) =  A_{\rm box}(1,2,3,4,5) + A_{\rm pent}(1,2,3,4,5|\ell) \ ,
}
where $A_{\rm box}(\ldots)$ is independent on $\ell$ and $A_{\rm
pent}(\ldots)$ carries at most linear $\ell$-dependence. For the numerators of the boxes, the
requirement \principle\ yields a natural pure spinor superspace representation seen in \FigBoxOneTwo. Since triangles in the
BRST variation cannot be compensated by any other $p$-gon diagram with $p\geq 4$, their BRST variation
must cancel the propagator $(k_{i}+k_{i+1})^2$ of their external tree-level subdiagram:
\eqnn\Boxfive
$$\eqalignno{
A_{\rm box}(1,2,3,4,5) &= {V_{12} T_{3,4,5} \over (k_1+k_2)^2 \ell^2  (\ell-k_{12})^2 (\ell-k_{123})^2 (\ell-k_{1234})^2} \cr
&+  {V_{1} T_{23,4,5} \over (k_2+k_3)^2 \ell^2 (\ell-k_1)^2 (\ell-k_{123})^2 (\ell-k_{1234})^2} \cr
&+  {V_{1} T_{2,34,5} \over (k_3+k_4)^2 \ell^2 (\ell-k_1)^2 (\ell-k_{12})^2  (\ell-k_{1234})^2}  & \Boxfive \cr
&+  {V_{1} T_{2,3,45} \over (k_4+k_5)^2 \ell^2 (\ell-k_1)^2 (\ell-k_{12})^2 (\ell-k_{123})^2 }\cr
&+  {V_{51} T_{2,3,4} \over (k_1+k_5)^2 (\ell-k_1)^2 (\ell-k_{12})^2 (\ell-k_{123})^2 (\ell-k_{1234})^2} \ .
}$$
These expressions can be thought of as descending from a string calculation where particle
one enters through an unintegrated vertex $V_1$. That is why the first leg always enters in the form
$V_{1\ldots}$ and ambiguities such as $V_{1} T_{23,4,5}\leftrightarrow V_{23} T_{1,4,5}$ do not arise.

On the other hand, the pentagon numerator $N^{(5)}_{1|2,3,4,5}(\ell)$ in
\eqnn\Pentfive
$$\eqalignno{
A_{\rm pent}(1,2,3,4,5|\ell) &= { N^{(5)}_{1|2,3,4,5}(\ell) \over \ell^2 (\ell-k_1)^2 (\ell-k_{12})^2 (\ell-k_{123})^2 (\ell-k_{1234})^2} &\Pentfive
}$$
must be designed such that $\ell$-dependent propagators cancel in its BRST variation. This interlocks
the vector $\ell_m V_1 T^m_{2,3,4,5}$ with the scalars in
\eqnn\specPent
$$\eqalignno{
N^{(5)}_{1|2,3,4,5}(\ell) &\equiv
\ell_m  V_1 T^m_{2,3,4,5} + {1\over 2} \big[V_{12} T_{3,4,5} + (2\leftrightarrow 3,4,5) \big] &\specPent\cr
&\quad{} + {1\over 2} \big[V_{1} T_{23,4,5} + (2,3|2,3,4,5) \big]\,.
}$$
In contrast to the box numerators of the form $V_A T_{B,C,D}$, the pentagon numerator \specPent\
depends on the ordering of the external legs $2, 3,4,5$ through the signs in the scalar part. The variations \BRSTTs\ and \BRSTTv\ of the
scalar and vectorial building blocks imply that
\eqnn\QPent
$$\eqalignno{
Q&N^{(5)}_{1|2,3,4,5}(\ell) =  {1\over 2} V_1 V_2 T_{3,4,5} \big[ (\ell-k_{12})^2
- (\ell-k_1)^2 \big]+{1\over 2} V_1 V_3 T_{2,4,5} \big[ (\ell-k_{123})^2 - (\ell-k_{12})^2 \big]
\cr
& \ \ \ \ \ \quad{} + {1\over 2} V_1 V_4 T_{2,3,5} \big[ (\ell-k_{1234})^2 - (\ell-k_{123})^2
\big] +{1\over 2} V_1 V_5 T_{2,3,4} \big[\ell^2 - (\ell-k_{1234})^2 \big] &\QPent
}$$
is compatible with \principle\ and precisely cancels the BRST variation of the boxes in \Boxfive:
\eqnn\Qboxes
$$\eqalignno{
QA_{\rm box}(1,2,3,4,5) &= { V_1 V_2 T_{3,4,5} \over 2 \ell^2 (\ell-k_{123})^2 (\ell-k_{1234})^2} \left( {1\over (\ell-k_{12})^2} - {1\over (\ell-k_{1})^2} \right)
\cr
&+
{ V_1 V_3 T_{2,4,5} \over 2 \ell^2 (\ell-k_{1})^2 (\ell-k_{1234})^2} \left( {1\over (\ell-k_{123})^2} - {1\over (\ell-k_{12})^2} \right)
\cr
&+
{ V_1 V_4 T_{2,3,5} \over 2 \ell^2 (\ell-k_{1})^2 (\ell-k_{12})^2} \left( {1\over (\ell-k_{1234})^2} - {1\over (\ell-k_{123})^2} \right)
\cr
&+
{ V_1 V_5 T_{2,3,4} \over 2 (\ell-k_1)^2 (\ell-k_{12})^2 (\ell-k_{123})^2} \left( {1\over \ell^2} - {1\over (\ell-k_{1234})^2} \right) \ .
&\Qboxes\cr
}$$
Hence, the interplay between boxes and pentagons renders the superspace integrand \OneLoopFive\ BRST
invariant, $QA(1,2,3,4,5|\ell) = 0$. Note the factor of ${1\over 2}= {s_{ij} \over (k_i+k_j)^2}$ by the
convention \mandconv\ for Mandelstam invariants. In section \secfivetwo, we will present an alternative
representation of the five-point amplitude where BRST invariance is manifest. And it will be shown in
appendix~A that the integrand \OneLoopFive\ is confirmed by the
field-theory limit of the superstring amplitude.

\subsec Shorthand notations

\subseclab\secfourthree

\noindent The length of the five-point box contribution in \Boxfive\ and the superspace presentation of the pentagon
numerator \specPent\ motivate to introduce a compact notation for both numerators and $\ell$-dependent
propagators before addressing the six--point amplitude. 


\break

\subsubsec $\ell$-dependent propagators

\medskip
\figflow{-1.4 truein}{1.6 truein}{{\epsfxsize=1.00\hsize\epsfbox{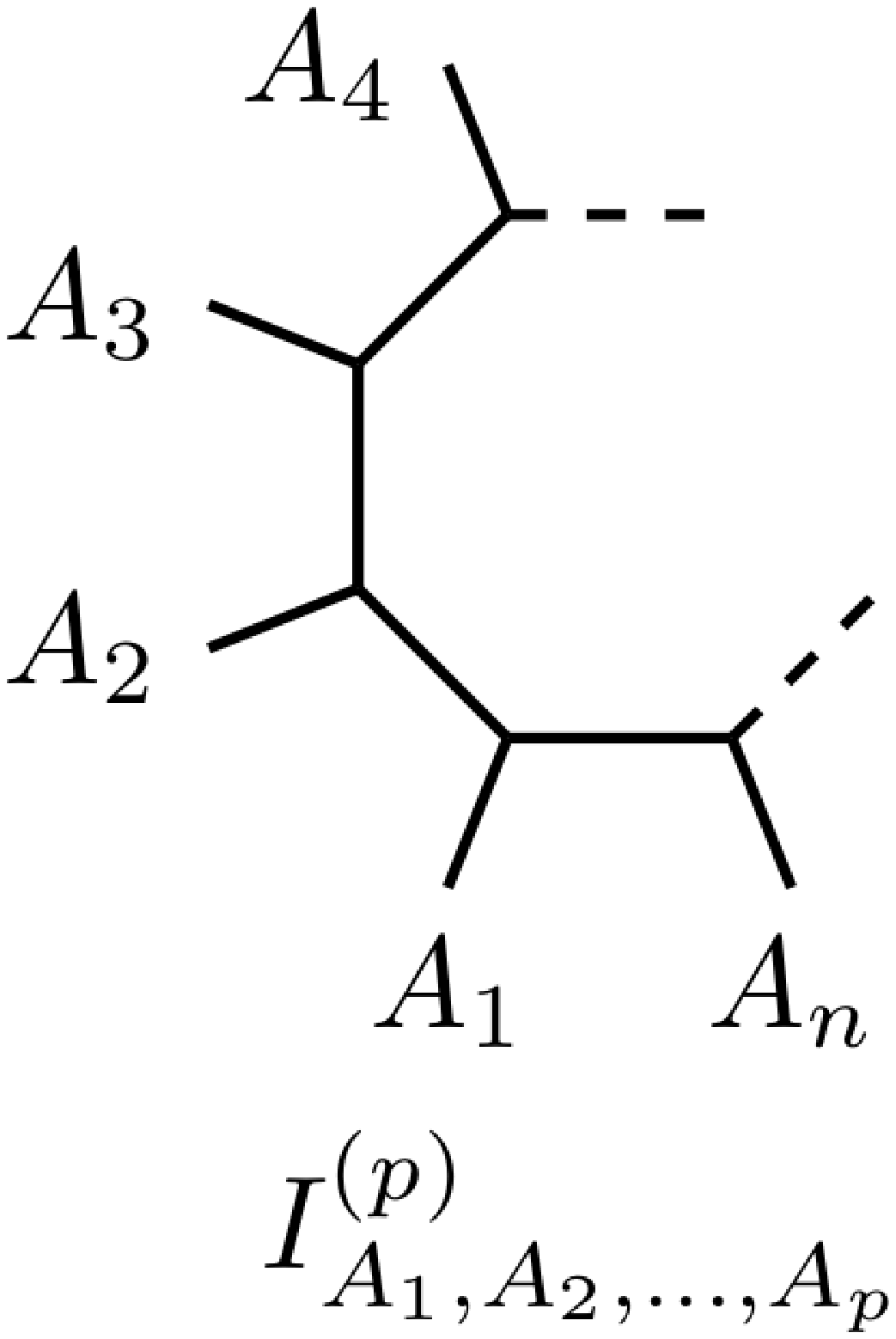}}\vfill}
\noindent A general $p$-gon diagram involves multiparticle tree subdiagrams $A_1,A_2,\ldots,A_p$ in its corners
and we define a shorthand
$I^{(p)}_{A_1,A_2,\ldots,A_p}$ for the $p$ propagators which depend on $\ell$.
The figure on the right does not yet specify the position of the loop momentum $\ell$ in the diagram. The five-point
amplitude \OneLoopFive\ was presented with uniform propagators
$\ell^2,(\ell-k_1)^2,\ldots,(\ell-k_{1234})^2$, i.e. without any shifts $\ell \rightarrow \ell+k_i$ of
integration variables between different diagrams. This was crucial to demonstrate BRST invariance at
the level of the integrand. Hence, we pick the following convention to freeze the freedom of redefining
$\ell$:
\eqn\freeze{
\hbox{\it the only $\ell$-dependent propagators  in $A(1,2,\ldots,n|\ell)$  are  $(\ell-k_{12\ldots j})^2$
with $1\leq j \leq n$.}
}
The explicit formula for the $I^{(p)}_{A_1,A_2,\ldots,A_p}$ in the above figure therefore requires to
specify the position of leg 1 within the first massive corner $A_1\equiv B1C$ :
\eqn\pgon{
I^{(p)}_{B1C,A_2,A_3,\ldots,A_p} \equiv {1\over (\ell-k_{1C})^2(\ell-k_{1CA_2})^2(\ell-k_{1CA_2A_3})^2 \ldots (\ell-k_{1CA_2 A_3\ldots A_p})^2} \ .
}
Whenever the first corner $A_1$ does not contain particle $n$ and starts with particle $1$, we have $B= \emptyset$ and obtain
$(\ell-k_{1CA_2 A_3\ldots A_p})^2 = \ell^2$ by momentum conservation, e.g.
\eqn\simplegon{
I^{(4)}_{12,34,5,6} \equiv {1\over \ell^2 (\ell-k_{12})^2(\ell-k_{1234})^2(\ell-k_{12345})^2} \ . 
}
However, in cases with $B\neq \emptyset$ where a massive corner encompasses both legs 1 and $n$,
absence of the $\ell^2$ propagator will later
on play a crucial role for the hexagon anomaly, e.g.
\eqn\diffigon{
I^{(5)}_{61,2,3,4,5} \equiv {1\over (\ell-k_1)^2 (\ell-k_{12})^2 (\ell - k_{123})^2 (\ell-k_{1234})^2(\ell-k_{12345})^2} \ .
}

\subsubsec Box and pentagon numerators

The form of the box numerators in \Boxfive\ strongly suggests the general pattern when arbitrary tree
subdiagrams are attached to the four corners. Multiparticle labels such as $A=a_1 a_2\ldots a_p$ allow
for the following general formula,
\eqn\genBox{
N^{(4)}_{A|B,C,D} \equiv V_{A} T_{B,C,D} \ ,
}
with $T_{B,C,D}$ given by \Tsc. The interpretation of all the multiparticle superfields in $V_{A} T_{B,C,D}$ as off-shell tree
subdiagrams as seen in \figone\ leads to the desired box diagram for the right-hand side of \genBox.

\ifig\figFive{On-shell diagrams represented by $V_AT_{B,C,D}$ connect four off-shell subdiagrams.
Propagators such as $s_{a_1a_2}$ and $s_{a_1a_2a_3}$ are suppressed on the left-hand side.}
{\epsfxsize=0.65\hsize\epsfbox{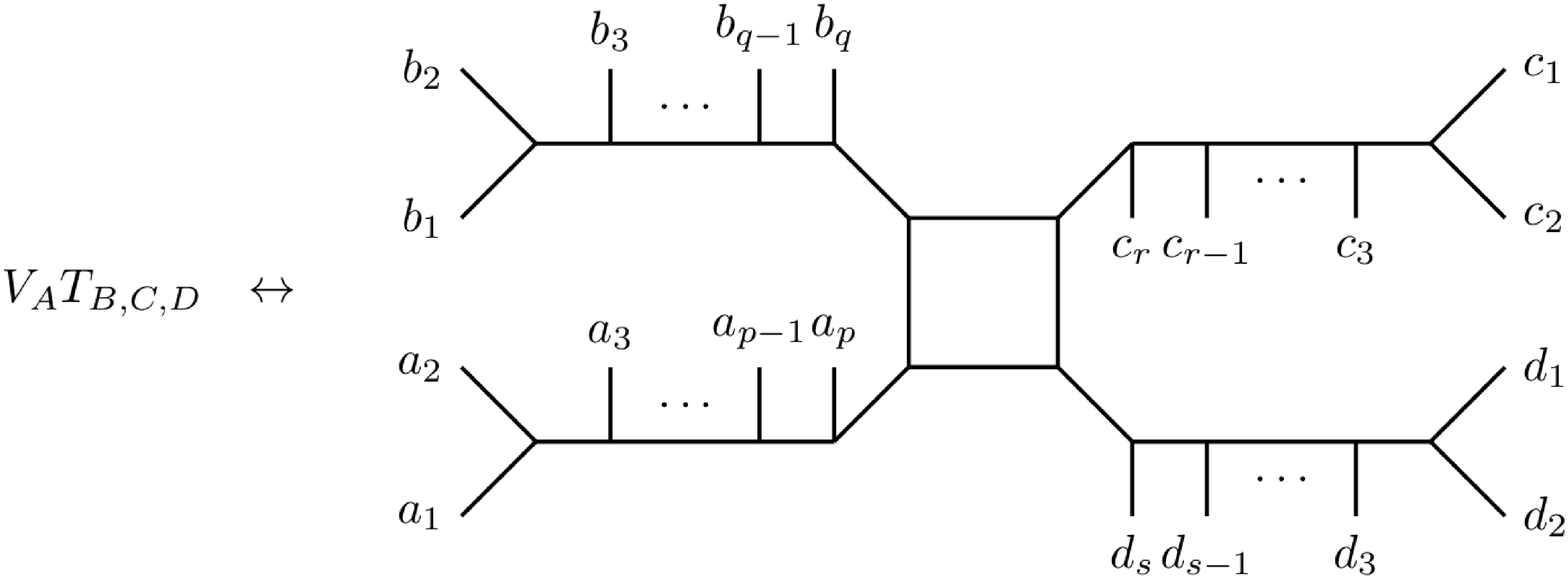}}

According to the BRST covariant transformation of both $V_A$ and $T_{B,C,D}$ -- see \QBRSTV\
and \BRSTTs\ for examples -- the expression \genBox\ for box numerators is compatible with both
\principle\ and the no-triangle property.

A uniform description of pentagon numerators can be achieved using the bracketing convention of BRST
blocks explained in appendix A of \eombbs: Multiparticle indices $B=b_1b_2\ldots b_p$ associated with
the local superfields $[A_\alpha^B,A_B^m,W^\alpha_B,F^{mn}_B]$ can be combined through an antisymmetric
bracket $[B_1,B_2] \rightarrow B_3$, e.g.
\eqn\bracket{
V_{[1,2]}\equiv V_{12} \ , \ \ \ \ V_{[12,3]}\equiv V_{123} \ , \ \ \ \ V_{[123,4]}\equiv V_{1234} \ , \ \ \ \ V_{[12,34]}\equiv V_{1234}-V_{1243} \ .
}
The diagram associated with the superfield $V_{[B_1,B_2]}$ is obtained by connecting the off-shell legs
of $B_1,B_2$ through a cubic vertex,
\smallskip
\centerline{{\epsfxsize=0.70\hsize\epsfbox{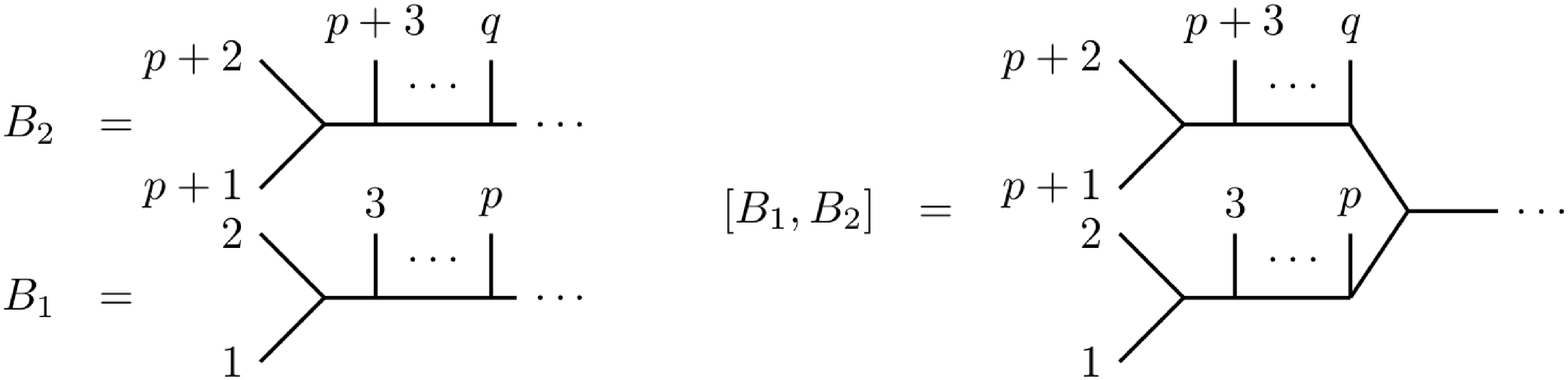}}}

\noindent In this convention, the five-point pentagon numerator \specPent\ can be generalized to
\eqnn\genPent
$$\eqalignno{
N^{(5)}_{A|B,C,D,E}(\ell) &\equiv
\ell_m  V_A T^m_{B,C,D,E} + {1\over 2} \big[V_{[A,B]} T_{C,D,E} + (B\leftrightarrow C,D,E) \big] &\genPent\cr
&\quad{} + {1\over 2} \big[V_{A} T_{[B,C],D,E} + (B,C|B,C,D,E) \big]\,.
}$$
In contrast to the box numerator \genBox, this pentagon numerator depends on the ordering of the
external trees $B,C,D,E$. The antisymmetric components such as
\eqn\penttobox{
N^{(5)}_{A|B,C,D,E}(\ell) - N^{(5)}_{A|C,B,D,E}(\ell) = N^{(4)}_{A|[B,C],D,E}
}
reproduce box numerators \genBox\ with a bracket $[B,C]$ as exemplified in \bracket\ in one of the
multiparticle slots. This is in lines with the BCJ duality between color and kinematics \BCJ\ discussed
in section~\secsix.

\subsubsec Representing external tree-level propagators

In order to compactly describe the $\ell$-independent propagators in the external tree-level
subdiagrams, it is convenient to use the notation
\eqn\BGone{
{\cal N}_{12|3,\ldots,p+1}^{(p)} \equiv {N^{(p)}_{12|3,\ldots,p+1}\over s_{12}}  \ , \ \ \ \ 
{\cal N}_{1|23,\ldots,p+1}^{(p)} \equiv {N^{(p)}_{1|23,\ldots,p+1}\over s_{23}} 
\
}
for five-point boxes and six-point pentagons. Likewise, the presentation of six-point boxes benefits
from the shorthands such as
\eqnn\BGtwo
$$\eqalignno{
{\cal N}_{123|4,5,6}^{(4)}& \equiv {N_{123|4,5,6}^{(4)}\over s_{12}s_{123}} + {N_{321|4,5,6}^{(4)}\over s_{23}s_{123}}  \ , \ \ \ \ {\cal N}^{(4)}_{12|34,5,6} \equiv {N^{(4)}_{12|34,5,6}\over s_{12}s_{34}}  \cr
{\cal N}^{(4)}_{1|234,5,6}& \equiv {N^{(4)}_{1|234,5,6}\over s_{23}s_{234}} + {N^{(4)}_{1|432,5,6}\over s_{34}s_{234}}  \ , \ \ \ \ {\cal N}^{(4)}_{1|23,45,6} \equiv {N^{(4)}_{1|23,45,6}\over s_{23}s_{45}}  \ . &\BGtwo
}$$
They streamline the pairing of one-mass boxes and incorporate the concept of Berends--Giele currents
\eombbs, see \BGboxes. 

With the above shorthands, the five-point amplitude determined by \OneLoopFive\ to
\specPent\ can be cast into the compact form
\eqnn\compfive
$$\eqalignno{
A(1,2,3,4,5|\ell) &= N^{(5)}_{1|2,3,4,5}(\ell) I_{1,2,3,4,5}^{(5)} + \half \Big[ {\cal N}^{(4)}_{12|3,4,5} I_{12,3,4,5}^{(4)}  
+{\cal N}^{(4)}_{1|23,4,5}  I_{1,23,4,5}^{(4)}    \cr
&\ \ + {\cal N}^{(4)}_{1|2,34,5}  I_{1,2,34,5}^{(4)} +{\cal N}^{(4)}_{1|2,3,45}  I_{1,2,3,45}^{(4)} + {\cal N}^{(4)}_{51|2,3,4}  I_{51,2,3,4}^{(4)}   \Big]
 \ .  & \compfive  
}$$
The six-point amplitude will now be presented along similar lines.

\ifig\BGboxes{The pure spinor superspace description of two one-mass box graphs in the
 six-point amplitude is compactly
 captured by Berends--Giele numerators such as ${\cal N}^{(4)}_{123|4,5,6}$ defined in \BGtwo. The expansion of
 the above superspace expression is given in \twoBoxes.}
{\epsfxsize=0.60\hsize\epsfbox{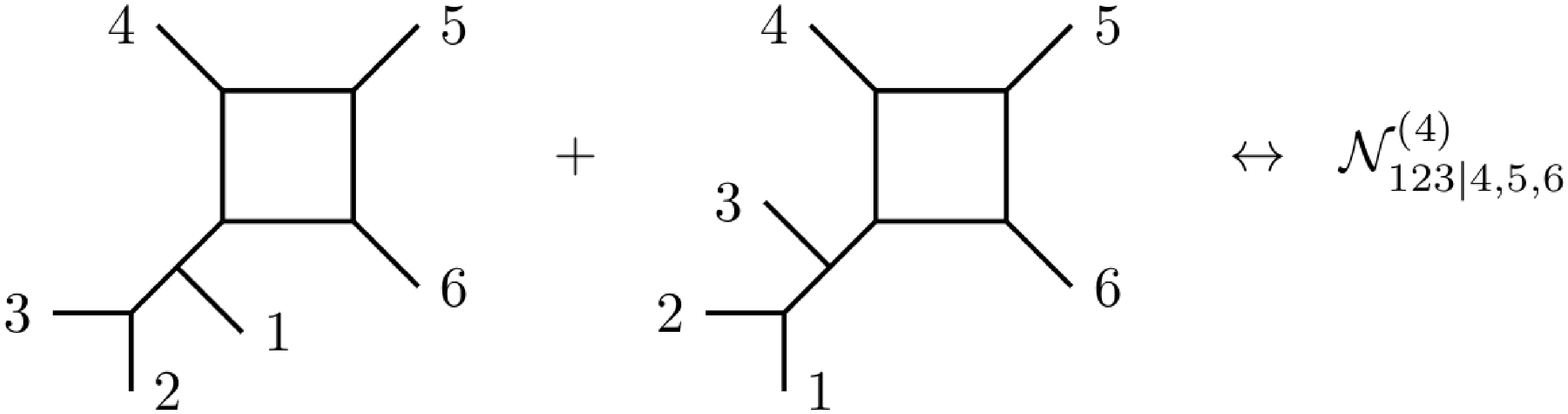}}


\subsec Local form of the one-loop six-point SYM integrand

\subseclab\secfourfour

\noindent
The color-ordered six-point SYM integrand will be constructed in a local form following the
propagator-cancellation principle \principle\ for all its numerators. Boxes, pentagons and the hexagon are analyzed
separately to get an overview of their BRST interplay,
\eqn\OneLoopSix{
A(1,2,\ldots,6|\ell) = A_{\rm box}(1,2,\ldots,6)
+ A_{\rm pent}(1,2,\ldots,6|\ell) + A_{\rm hex}(1,2,\ldots,6|\ell) \ .
}
The highest power of loop momentum in the numerators of $A_{\rm pent}(\ldots)$ and $ A_{\rm hex}(\ldots)$ is one and two, respectively.

\subsubsec Six-point boxes

The 21 boxes in the cubic-graph representation
of the six-point amplitude can be described by the following
15 Berends--Giele currents \BGtwo\ and box propagators \pgon,
\eqnn\sixBoxes
$$\eqalignno{
A_{\rm box}(1,2,\ldots,6)&= {1\over 4}\Big[
\Box123,4,5,6,
+ \Box1,234,5,6,
+ \Box1,2,345,6, \cr
&\quad{} + \Box1,2,3,456,
+ \Box561,2,3,4,
+ \Box612,3,4,5, \cr
&\quad{}+ \Box12,34,5,6,
+ \Box12,3,45,6,
+ \Box12,3,4,56,  &\sixBoxes\cr
&\quad{}+ \Box1,23,45,6,
+ \Box1,23,4,56,
+ \Box1,2,34,56, \cr
&\quad{}+ \Box61,23,4,5,
+ \Box61,2,34,5,
+ \Box61,2,3,45, \Big] \ .
}$$
Given the universal form of box numerators \genBox\ and their Berends--Giele currents \BGtwo, the superfield
and pole content of any term in \sixBoxes\ can be straightforwardly recovered. For example, locality is evident from
\eqn\twoBoxes{\eqalign{
{\cal N}^{(4)}_{123|4,5,6} I^{(4)}_{123,4,5,6}&=
\left( {V_{123}\over  s_{12}s_{123}} + {V_{321} \over s_{23}s_{123}} \right)
 {T_{4,5,6} \over \ell^2 (\ell-k_{123})^2(\ell-k_{1234})^2(\ell-k_{12345})^2}\cr
{\cal N}^{(4)}_{12|34,5,6} I^{(4)}_{12,34,5,6}&
= { V_{12 }T_{34,5,6}  \over s_{12} s_{34} \ell^2 (\ell-k_{12})^2 (\ell-k_{1234})^2 (\ell-k_{12345})^2 }
\ .
}}
The BRST variation of the boxes is most conveniently expressed in terms of
\eqn\BGcurr{
M_{12} \equiv {V_{12} \over s_{12}} \ , \ \ \ \  M_{23,4,5} \equiv  { T_{23,4,5} \over s_{23}} \   , \ \ \ \ M_1 \equiv V_1 \ , \ \ \ M_{2,3,4} \equiv T_{2,3,4}
}
and given by
\eqnn\VarSixBoxes
$$\eqalignno{
4&QA_{\rm box}(1,2,\ldots,6) = &\VarSixBoxes \cr
&\phantom{+}\, \; M_1M_2 M_{34,5,6}\big(
I^{(4)}_{12,34,5,6}
-I^{(4)}_{1,234,5,6}
\big)
+  M_1M_2 M_{45,3,6}\big(
I^{(4)}_{12,3,45,6}
-I^{(4)}_{1,23,45,6}
\big)\cr
&        + M_1 M_2 M_{56,3,4} \big(
           I^{(4)}_{12,3,4,56}
          - I^{(4)}_{1,23,4,56}
         \big)
       + M_1 M_3 M_{45,2,6} \big(
           I^{(4)}_{1,23,45,6}
          - I^{(4)}_{1,2,345,6}
          \big)\cr
&
	+ M_1 M_3 M_{56,2,4} \big(
           I^{(4)}_{1,23,4,56}
          - I^{(4)}_{1,2,34,56}
          \big)
       + M_1 M_4 M_{23,5,6} \big(
           I^{(4)}_{1,234,5,6}
          - I^{(4)}_{1,23,45,6}
          \big)\cr
&
       + M_1 M_4 M_{56,2,3} \big(
           I^{(4)}_{1,2,34,56}
          - I^{(4)}_{1,2,3,456}
          \big)
       + M_1 M_5 M_{23,4,6} \big(
           I^{(4)}_{1,23,45,6}
	- I^{(4)}_{1,23,4,56}
          \big)\cr
&
       + M_1 M_5 M_{34,2,6} \big(
           I^{(4)}_{1,2,345,6}
          - I^{(4)}_{1,2,34,56}
          \big)
       + M_1 M_6 M_{23,4,5} \big(
           I^{(4)}_{1,23,4,56}
          - I^{(4)}_{61,23,4,5}
          \big)\cr
&
	+ M_1 M_6 M_{34,2,5} \big(
           I^{(4)}_{1,2,34,56}
          - I^{(4)}_{61,2,34,5}
          \big)
       + M_1 M_6 M_{45,2,3} \big(
           I^{(4)}_{1,2,3,456}
          - I^{(4)}_{61,2,3,45}
          \big)\cr
&
       + M_1 M_{23} M_{4,5,6} \big(
            I^{(4)}_{123,4,5,6}
          - I^{(4)}_{1,234,5,6}
          \big)
       + M_1 M_{34} M_{2,5,6} \big(
            I^{(4)}_{1,234,5,6}
          - I^{(4)}_{1,2,345,6}
          \big)\cr
&
       + M_1 M_{45} M_{2,3,6} \big(
            I^{(4)}_{1,2,345,6}
          - I^{(4)}_{1,2,3,456}
          \big)
       + M_1 M_{56} M_{2,3,4} \big(
            I^{(4)}_{1,2,3,456}
          - I^{(4)}_{561,2,3,4}
          \big)\cr
&
       + M_2 M_{16} M_{3,4,5} \big(
            I^{(4)}_{612,3,4,5}
          - I^{(4)}_{61,23,4,5}
          \big)
	+ M_3 M_{12} M_{4,5,6} \big(
            I^{(4)}_{12,34,5,6}
          - I^{(4)}_{123,4,5,6}
          \big)\cr
&
       + M_3 M_{16} M_{2,4,5} \big(
            I^{(4)}_{61,23,4,5}
          - I^{(4)}_{61,2,34,5}
          \big)
       + M_4 M_{12} M_{3,5,6} \big(
            I^{(4)}_{12,3,45,6}
          - I^{(4)}_{12,34,5,6}
          \big)\cr
&
       + M_4 M_{16} M_{2,3,5} \big(
            I^{(4)}_{61,2,34,5}
          - I^{(4)}_{61,2,3,45}
          \big)
       + M_5 M_{12} M_{3,4,6} \big(
            I^{(4)}_{12,3,4,56}
          - I^{(4)}_{12,3,45,6}
          \big)\cr
&
       + M_5 M_{16} M_{2,3,4} \big(
            I^{(4)}_{61,2,3,45}
          - I^{(4)}_{561,2,3,4}
          \big)
       + M_6 M_{12} M_{3,4,5} \big(
            I^{(4)}_{612,3,4,5}
          - I^{(4)}_{12,3,4,56}
          \big) \ .
}$$
We will next see how this BRST variation of the boxes is cancelled by pentagons.

\subsubsec Six-point pentagons

The pentagon content of the six-point one-loop integrand is given by
\eqnn\gamo
$$\eqalignno{
A_{\rm pent}&(1,2,\ldots,6|\ell)  \equiv
\half \Big[ \Pentagon12,3,4,5,6,
+ \Pentagon1,23,4,5,6,
\cr
&\quad{} + \Pentagon1,2,34,5,6,+  \Pentagon1,2,3,45,6,
+  \Pentagon1,2,3,4,56,\cr
&\quad{} + I^{(5)}_{61,2,3,4,5}\,\big( {\cal N}^{(5)}_{61|2,3,4,5}(\ell+k_6) - V_1 J_{6|2,3,4,5}\big) \Big] \ . &\gamo
}$$
The new building block $J_{6|2,3,4,5}$ will be defined below, and the first five numerators follow the universal form \genPent\ of pentagons such as
\eqnn\firstpentagon
$$\eqalignno{
{\cal N}^{(5)}_{12|3,4,5,6}(\ell) &= {1 \over s_{12}} \Big\{
\ell_m V_{12} T^m_{3,4,5,6}
+ {1\over 2} \big[ V_{123} T_{4,5,6} + (3\leftrightarrow 4,5,6) \big]\cr
&{} \ \ \ \ \ \ \ \ + {1\over 2} \big[ V_{12} T_{34,5,6} + (3,4|3,4,5,6) \big] \Big\}\,.&\firstpentagon
}$$
The tree-level propagator stems from \BGone, and we identify $V_{123} \equiv V_{[12,3]}$. It is easy to check from \QBRSTV,  \BRSTTs\ and \BRSTTv\ that the BRST variation of \firstpentagon\ cancels propagators,
\eqnn\gamp
$$\eqalignno{
&Q{\cal N}^{(5)}_{12|3,4,5,6}(\ell) = - V_2 N^{(5)}_{1|3,4,5,6}(\ell)
+{1 \over 2} V_1 (V_{23} T_{4,5,6} \! +\! V_{24} T_{3,5,6} \!+\! V_{25} T_{3,4,6} \!+\! V_{26} T_{3,4,5} ) &\gamp\cr
&\, - M_{12} M_3 M_{4,5,6} \half \big[ (\ell-k_{12})^2 - (\ell- k_{123})^2 \big]- M_{12} M_4 M_{3,5,6}\half \big[ (\ell-k_{123})^2 - (\ell- k_{1234})^2 \big] \cr
&\, - M_{12} M_5 M_{3,4,6}\half \big[ (\ell-k_{1234})^2 - (\ell- k_{12345})^2 \big] - M_{12} M_6 M_{3,4,5}\half \big[ (\ell-k_{12345})^2 - \ell^2 \big] \ ,\cr
}$$
in lines with \principle.

\ifig\effell{Diagrammatic justification for the shift of loop momentum in
$N^{(5)}_{61|2,3,4,5}(\ell+k_6)$: The shifted momentum $\ell+k_6$ occurs in the pentagon edge adjacent
to the tree-level subdiagram subtending particles $6$ and $1$.}
{\epsfxsize=0.19\hsize\epsfbox{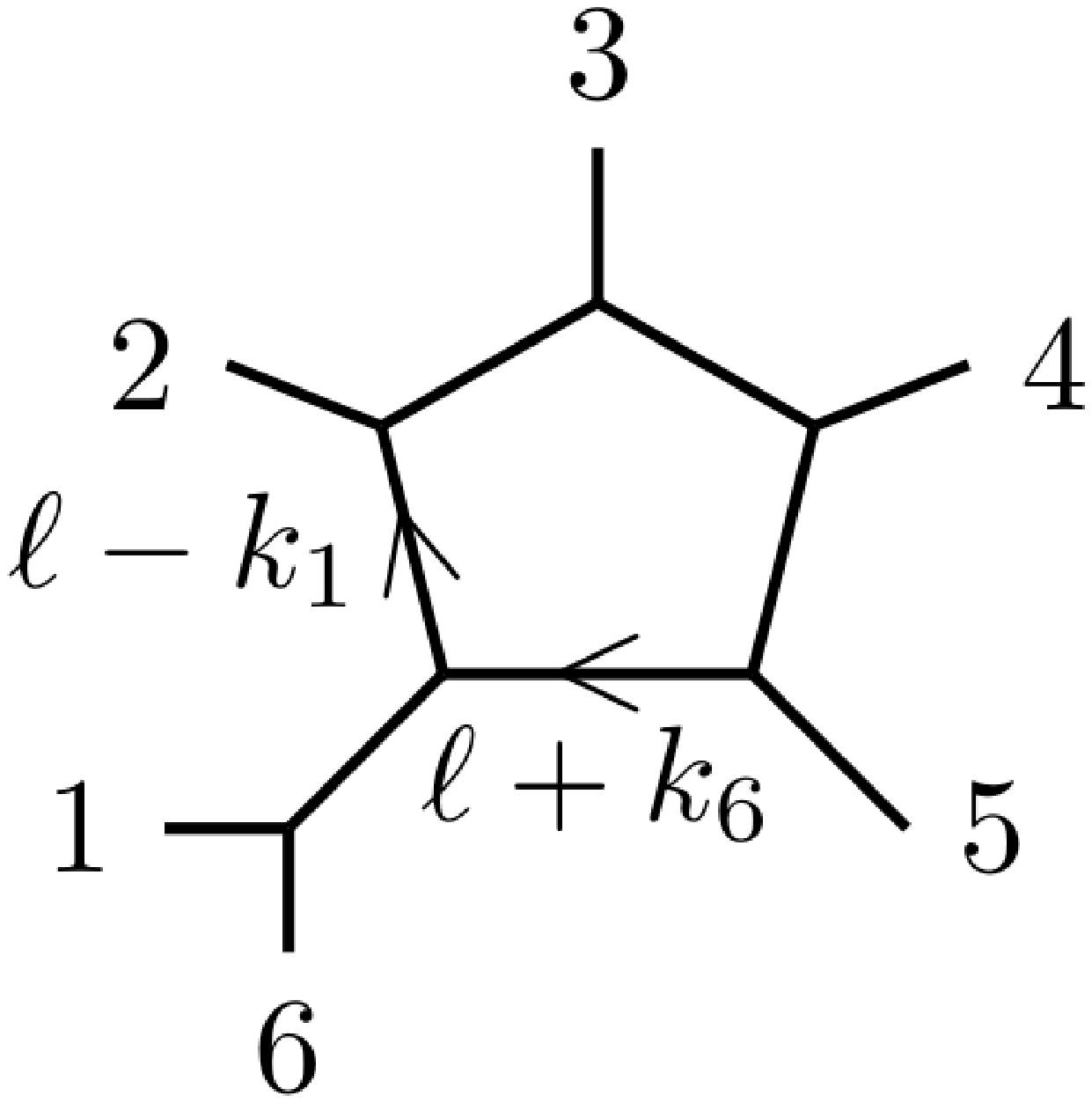}}

The last pentagon in \gamo\ requires further explanation since its numerator deviates from the naive
expectation $N^{(5)}_{61|2,3,4,5}(\ell)$ by $-V_{16}k_m^6 T^m_{2,3,4,5} - V_1 s_{16} J_{6|2,3,4,5}$
with
\eqn\defJ{
J_{6|2,3,4,5} \equiv {1\over 2} A_m^6 (T^m_{2,3,4,5} + W^m_{2,3,4,5}) \ .
}
The first extra term can be absorbed into a redefinition of the loop momentum to yield
$N^{(5)}_{61|2,3,4,5}(\ell+k_6)- V_1 s_{16} J_{6|2,3,4,5}$. As shown in \effell, the effective
loop momentum $\ell+k_6$ is determined by the pentagon edge adjacent to the cubic tree subdiagram
represented by $V_{61}$. The second extra term $\sim J_{6|2,3,4,5}$ contributes to the hexagon anomaly
via\foot{This identity was firstly noticed in \oneloopMichael\ to express the contractions of $V_6
T^m_{2,3,4,5} $ with external momenta in terms of scalar building blocks. In the five-point context of
this reference, the anomalous contribution $Y_{2,3,4,5,6}$ drops out by momentum conservation.}
\eqn\QJ{
QJ_{6|2,3,4,5} = Y_{2,3,4,5,6} + V_6 k_m^6 T^m_{2,3,4,5} + \big[ V_{62} T_{3,4,5} + (2\leftrightarrow 3,4,5) \big] \ ,
}
with anomaly superfield $Y_{2,3,4,5,6}$ defined in \BRSTTw. Together with the shift of loop momentum
$\sim V_{16}k_m^6 T^m_{2,3,4,5}$, the variation \QJ\ ensures that the overall pentagon numerator
satisfies the criterion \principle,
\eqnn\Qlastpent
$$\eqalignno{
&Q\big[ {\cal N}^{(5)}_{61|2,3,4,5}(\ell+k_6) - V_1 J_{6|2,3,4,5} \big] =  V_6 N^{(5)}_{1|2,3,4,5}(\ell) &\Qlastpent \cr
&+{1 \over 2} V_1 (V_{62} T_{3,4,5} \!+\! V_{63} T_{2,4,5}\! +\! V_{64} T_{2,3,5} \! +\! V_{65} T_{2,3,4} )+ M_{16} M_2 M_{3,4,5} \half \big[ (\ell-k_{1})^2 - (\ell- k_{12})^2 \big]\cr
&\quad{} + M_{16} M_3 M_{2,4,5}\half \big[ (\ell-k_{12})^2 - (\ell- k_{123})^2 \big]  + M_{16} M_4 M_{2,3,5}\half \big[ (\ell-k_{123})^2 - (\ell- k_{1234})^2 \big]\cr
&\quad{}+ M_{16} M_5 M_{2,3,4}\half \big[ (\ell-k_{1234})^2 - (\ell-k_{12345})^2 \big] \ ,
}$$
which would be violated by the naive choice $N^{(5)}_{61|2,3,4,5}(\ell)$. Generally speaking, the
choice of loop momentum \freeze\ requires redefinitions in any $(n\geq 5)$-gon numerator where an
external tree involves particles $\ldots n1\ldots$ and thereby removes the $\ell^2$ propagator. The
superfield $J_{6|2,3,4,5}$ can be viewed as the BRST completion of the shift of loop momentum.

The BRST variation of the remaining pentagons takes a form similar to \gamp\ and \Qlastpent. This
compensates for the BRST variation \VarSixBoxes\ of the boxes by cancellation of propagators such as
\eqn\fewPents{
(\ell - k_{12})^2 I^{(5)}_{12,3,4,5,6} = I^{(4)}_{123,4,5,6},\quad
(\ell - k_{123})^2 I^{(5)}_{12,3,4,5,6} = I^{(4)}_{12,34,5,6}\,.
}
By adding up the contributions of all box and pentagon diagrams, we find
\eqn\gamt{
4QA_{\rm pent}(1,2,3,4,5,6|\ell) = - 4QA_{\rm box}(1,2,3,4,5,6)  \ \ \ \ \ \ \ \ \ \ \ \ \ \ \ \ \ \ \ \ \ \ \ \ \ \ \ \ \ \  \ \ \ \ \  \ \ \ \ \  \ \ \ \ \
}
$$\abovedisplayskip=-2pt
\eqalignno{
& + 2 V_1 Y_{2,3,4,5,6} I^{(5)}_{61,2,3,4,5}
  + 2 V_2 N^{(5)}_{1|3,4,5,6}(\ell) \big[ I^{(5)}_{1,23,4,5,6} - I^{(5)}_{12,3,4,5,6} \big]\cr
& + 2 V_3N^{(5)}_{1|2,4,5,6}(\ell) \big[ I^{(5)}_{1,2,34,5,6} - I^{(5)}_{1,23,4,5,6} \big]
  + 2 V_4 N^{(5)}_{1|2,3,5,6}(\ell) \big[ I^{(5)}_{1,2,3,45,6} - I^{(5)}_{1,2,34,5,6} \big] \cr
& + 2V_5 N^{(5)}_{1|2,3,4,6}(\ell) \big[ I^{(5)}_{1,2,3,4,56} - I^{(5)}_{1,2,3,45,6} \big]
  + 2 V_6 N^{(5)}_{1|2,3,4,5}(\ell) \big[ I^{(5)}_{61,2,3,4,5} - I^{(5)}_{1,2,3,4,56} \big]  \cr
& + V_1 V_{23} T_{4,5,6} \big[ I^{(5)}_{12,3,4,5,6} - I^{(5)}_{1,2,34,5,6} \big]
  +  V_1 V_{34} T_{2,5,6} \big[ I^{(5)}_{1,23,4,5,6} - I^{(5)}_{1,2,3,45,6} \big] \cr
& +  V_1 V_{45} T_{2,3,6} \big[ I^{(5)}_{1,2,34,5,6} - I^{(5)}_{1,2,3,4,56} \big]
  +   V_1 V_{56} T_{2,3,4} \big[ I^{(5)}_{1,2,3,4,56} - I^{(5)}_{61,2,3,4,5} \big]  \cr
& +  V_1 V_{24} T_{3,5,6} \big[ I^{(5)}_{12,3,4,5,6} - I^{(5)}_{1,23,4,5,6} + I^{(5)}_{1,2,34,5,6} - I^{(5)}_{1,2,3,45,6} \big] \cr
& +  V_1 V_{25} T_{3,4,6} \big[ I^{(5)}_{12,3,4,5,6} - I^{(5)}_{1,23,4,5,6} + I^{(5)}_{1,2,3,45,6} - I^{(5)}_{1,2,3,4,56} \big] \cr
& +   V_1 V_{26} T_{3,4,5} \big[ I^{(5)}_{12,3,4,5,6} - I^{(5)}_{1,23,4,5,6} + I^{(5)}_{1,2,3,4,56} - I^{(5)}_{61,2,3,4,5} \big] \cr
& +   V_1 V_{35} T_{2,4,6} \big[ I^{(5)}_{1,23,4,5,6} - I^{(5)}_{1,2,34,5,6} + I^{(5)}_{1,2,3,45,6} - I^{(5)}_{1,2,3,4,56} \big] \cr
& +  V_1 V_{36} T_{2,4,5} \big[ I^{(5)}_{1,23,4,5,6} - I^{(5)}_{1,2,34,5,6} + I^{(5)}_{1,2,3,4,56} - I^{(5)}_{61,2,3,4,5} \big] \cr
& +   V_1 V_{46} T_{2,3,5} \big[ I^{(5)}_{1,2,34,5,6} - I^{(5)}_{1,2,3,45,6} + I^{(5)}_{1,2,3,4,56} -
I^{(5)}_{61,2,3,4,5} \big]\,.
}$$
As mentioned before, the above terms beyond the first line come from cancellations of external tree propagators and
must be cancelled by the BRST variation of the hexagon.

\subsubsec Six-point hexagon

The six-point hexagon whose BRST variation cancels the terms in \gamt\ is given by,
\eqn\SixHexagon{
A_{\rm hex}(1,2,3,4,5,6|\ell) = I^{(6)}_{1,2,3,4,5,6}\,N^{(6)}_{1|2,3,4,5,6}(\ell) \,,
}
where $N^{(6)}_{1|2,3,4,5,6}(\ell) \equiv n^{(6)}_{1|2,3,4,5,6}(\ell) + n^{(6)}_{1|2,3,4,5,6}$
and all the $\ell$-dependent part of the local hexagon numerator is represented by
\eqnn\hexNum
$$\eqalignno{
n^{(6)}_{1|2,3,4,5,6}(\ell) &\equiv \half \ell_m \ell_n V_1 T^{mn}_{2,3,4,5,6}
+ \half \ell_m \big[ V_{12} T^m_{3,4,5,6} + (2\leftrightarrow 3,4,5,6 )\big] &\hexNum\cr
& \ \ \ \ \ + \half \ell_m V_1 \big[ T_{23,4,5,6}^m + (2,3|2,3,4,5,6) \big] \ .\cr
}$$
This is analogous to the five-point pentagon \specPent\ with building blocks of higher ranks and
an additional contraction with $\ell_m$. Note that the tensor building block $T^{mn}_{2,3,4,5,6}$ defined in
\Ttens\ introduces an anomalous contribution $\sim Y_{2,3,4,5,6}$ to the BRST variation due to \BRSTTv.

The scalar hexagon, on the other hand, is determined by \principle: The $Q$ variation of \SixHexagon\
must be expressible in terms of hexagon propagators $(\ell-k_{12\ldots j})^2$ with $j=0,1,\ldots,5$.
Apart from $Y_{2,3,4,5,6}$, this is only possible if any factor of $(\ell\cdot k_j)$ in the variation
of \hexNum\ is accompanied by $-(s_{1j}+s_{2j}+\ldots+s_{j-1,j})$ to build up the difference 
\eqn\propdiff{
(\ell-k_{12\ldots j})^2 - (\ell-k_{12\ldots j-1})^2 = - 2(\ell \cdot k_j) + 2 (s_{1j}+s_{2j}+\ldots+s_{j-1,j}) \ .
}
The unique local superfield which is compatible with this requirement and constructed out of the
building blocks in section \sectwothree\ reads
\eqnn\scalarHex
$$\eqalignno{
&n^{(6)}_{1|2,3,4,5,6} \equiv {1\over 4}\big[V_1 T_{23,45,6} + (2,3|4,5|2,3,4,5,6) \big]+ {1\over 4}\big[ V_{12} T_{34,5,6} + (2|3,4|2,3,4,5,6)\big] 
   \cr
   &+ {1\over 6}\big[ (V_1 T_{234,5,6} + V_1 T_{432,5,6}) + (2,3,4|2,3,4,5,6)\big] - {1\over 12}\big[ (k^1_m\! -\! k^2_m) V_{12} T^m_{3,4,5,6} + (2 \leftrightarrow 3,4,5,6)\big]\cr
&+ {1\over 6}\big[ (V_{123} T_{4,5,6} + V_{321} T_{4,5,6}) + (2,3|2,3,4,5,6)\big] - {1\over 12}\big[ (k^2_m\! -\! k^3_m) V_1 T^m_{23,4,5,6} + (2,3|2,3,4,5,6)\big]\cr
&- {1\over 24}V_1 T^{mn}_{2,3,4,5,6}\big[k^1_m k^1_n + (1\leftrightarrow 2,3,4,5,6)\big]\ . &\scalarHex
}$$
The notation $(2,3|4,5|2,3,4,5,6)$ instructs to sum all possible ways to distribute the set of labels
$\{2,3, \ldots,6\}$ into two {\it ordered\/} sets $\{2,3\}$ and $\{4,5\}$ without double counting where
the ordering is with respect to the set $\{2,3, \ldots,6\}$. For example, $\{2,4\},\{5,3\}$ is not an
allowed distribution because it violates the ordering in the second set, and only one of
$\{2,5\},\{3,6\}$ and $\{3,6\},\{2,5\}$ enters \scalarHex\ to avoid overcounting. A similar ordering convention
holds for permutations of the set $\{3,4\}$ in $(2|3,4|2,3,4,5,6)$.

A long but straightforward analysis shows that the hexagon \SixHexagon\ has the required properties
to make the whole six-point one-loop integrand \OneLoopSix\ (naively) BRST invariant,
\eqn\NaiveSix{
QA(1,2,\ldots,6|\ell)  = \half V_1 Y_{2,3,4,5,6} ( I^{(5)}_{61,2,3,4,5} - \ell^2 I^{(6)}_{1,2,3,4,5,6}) \,.
}
The right-hand side keeps track of the anomalous contributions due to the tensor hexagon $Q\half \ell_m
\ell_n V_1 T^{mn}_{2,3,4,5,6}$ and the second line of the pentagon variation in \gamt. By inserting the
propagators in \diffigon\ and $I^{(6)}_{1,2,3,4,5,6}=\prod_{j=0}^5 (\ell-k_{12\ldots j})^{-2}$, the
variation \NaiveSix\ appears to cancel at the level of the integrand. However, the logarithmically
divergent nature of the ten-dimensional integral over \NaiveSix\ leads to subtleties to be resolved below.

\subsec The gauge anomaly of the six-point amplitude

\subseclab\secfourfive

\noindent In this subsection, we perform a worldline analysis of the anomalous BRST variation
\NaiveSix\ of the one-loop SYM six-point amplitude. The string-based formalism gives rise to the
following worldline representation of a $N$-gon integral in $D$ dimensions \worldline
\eqnn\wlone
$$\eqalignno{
\int   &  { d^D\ell\   (p + q_m \ell^m + r_{mn} \ell^m \ell^n)  \over \ell^2 (\ell-k_1)^2 (\ell-k_{12})^2 \ldots   (\ell-k_{12\ldots N-1})^2 }  = \int d^D\ell\  I^{(N)}_{1,2,\ldots,N}(p + q_m \ell^m + r_{mn} \ell^m \ell^n)&\wlone \cr
& = \pi^N  \! \int_{0}^{\infty}  \! {d t\over t} \, t^{N-D/2}  \! \! \! \! \! \! \! \!  \! \!  \! \! \int \limits_{0\leq \nu_i \leq \nu_{i+1} \leq 1}  \! \! \! \! \! \! \! \! \! \! \!  \! d \nu_2 \, d \nu_3 \, \ldots \, d \nu_n   \, \Big( p + q_m L^m + r_{mn} \Big[ L^m L^n +  { \delta^{mn}  \over 2\pi t} \Big]\Big) e^{-\pi t Q_N} \, \Big|_{\nu_1=0}
}$$
where $p,q_m$ and $r_{mn}$ are arbitrary scalars, vectors and tensors independent on $\ell$. We use the
following shorthands for the shift in loop momentum $L^m$ and the exponent $Q_N$,
\eqn\wlthree{
L^m \equiv -\sum_{i=1}^N k_i^m \nu_i
 \ , \ \ \ \ \ \ 
Q_N \equiv \sum_{1\leq i<j}^N s_{ij} (\nu_{ij}^2 - |\nu_{ij}|)  \ ,
}
where $\nu_{ij} \equiv \nu_i-\nu_j$ and $\nu_1=0$ is implicit from now on. To make contact
with the six-point anomaly, consider the following boundary term in $D=10$ dimensions,
\eqn\wlfive{
B_6 \equiv - \pi^{5} \int_{0}^{\infty} d t \ { \partial \over \partial t} \Big\{ t^{5-D/2}\! \! \! \!  \! \! \!  \! \!  \int \limits_{0\leq \nu_i\leq\nu_{i+1}\leq 1}  \! \! \! \! \! \! \! \! \!   d \nu_2 \, d\nu_3 \, \ldots \, d \nu_6 \,  e^{-\pi t Q_6} \Big\} \Big|_{D=10} = {\pi^5\over 5!} \ ,
}
which yields the volume of a five-simplex from the lower integration limit $t=0$. For
hexagons at $N=6$, the exponential in \wlone\ satisfies the differential equation
\eqn\wlfour{
-{1\over \pi } \, {\partial\over \partial t } e^{-\pi t Q_6} 
= \Big( L_m L^m + {5 \over \pi t} \Big) e^{-\pi t Q_6} + {1\over \pi t} \sum_{p=2}^6  (\partial_{\nu_p} \nu_{1p}  e^{-\pi t Q_6} )  \ ,
}
which allows for the following rewriting of the boundary term $B_6$:
\eqn\wlsix{
B_6 =  \pi^6 
 \int_{0}^{\infty} d t \  t^{5-D/2}  \! \! \!  \! \! \!  \! \! \int \limits_{0\leq\nu_i \leq \nu_{i+1} \leq 1}  \! \! \! \! \! \! \! \! \!   d \nu_2 \, \ldots \, d \nu_6 \,  \Big\{  L^2 + {5 \over  \pi t} +{1 \over \pi t} \sum_{p=2}^n \partial_{\nu_p} \nu_{1p} \Big\} \, e^{-\pi t Q_6}  \ .
}
The $\partial_{\nu_p}$ derivatives in the last term are understood to also act on $e^{-\pi t Q_6}$.
They can be evaluated as a series of boundary terms $\nu_i \rightarrow \nu_{i\pm 1}$ with $\nu_1=0$
and $\nu_7=1$ where only the upper limit of the $\nu_6$ integration remains uncancelled.
With $\nu_{17}=-1$, the result is
\eqn\wlseven{
B_6 = \pi^6 
 \int_{0}^{\infty} dt \  t^{5-D/2}  \! \! \!  \! \! \!  \! \! \int \limits_{0\leq \nu_i \leq \nu_{i+1} < 1}  \! \! \! \! \! \! \! \! \!   d \nu_2 \, \ldots \,d \nu_6 \,  \, \Big\{  L^2 + {5 \over  \pi t} -{\delta(\nu_6-1) \over \pi t} \Big\} \, e^{-\pi t Q_6}  \ .
}
The first two terms can be recognized as a tensor hexagon with $r_{mn}=\delta_{mn}$, see \wlone\ at $N=6, \ D=10$ and $p=q_m=0$. The last term, on the other hand, describes a scalar pentagon, hence we recover both integrals in the anomalous BRST variation \NaiveSix:
\eqn\wleight{
B_6 = \int d^D\ell \, \big( \ell^2 I^{(6)}_{1,2,3,4,5,6} - I^{(5)}_{61,2,3,4,5} \big)
}
Using the value of $B_6$ found in \wlfive, the BRST variation of the six-point amplitude turns out to
be a rational function in external momenta,
\eqn\wlnine{
Q \int d^D\ell\, A(1,2,3,4,5,6|\ell) = -{1\over 2} V_1Y_{2,3,4,5,6}  B_6 = -   { \pi^5 \over 240} V_1Y_{2,3,4,5,6} \ .
}
Note that the BRST anomaly $Q A(1,2,3,4,5,6|\ell) \sim V_1Y_{2,3,4,5,6}$ is equivalent to the
anomalous gauge variation $\delta_1 A(1,2,3,4,5,6|\ell) \sim \langle \Omega_1 Y_{2,3,4,5,6} \rangle$ under $\delta_1 V_1 = Q \Omega_1$ with scalar superfield $\Omega_1$,
see the appendix of \partI. This reproduces the anomaly analysis of the pure spinor superstring \anomaly.

The discussion of \ChenEVA\ is helpful to shed further light on the apparent paradox between the
formally vanishing integrand in \wleight\ and the finite result in \wlfive: In a dimensional
regularization scheme $D\rightarrow D-2\varepsilon$, anomalies in chiral gauge theories arise from
components of the loop momentum in the fractional $-2\varepsilon$ dimensions. The idea is to formally
split the $D-2\varepsilon$ dimensional loop momentum into $\ell^2_{D-2\varepsilon} = \ell^2_D+\mu^2$
with $D$ dimensional part $\ell_D$ and ``$-2\varepsilon$ dimensional'' component $\mu$. In $D-2\varepsilon$ dimensions, the tensor hexagon numerator along with $ I^{(6)}_{1,2,3,4,5,6}$ remains $\ell_D^m \ell_D^n$ rather than $\ell_{D-2\varepsilon}^m \ell_{D-2\varepsilon}^n$ since the loop momenta are contracted into the $D$ dimensional polarization tensors from the
{\it external\/} states in $T^{mn}_{2,3,4,5,6}$. The {\it internal\/} states propagating through the loop, however, give rise to
momenta $\ell_{D-2\varepsilon}$ in the propagators, so the integrand
\NaiveSix\ is proportional to $\ell_D^2 - \ell_{D-2\varepsilon}^2 = -\mu^2$.

This argument based on dimensional regularization also explains why none of the other cancellations
among $\ell$-dependent propagators in $QA(1,2,\ldots,n|\ell)$ introduces rational terms: Since
$(\ell_D\cdot k_j)=(\ell_{D-2\varepsilon}\cdot k_j)$ for $D$-dimensional external momenta, we can still rewrite
\eqn\elleps{
(\ell_D\cdot k_j) = \half \big[ (\ell_{D-2\varepsilon} - k_{12\ldots j-1})^2 - (\ell_{D-2\varepsilon}-k_{12\ldots j})^2 \big] + s_{1j}+s_{2j}+\ldots+s_{j-1,j}
}
and cancel propagators with $(D-2\varepsilon)$-dimensional loop momenta on the right-hand side.

\newsec Manifestly BRST pseudo-invariant SYM integrands

\seclab\secfive

\noindent In this section, we manifest the BRST and cyclicity properties of the above SYM integrands
by rewriting the kinematic numerators in terms of (almost) BRST invariant superfields where only the
fingerprints of the hexagon anomaly appear in the $Q$ variation.

\subsec Manifesting BRST pseudo-invariance

\subseclab\secfiveone

\noindent The above cohomology construction of the six-point amplitude suggests that anomalous superfields such
as $Y_{A,B,C,D,E}$ in \BRSTTw\ have to be treated separately in the analysis of BRST properties. This
led to call a superfield {\it BRST-pseudo-invariant} if each term in its BRST variation contains an
anomalous factor of $Y_{A,B,C,D,E}$ \partI. 

A procedure is described in \partI\ to recursively construct BRST pseudo-invariants from superfields $V_A
T^{mn\ldots}_{B_1,B_2,B_3,\ldots}$ as defined in section \secthreetwo. The setup in \partI\ also includes $J_{1|2,3,4,5}$
in \defJ\ as well as generalizations to arbitrary rank and multiplicity. These pseudo-invariant are
denoted by $C^{mn\ldots}_{1|B_1,B_2,\ldots}$ or $P_{1|6|2,3,4,5}$ and classified by a term $V_1
T^{mn\ldots}_{B_1,B_2,B_3,\ldots}$ or $V_1J_{6|2,3,4,5}$ with a single particle representative $V_1$ of
the unintegrated vertex. The pseudo-invariant completion of these terms is furnished by multiparticle versions of
$V_{12\ldots p}$ with $p \geq 2$ and most conveniently described in the basis of Berends--Giele currents
such as \BGcurr. Similar to $T^{mn\ldots}_{B_1,B_2,B_3,\ldots}$ and $J_{6|2,3,4,5}$, the pseudo-invariants
$C^{mn\ldots}_{1|B_1,B_2,\ldots}$ and $P_{1|6|2,3,4,5}$ are symmetric under exchange
of slots $B_j$ which are separated by a comma.

At $n \leq 5$ points, for instance, it is straightforward to show that
\eqnn\firstCs
$$\eqalignno{
C_{1|2,3,4} &\equiv M_1 M_{2,3,4} \cr
C_{1|23,4,5} &\equiv M_{1} M_{23,4,5} + M_{12} M_{3,4,5} - M_{13} M_{2,4,5}  &\firstCs \cr
C^m_{1|2,3,4,5} &\equiv M_1 T^m_{2,3,4,5} + \big[ k_2^mM_{12} M_{3,4,5} + (2\leftrightarrow 3,4,5) \big]
}$$
are BRST closed. For the six-point BRST invariants $C_{1|234,5,6}, C_{1|23,45,6}$ and
$C^m_{1|23,4,5,6}$, analogous superfield expansions can be found in \refs{\eombbs,\partI}.
Reference \partI\ also displays the first pseudo-invariants $C^{mn}_{1|2,3,4,5,6}$ as well as
\eqn\ppsinv{
P_{1|6|2,3,4,5} \equiv V_1 J_{6|2,3,4,5} + M_{16} k_m^6 T^m_{2,3,4,5} + \big[ M_{162} T_{3,4,5} + (2\leftrightarrow 3,4,5) \big] 
}
with $M_{123} = {1\over s_{123}}( {V_{123}\over s_{12}}+ {V_{321}\over s_{23}})$ subject to
\eqn\psinv{
Q C^{mn}_{1|2,3,4,5,6} = - \delta^{mn} V_1 Y_{2,3,4,5,6} \ , \ \ \ \ Q P_{1|6|2,3,4,5} = - V_1 Y_{2,3,4,5,6} \ .
}
For external gluons, the explicit component expansions can be downloaded from \WWW.
Note that any scalar invariant $C_{1|A,B,C}$ can be expanded in terms of SYM tree amplitudes using the general formula
given in the appendix of \eombbs, see the five-point examples below.

As experimentally observed in \partI, one can rewrite BRST pseudo-invariant expressions in terms of manifestly BRST pseudo-invariant building blocks by mapping
\eqnn\InvMap
$$\eqalignno{
V_{12\ldots p} \big|_{p \geq 2}&\rightarrow 0 \ , \ \ \ \ V_1 T_{23,4,5} \rightarrow s_{23} C_{1|23,4,5} \ , \ \ \ \ V_1 T^m_{2,3,4,5} \rightarrow C^m_{1|2,3,4,5}  \cr
V_1 T_{234,5,6} &\rightarrow s_{23} (s_{34} C_{1|234,5,6} - s_{24} C_{1|324,5,6}) \ , \ \ \ \ V_1 T_{23,45,6} \rightarrow s_{23} s_{45} C_{1|23,45,6} &\InvMap \cr
V_1 T^m_{23,4,5,6} &\rightarrow s_{23} C^m_{1|23,4,5,6} \ , \ \ \ \ V_1 T^{mn}_{2,3,4,5,6} \rightarrow C^{mn}_{1|2,3,4,5,6} \ , \ \ \ \ V_1 J_{6|2,3,4,5} \rightarrow P_{1|6|2,3,4,5} \ .
}$$
Any appearance of $V_1$ signals a pseudo-invariant, whereas the multiparticle instances of $V_{12\ldots
p}$ with $p\geq 2$ are absorbed into the BRST completion of the former. When manifesting BRST pseudo-invariance
of the superspace integrand $A(1,2,\ldots,n|\ell)$ at $n=5,6$, the prescription \InvMap\ allows to
foresee the result of algebraic manipulations among the (mostly $\ell$ dependent) propagators.
However, the different kinematic poles in the expressions \firstCs\ for $C_{1|23,4,5}$ and
$C^m_{1|2,3,4,5}$ exemplify that locality is obscured when the five-point amplitude is expressed in
terms of BRST invariants. Hence, the representations for $A(1,2,\ldots,n|\ell)$ discussed in the
subsequent trade manifest locality for manifest BRST pseudo-invariance.

\subsec Five-point one-loop integrand

\subseclab\secfivetwo

\noindent
Applying the map in \InvMap\ to the five-point integrand \OneLoopFive\ leads to
\eqnn\BRSTFive
$$\displaylines{
A(1,2,3,4,5|\ell) =
{{1\over 2 } C_{1|23,4,5} \over \ell^2 (\ell-k_1)^2 (\ell-k_{123})^2 (\ell-k_{1234})^2 }
+ {{1\over 2}  C_{1|34,2,5} \over \ell^2 (\ell-k_1)^2 (\ell-k_{12})^2 (\ell-k_{1234})^2 } \cr
+ {{1\over 2} C_{1|45,2,3} \over \ell^2 (\ell-k_1)^2 (\ell-k_{12})^2 (\ell-k_{123})^2 }
+  { C_{1|2;3;4;5} + \ell_m C^m_{1|2,3,4,5}  \over \ell^2 (\ell-k_1)^2 (\ell-k_{12})^2  (\ell-k_{123})^2(\ell-k_{1234})^2 } \ ,
\hfil\BRSTFive\hfilneg
}$$
see \firstCs\ for $C_{1|23,4,5}$ and $C^m_{1|2,3,4,5}$. The scalar pentagon is represented by the shorthand
\eqn\scpent{
C_{1|2;3;4;5} \equiv {1\over 2} \big[s_{23} C_{1|23,4,5} + (2,3|2,3,4,5)\big] 
}
and can be obtained from the scalar part of $N_{1|2,3,4,5}(\ell)$ under \InvMap. Using the expansion of the invariants given in \firstCs, it is a matter of algebraic manipulations to check that \BRSTFive\ agrees with the local representation in section \secfourtwo\ at the level of the integrand. For example, the massive box $I_{12,3,4,5}^{(4)}$ can be eliminated using
\eqn\massbox{
I^{(4)}_{12,3,4,5} = I^{(4)}_{1,23,4,5} + 2 I^{(5)}_{1,2,3,4,5} \big[ (\ell \cdot k_2) - s_{12} \big]\ .
}
The scalar invariants in \BRSTFive\ are related to SYM tree subamplitudes through
\eqn\scAYM{
\eqalign{
\langle C_{1|23,4,5} \rangle&= s_{45}\big[  s_{24}  A^{{\rm tree}}(1,3,2,4,5) - s_{34} A^{{\rm tree}}(1,2,3,4,5) \big]
\cr
\langle C_{1|2;3;4;5} \rangle&= {s_{23} s_{45} \over s_{14}}  \big[  s_{12} s_{34} \, A^{{\rm tree}}(1,2,3,4,5) - s_{24}(s_{12}+s_{15}) A^{{\rm tree}}(1,3,2,4,5)  \big]
\ ,
}}
and integrals over $\ell_m C^m_{1|2,3,4,5}$ boil down to permutations of
\eqn\vecAYM{
\langle k_m^4 C^m_{1|2,3,4,5} \rangle = - s_{24} s_{34} s_{45} \big[ A^{{\rm tree}}(1,2,3,4,5)+A^{{\rm tree}}(1,3,2,4,5)\big] \ ,
}
see \vecpent\ for the Schwinger parametrizaton of the vector integral. These reductions to trees furnish the five-point generalization of $\langle
C_{1|2,3,4} \rangle = s_{12} s_{23} A^{{\rm tree}}(1,2,3,4)$ relevant for \loopSYM.

\subsec Six-point one-loop integrand

\subseclab\secfivethree

\noindent
Similarly, applying the map \InvMap\ to the six-point expression \OneLoopSix\ yields,
\eqnn\BRSTSix
$$\eqalignno{
A&(1,2,\ldots,6|\ell) = {1\over 4} \Big[ C_{1|234,5,6}  I^{(4)}_{1,234,5,6} + C_{1|2,345,6}  I^{(4)}_{1,2,345,6}  + C_{1|2,3,456}   I^{(4)}_{1,2,3,456} \cr
&\ \ \ + C_{1|23,45,6}   I^{(4)}_{1,23,45,6} + C_{1|23,4,56}  I^{(4)}_{1,23,4,56}  +C_{1|2,34,56}   I^{(4)}_{1,2,34,56} \Big] &\BRSTSix \cr
&\ \ \ + {1\over 2}\Big[  (C_{1|23;4;5;6} + \ell_m C^m_{1|23,4,5,6})   I^{(5)}_{1,23,4,5,6}  + (C_{1|2;34;5;6} + \ell_m C^m_{1|2,34,5,6})  I^{(5)}_{1,2,34,5,6} \cr
&\ \ \ + (C_{1|2;3;45;6} + \ell_m C^m_{1|2,3,45,6})  I^{(5)}_{1,2,3,45,6}+ (C_{1|2;3;4;56} + \ell_m C^m_{1|2,3,4,56})  I^{(5)}_{1,2,3,4,56}\Big] \cr
&\ \ \ +\big( C_{1|2;3;4;5;6} + \ell_m C^m_{1|2;3;4;5;6}  + {1\over 2} \ell_m \ell_n C_{1|2,3,4,5,6}^{mn}  \big)  I^{(6)}_{1,2,3,4,5,6}  -{1\over 2} P_{1|6|2,3,4,5}   I^{(5)}_{61,2,3,4,5}  \ .
}$$
The pseudo-invariants $C_{1|A,B,C},C^m_{1|A,B,C,D},C_{1|2,3,4,5,6}^{mn}$ and $P_{1|6|2,3,4,5}$ are defined in \partI\ and \ppsinv\ whereas\foot{For completeness, the remaining scalar pentagons are given by
$$\eqalignno{
2C_{1|2;34;5;6} &\equiv   s_{25} C_{1|25,34,6} + s_{26} C_{1|26,34,5} + s_{56} C_{1|2,34,56} \cr
&\ \ \ \ \ \ \ + s_{23} C_{1|234,5,6}-s_{24} C_{1|243,5,6}+ \big[ s_{45} C_{1|2,345,6} - s_{35} C_{1|2,435,6} + (5\leftrightarrow 6) \big]  \cr
2C_{1|2;3;45;6} &\equiv    s_{23} C_{1|23,45,6} + s_{26} C_{1|26,45,3} + s_{36} C_{1|2,36,45} \cr
&\ \ \ \ \ \ \ + \big[ s_{24} C_{1|245,3,6} - s_{25} C_{1|254,3,6} + (2\leftrightarrow 3) \big] + s_{56} C_{1|2,3,456} - s_{46} C_{1|2,3,546} \cr
2C_{1|2;3;4;56} &\equiv    s_{23} C_{1|23,4,56} + s_{24} C_{1|24,3,56} + s_{34} C_{1|2,34,56}  \cr
&\ \ \ \ \ \ \ + \big[ s_{45} C_{1|2,3,456} - s_{46} C_{1|2,3,465} + (4\leftrightarrow 2,3) \big] \ .
}$$
}
$C_{1|23;4;5;6}$ and $C_{1|2;3;4;5;6}^m$ are shorthands for subleading powers of $\ell$:
\eqnn\NewCSixOne
\eqnn\NewCSixThree
$$\eqalignno{
C_{1|23;4;5;6} &\equiv  {1 \over 2 } \Big( s_{45} C_{1|23,45,6} 
+ s_{46} C_{1|23,46,5} + s_{56} C_{1|23,56,4} &\NewCSixOne\cr
&\quad{} + \big[ s_{34} C_{1|234,5,6} - s_{24} C_{1|324,5,6} + (4\leftrightarrow 5,6) \big] \Big) \cr
C^m_{1|2;3;4;5;6} &\equiv  {1 \over 2} \big(s_{23}C^m_{1|23,4,5,6} + (2,3|2,3,4,5,6)\big)  \ . &\NewCSixThree 
}$$
Moreover, the BRST closed version of the scalar hexagon numerator in \scalarHex\ is given by:
\eqnn\NewCSixTwo
$$\eqalignno{
&C_{1|2;3;4;5;6} \equiv  {1\over 4} s_{23} s_{45} C_{1| 23,45,6} + (2,3|4,5|2,3,4,5,6) &\NewCSixTwo\cr
& + {1\over 6} \Big[s_{23} \big( s_{34} C_{1|234,5,6} -  s_{24} C_{1|324,5,6} \big)
+ s_{43} \big( s_{32} C_{1|432, 5,6} -  s_{24} C_{1|342,5,6} \big)  + (2,3,4|2,3,4,5,6) \Big]\cr
&+ {1\over 12} \Big[ (k^3_m \!- \!k^2_m) s_{23} C^m_{1|23,4,5,6} + (2,3|2,3,4,5,6) \Big]
- {1\over 24} C^{mn}_{1|2,3,4,5,6} \big[ k^1_m k^1_n  + (1\leftrightarrow 2,3,4,5,6) \big] \ .
}$$
The expansion of the pseudo-invariants in terms of local numerators can be found in \partI. On their
basis, it is a matter of algebraic relations similar to \massbox\ to verify agreement between the
manifestly BRST pseudo-invariant and the manifestly local representation of $A(1,2,3,4,5,6|\ell)$.
In upcoming work \wipNpt, the SYM integrand \BRSTSix\ will be shown to follow in the field-theory limit of
the open superstring.

Note that the anomalous BRST variation \psinv\ of $C_{1|2,3,4,5,6}^{mn}$ and $P_{1|6|2,3,4,5}$ allows
to reproduce the hexagon anomaly \NaiveSix\ from \BRSTSix.

\subsec Cyclicity of the five- and six-point integrands

\subseclab\secfivefour

\noindent As another virtue of the manifestly pseudo-invariant representations \BRSTFive\ and
\BRSTSix\ of $A(1,2,\ldots,n|\ell)$, their cyclicity can be analyzed in superspace. Strictly speaking,
only the integrated subamplitude in \defintegrand\ is cyclically pseudo-invariant because the hexagon
anomaly turns out to obstruct cyclic symmetry of the six-point amplitude. Starting point of the
cyclicity analysis is the rewriting of the $n$-point amplitude \defintegrand\ as
\eqnn\UUg
$$\eqalignno{
A(1,2,\ldots,n) &\equiv \int {d^D\ell \ {\widehat A}(1,2,\ldots,n|\ell) \over  \ell^2 (\ell-k_1)^2 (\ell-k_{12})^2   \ldots (\ell-k_{12\ldots n-1})^2 }  \ 
&\UUg
}$$ 
with {\it stripped integrand} ${\widehat A}(1,2,\ldots,n|\ell)$. Under cyclic shifts $i \rightarrow i+n$ mod
$n$ of all the labels, the $n$-gon denominator in \UUg\ transforms to $\ell^2(\ell-k_2)^2
(\ell-k_{23})^2 \ldots (\ell - k_{234n})^2$ which can be undone by change of integration variables
$\ell \rightarrow \ell - k_1$. Since this is the cyclic image of the shift $\ell \rightarrow \ell -
k_n$, the integrated amplitude \UUg\ is cyclically invariant if the stripped integrand satisfies
\eqnn\UUf
$$\eqalignno{
{\widehat A}(1,2,\ldots,n|\ell-k_n) \big|_{i \rightarrow i+1 \ {\rm mod} \ n} &= {\widehat A}(1,2,\ldots,n|\ell) \ . &\UUf
}
$$
The method is most conveniently illustrated at the five-point level. The stripped integrand
\eqnn\UUe
$$\eqalignno{
&{\widehat A}(1,2,3,4,5|\ell) =
 \bigg \langle {1\over 2} (s_{23} + 2 s_{345} )C_{1|23,4,5}  + {1\over 2} (s_{34}+2 s_{45}) C_{1|34,2,5} + {1\over 2} s_{45} C_{1|45,2,3}\cr
 &\ \ +{1\over 2} s_{24} C_{1|24,3,5}+ {1\over 2} s_{25} C_{1|25,3,4}+ {1\over 2} s_{35} C_{1|35,2,4} 
 + \ell_m C^m_{1|2,3,4,5} +(\ell \cdot k_3) C_{1|23,4,5}&\UUe  \cr
 &\ \ + (\ell \cdot k_4) (C_{1|23,4,5}+C_{1|34,2,5})+ \Big[  (\ell \cdot k_5)+{1\over 2} \ell^2 \Big] (C_{1|23,4,5}+C_{1|34,2,5}+C_{1|45,2,3}) 
 \bigg \rangle 
}$$
associated with \BRSTFive\ introduces permuted invariants such as $C_{2|34,1,5}$ and $C^m_{2|3,4,5,1}$ after the cyclic shift $i \rightarrow i+1 \ {\rm mod} \ 5$. They are no longer in the canonical form $C_{1|\ldots}$ and $C^m_{1|\ldots}$, but a procedure to restore it is described in section 11 of \partI. After discarding appropriate BRST-exact terms, the scalar invariants in \UUe\ are found to transform to
\eqnn\UUd
$$\eqalignno{
\langle C_{2|34,5,1} \rangle &= \langle C_{1|34,2,5} + C_{1|23,4,5} - C_{1|24,3,5}\rangle \ , \ \ \ \ \langle C_{2|31,4,5} \rangle = \langle C_{1|23,4,5} \rangle \cr
\langle C_{2|35,4,1} \rangle &= \langle C_{1|35,2,4} + C_{1|23,4,5} - C_{1|25,3,4} \rangle \ , \ \ \ \ \langle C_{2|41,3,5} \rangle = \langle C_{1|24,3,5}\rangle  &\UUd \cr
\langle C_{2|45,3,1} \rangle &= \langle C_{1|45,2,3} + C_{1|24,3,5} - C_{1|25,3,4} \rangle  \ , \ \ \ \ \langle C_{2|51,3,4} \rangle = \langle C_{1|25,3,4} \rangle 
}$$
under $i\rightarrow i+5 \ {\rm mod} \ 5$, and the vector invariant is mapped to
\eqn\UUdd{
\langle C^m_{2|3,4,5,1} \rangle= \langle C^m_{1|2,3,4,5}
+ \big[k^m_3 C_{1|23,4,5} + (3\leftrightarrow 4,5)\big] \rangle \ .
}
On these grounds, it is straightforward to verify cyclicity of the five-point amplitude via
\eqn\UUfivefail{
{\widehat A}(1,2,3,4,5|\ell-k_5)   \big|_{i \rightarrow i+1 \ {\rm mod} \ 5} - {\widehat A}(1,2,3,4,5|\ell) = 0 \ .
}
At six points, the relevant cyclic shifts $i\rightarrow i+6 \ {\rm mod} \ 6$ such as
\eqnn\UUx
\eqnn\UUy
$$\eqalignno{
\langle P_{2|1|3,4,5,6} \rangle &= \langle P_{1|2|3,4,5,6} + {\cal Y}_{12,3,4,5,6} \rangle &\UUx \cr
\langle C^{mn}_{2|3,4,5,6,1} \rangle &= \langle \delta^{mn} {\cal Y}_{12,3,4,5,6} + C^{mn}_{1|2,3,4,5,6} + \big[ 2 k_{3}^{(m} C^{n)}_{1|23,4,5,6} + (3\leftrightarrow 4,5,6) \big] \cr
& + \big[ 2 k_{3}^{(m} k_4^{n)} (C_{1|234,5,6}+C_{1|243,5,6}) + (3,4|3,4,5,6) \big]  \rangle
&\UUy
}$$
can again be found in section 11 and the appendix of \partI. The anomalous superfield 
\eqnn\UUw
$$\eqalignno{
{\cal Y}_{12,3,4,5,6} &\equiv {1\over s_{12}} Y_{12,3,4,5,6}&\UUw
}$$
with $Y_{12,3,4,5,6}$ defined in \BRSTTw\ measures the response of the hexagon anomaly to a cyclic
shift. It has parity-odd bosonic components (with gluon polarizations $e_i^p$)  \PSS
\eqn\UUv{
\langle {\cal Y}_{12,3,4,5,6} \rangle = -\epsilon_{p_3 p_4 p_5 p_6 q_1 q_2 \ldots q_6} k_{3}^{p_3}
k_{4}^{p_4}k_{5}^{p_5}k_{6}^{p_6} e_1^{q_1} e_2^{q_2}  \cdots e_6^{q_6} \ .
}
Apart from the subtleties associated with ${\cal Y}_{12,3,4,5,6}$, it is straightforward to show that
the stripped integrand \UUg\ associated with \BRSTSix\ satisfies \UUf. The anomalous obstruction
\eqnn\UUz
$$\eqalignno{
{\widehat A}(1,2,3,4,5,6|\ell-k_6) & \big|_{i \rightarrow i+1 \ {\rm mod} \ 6} - {\widehat A}(1,2,3,4,5,6|\ell) \cr
& = \half \langle {\cal Y}_{12,3,4,5,6} \rangle  \big( \delta_{mn} \ell^m \ell^n - \ell^2 \big)
&\UUz
}$$
integrates to a rational term as described in section \secfourfive, i.e. the failure of cyclic invariance is given by
\eqn\UUfail{
A(2,3,4,5,6,1) - A(1,2,3,4,5,6) = {1\over (2\pi)^{10}} {\pi^5 \over 240} \langle {\cal Y}_{12,3,4,5,6} \rangle\,.
}
We have kept the Kronecker delta in \UUz\ explicit which stems from the cyclic transformation of the tensor-pseudoinvariant \UUy. As explained at the end of section \secfourfive, it can be understood from dimensional reduction that the formally vanishing integrand in \UUz\ integrates to the rational expression \UUfail\ in external momenta.

\newsec One-loop color-kinematics duality

\seclab\secsix

\noindent
In 2008, Bern, Carrasco and Johansson (BCJ) proposed an organization scheme for tree-level gauge and gravity
amplitudes based on cubic vertices where color and
kinematics enter on completely symmetric footing \BCJ. Color tensors $c_i$ are naturally associated 
with cubic diagrams by dressing each vertex with structure constants $f^{abc}$ of some gauge group.
Triplets of color tensors $c_i,c_j,c_k$ associated 
with the diagrams shown in \figtriplet\
vanish due to the Jacobi identity
\eqn\Jacobi{
f^{abe}f^{cde} + f^{bce}f^{ade} + f^{cae} f^{bde} = 0
}
valid for any gauge group.

\ifig\figtriplet{The vanishing of the color factors associated to the above triplet of cubic
 graphs, $c_i + c_j + c_k = 0$ is a consequence of the Jacobi identity. In the above diagrams, the legs
 $a$, $b$, $c$ and $d$ may represent arbitrary subdiagrams. The BCJ duality
 states that their corresponding kinematic numerators $N_i$ can be chosen such that $N_i + N_j + N_k=0$.}
{\epsfxsize=0.70\hsize\epsfbox{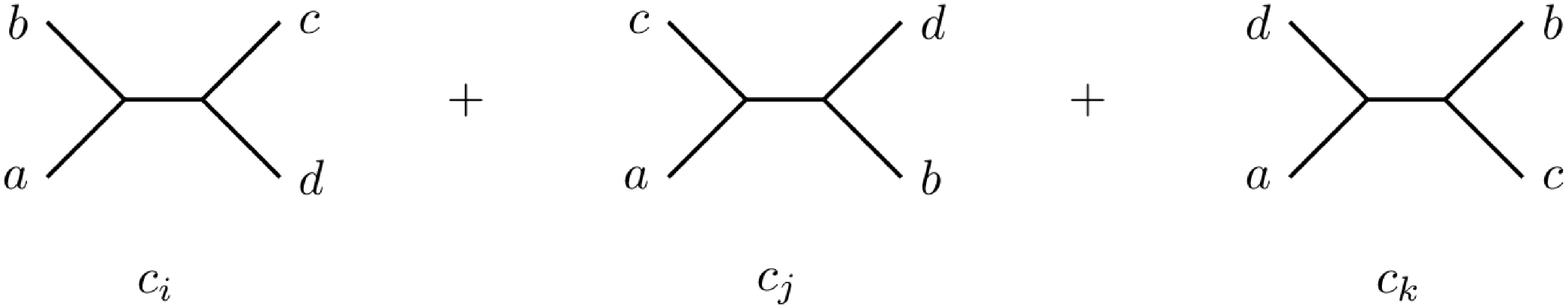}}


The BCJ conjecture states that amplitudes can be represented such that for any vanishing color triplet
$c_i + c_j + c_k$, the corresponding kinematic decorations $N_i + N_j + N_k$ of diagrams $i,j,k$ vanish
as well. The tree-level BCJ duality was later extended to loops in \BCJloop\ and successfully applied
to manifest UV properties of $4 \leq {\cal N}\leq 8$ supergravities up to four loops \BernUF\foot{Further recent work for situations with reduced or without supersymmetry includes \OchirovXBA, also see references therein.}.  However, the vanishing triplet of kinematic numerators associated to
the color triplet $c_i + c_j + c_k$ may depend on loop momenta as well, $N_i(\ell) + N_j(\ell) +
N_k(\ell) = 0$. The statement that the kinematic numerators satisfy the same relations as their
associated color factors is referred to as the {\it color-kinematic duality\/} or simply the {\it BCJ
duality}.

\subsec The five-point pure spinor representation and BCJ duality

In this section the pure spinor superspace representation of the five-point one-loop amplitude will be
shown to satisfy the BCJ color-kinematics duality. This generalizes the BCJ-satisfying five-point
numerators given in \CJfive\ using four-dimensional spinor helicity variables to ten dimensions, see
also \MonteiroOx.

\subsubsec Verifying Jacobi kinematic identities among boxes and pentagons

The BCJ identities associated to the external tree subdiagrams are trivially satisfied in the pure spinor
superspace representation since they are represented by BRST blocks which manifestly satisfy the
symmetries of their associated color factors\foot{For a detailed discussion of the general symmetries
of BRST blocks and their compatibility with BCJ identities, see appendix A of \eombbs.}. In the
five-point case of this subsection, this amounts to the antisymmetry of the massive legs in the boxes
and is manifestly satisfied by the rank-two BRST blocks.

\ifig\FiveBCJtriangle{Kinematic Jacobi relations relate box and triangle numerators. Since
maximally SYM amplitudes are not expected to contain triangles (nor bubbles and tadpoles), the left-hand
side must vanish.}
{\epsfxsize=0.55\hsize\epsfbox{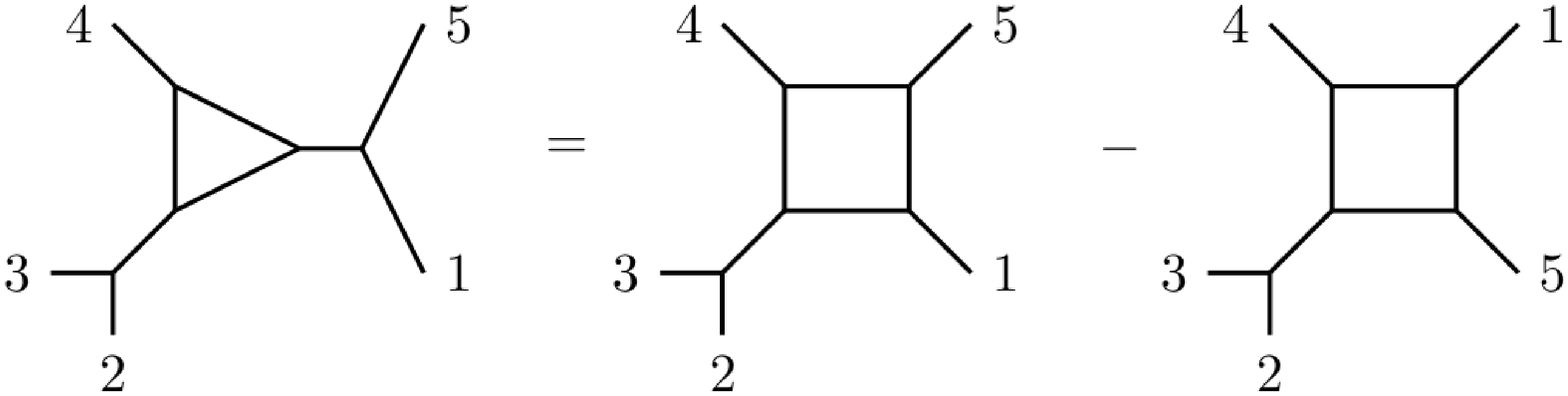}}

\ifig\FiveBCJFig{Kinematic Jacobi relations relate pentagon numerators to box numerators.}
{\epsfxsize=0.55\hsize\epsfbox{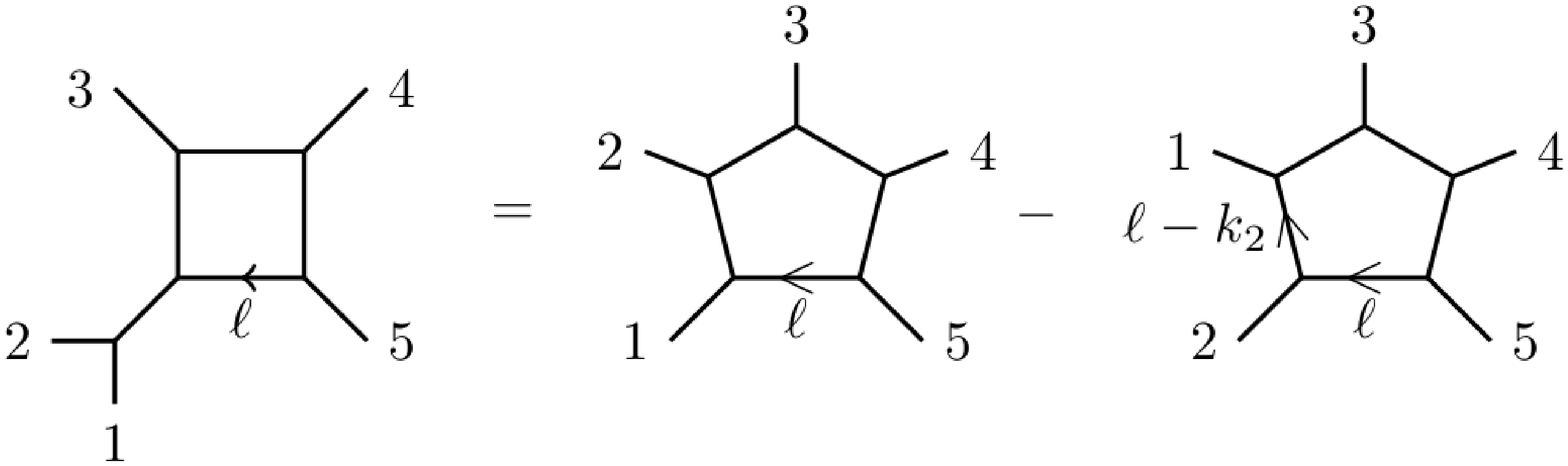}}

As discussed in \CJfive, 
a representation of the five-point one-amplitude satisfies the BCJ duality
if certain relations among the box and pentagon kinematic numerators hold.
For example, an antisymmetrization of any two corners of a box yields a triangle numerator as shown in
\FiveBCJtriangle. Since there are no triangles in maximally supersymmetric SYM amplitudes, this
antisymmetrization of boxes must vanish if BCJ is to be obeyed. And indeed, they vanish for the pure spinor representation of
section \secfourtwo.
For example, the kinematic numerators which correspond to the diagrams of the
right-hand side of
\FiveBCJtriangle\ are easily seen to cancel each other,
\eqn\triangleVanish{
\langle N^{(4)}_{1|23,4,5} - N^{(4)}_{1|5,23,4}\rangle = \langle V_1 (T_{23,4,5} - T_{5,23,4})\rangle =0,
}
since $T_{A,B,C}$ is symmetric in $A,B,C$. For any choice of massive corner in a five-point box, there
is only one possible superfield assignment -- either $\langle V_{12} T_{3,4,5} \rangle $ and
$(2\leftrightarrow 3,4,5)$ if leg one is part of the massive corner or $\langle V_1 T_{23,4,5} \rangle$
and $(2,3|2, 3,4,5)$ otherwise. Hence, the numerators cannot contain information about the ordering of
the box and every possible way to generate a triangle numerator vanishes with these superspace
representatives.


There is another class of kinematic Jacobi identities that needs to be checked whose depiction
is presented in \FiveBCJFig; the antisymmetrization of adjacent legs in the pentagon must give rise to
a box numerator \CJfive. Given the special role played by the particle label one in the pure spinor representation described in
section \secfourtwo, the relations which involve its participation in a non-trivial way will be
discussed separately.
\smallskip
\figflow{-1.6 truein}{1.8 truein}{{\epsfxsize=1.00\hsize\epsfbox{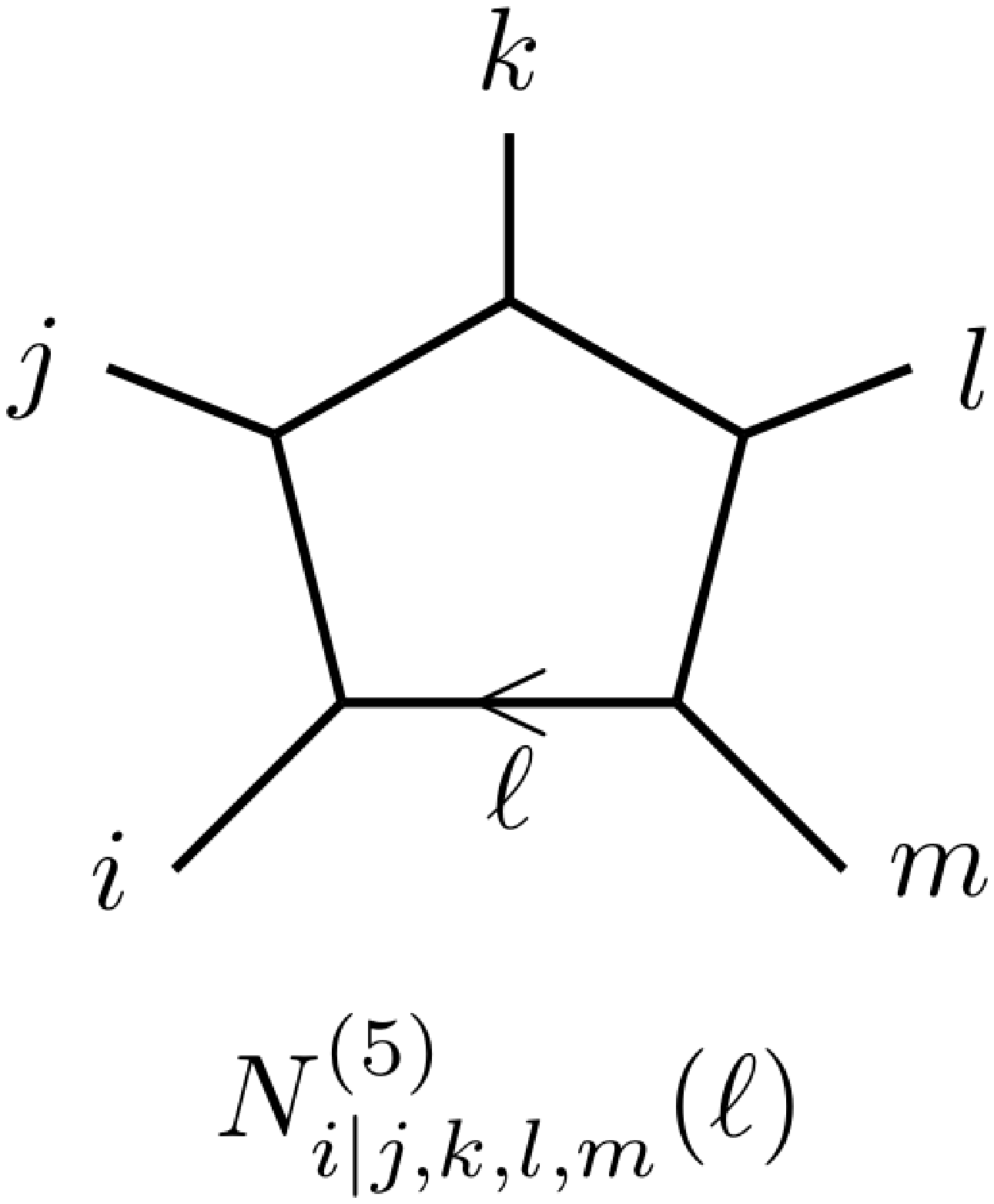}}\vfill}
\noindent Furthermore, the presence of the loop momentum in the pentagon
numerator requires a precise convention for the mapping between a
pentagon cubic graph and its pure spinor superspace representation: The loop momentum $\ell$ in
the pentagon numerator must be chosen such that the four legs involved in the
Jacobi identity have the same momentum in the three diagrams.
Our convention is that
for a given $N^{(5)}_{i|j,k,l,m}(\ell)$ labelling, $\ell$ is the momentum of
the $n$-gon edge between the vertices with external legs $i$ and $m$.
\noindent For example, the kinematic Jacobi identity generated upon
antisymmetrizing the legs $2$ and $3$ of the pentagon is given by
\eqn\BCJbox{
\langle N^{(4)}_{1|23,4,5}\rangle = \langle N^{(5)}_{1|2,3,4,5}(\ell) -
N^{(5)}_{1|3,2,4,5}(\ell)\rangle\,,
}
and it is easily verified by their superspace representations
for the pentagon given in \specPent. In \BCJbox, the expression for
$N^{(5)}_{1|2,3,4,5}(\ell)$ is projected to the antisymmetric part with respect to legs 2 and 3 which
amounts to $\half V_1 T_{23,4,5}-\half V_1 T_{32,4,5}=N^{(4)}_{1|23,4,5}$. The same argument applies to
antisymmetrizations of $\langle N^{(5)}_{1|2,3,4,5}(\ell) \rangle$ in 3,4 or in 4,5.

When the Jacobi identity involves leg number one as seen in \FiveBCJFig\ the analysis is a bit longer.
Using the pentagon convention above, the second diagram in \FiveBCJFig\ is written as\foot{This
is equivalent to the  ``dihedral'' symmetry condition of the pentagon numerator \CJfive.}
$N^{(5)}_{1|3,4,5,2}(\ell-k_2)$ and the kinematic Jacobi identity to verify is \MonteiroOx
\eqn\newBCJbox{
\langle N^{(4)}_{12|3,4,5}\rangle = \langle N^{(5)}_{1|2,3,4,5}(\ell) -
N^{(5)}_{1|3,4,5,2}(\ell-k_2)\rangle.
}
Plugging in the explicit five-point box and pentagon numerators of \genBox\ and \genPent\ translates \newBCJbox\ into the following superspace statement\foot{That a external momentum contracted with the vector pentagon numerator gives rise
to a sum of boxes like in \BCJcheck\ has already been derived in \MonteiroOx.}
\eqn\BCJcheck{
\langle k^2_m V_{1}T^m_{2,3,4,5} + V_{21}T_{3,4,5} + V_1 T_{23,4,5} + V_1 T_{24,3,5} + V_1 T_{25,3,4}\rangle = 0\,.
}
And indeed, one can show that \BCJcheck\ is BRST-trivial for five-point kinematics and therefore vanishes in the cohomology
as computed by the pure spinor brackets $\langle \cdots \rangle$. 

To see this recall that the superfield $D_{1|2|3,4,5}$ defined in equation (8.16) of \partI,
\eqn\Ddef{
D_{1|2|3,4,5} \equiv J_{2|1,3,4,5} + k^2_m M^m_{12,3,4,5} + \big[ s_{23} M_{123,4,5} + (3\leftrightarrow 4,5)\big]
}
where $M^m_{12,3,4,5} = (1/s_{12}) T^m_{12,3,4,5}$ was shown to satisfy
\eqn\trivial{
Q D_{1|2|3,4,5}  =Y_{1,2,3,4,5} + k^2_m C_{1|2,3,4,5}^m +  \big[ s_{23} C_{1|23,4,5} + (3
\leftrightarrow 4,5) \big].
}
Using the expansion in \firstCs\ for the BRST invariants, $\langle Q D_{1|2|3,4,5}\rangle = 0$ implies that
\eqn\conclu{
0 = \langle Y_{1,2,3,4,5}
+  k^2_m  V_{1}T^m_{2,3,4,5}
- {s_{23} + s_{24} + s_{25}\over s_{12}} V_{21} T_{3,4,5}
+ V_1 T_{23,4,5} + V_1 T_{24,3,5} + V_1 T_{25,3,4}\rangle.
}
Since $\langle Y_{1,2,3,4,5}\rangle \propto \e_{10}F^5$ vanishes by momentum conservation \anomaly\ and
$s_{23} + s_{24} + s_{25} = - s_{12}$, the proof of \BCJcheck\ is complete. Therefore \OneLoopFive\
furnishes a {\it local\/} BCJ-satisfying representation of the five-point one-loop SYM
amplitude.

One can also show that the manifestly BRST-invariant (and non-local) five-point
representation of section \secfivetwo\ satisfies the BCJ duality conditions. Box numerators with
particle one in the massive corner then vanish and the nonzero instances follow from permutations of
$N^{(4)}_{1|23,4,5}=s_{23} C_{1|23,4,5}$ in $2,3,4,5$.

\subsec The five-point supergravity amplitude

Once a SYM amplitude has been presented in a form which satisfies all kinematic Jacobi identities
$N_i(\ell) + N_j(\ell) + N_k(\ell)=0$, it can be transformed into a gravity amplitude by trading color
tensors for a second copy of the kinematic numerators, $c_i \rightarrow \tilde N_i(\ell)$
\refs{\BCJ,\BCJloop,\yutin}.
This allows to assemble the five-point supergravity amplitude at one-loop from the above box and
pentagon numerators:
\eqn\gravAmpBCJ{
M_5 = \int {d^D\ell \over (2\pi)^D}\sum_{\Gamma_i}{\langle N_i(\ell) \tilde N_i(\ell) \rangle\over \prod_k P_{k,i}(\ell)}
}
The combinatorics of the graph sum is explicit in the following representation
\eqnn\explicit
$$\eqalignno{
M_5 &= \int {d^D\ell \over (2\pi)^D} \Big\{ \big[ I^{(5)}_{1,2,3,4,5} N^{(5)}_{1|2,3,4,5}(\ell) \tilde N^{(5)}_{1|2,3,4,5}(\ell) + {\rm symm}(2,3,4,5)\big]  \cr
&+ \big[ (I^{(4)}_{12,3,4,5} +{\rm symm}(3,4,5))   N^{(4)}_{12|3,4,5}  \tilde N^{(4)}_{12|3,4,5} + (2\leftrightarrow 3,4,5) \big] &\explicit \cr 
&+ \big[ (I^{(4)}_{1,23,4,5} +{\rm symm}(23,4,5))   N^{(4)}_{1|23,4,5}  \tilde N^{(4)}_{1|23,4,5} + (2,3|2,3,4,5) \big] \Big\}
}$$
which can be confirmed by taking the field-theory limit\foot{The RNS derivation of the field-theory
limit can be found in \BjerrumBohrVC, the techniques are similar to the material of appendix~A combined
with the tensor integral \wlone.} of the five-point closed-string amplitude in pure spinor superspace
\oneloopMichael.

Depending on the relative chirality of the left- and right-moving superfields, the amplitude describes
type IIA or type IIB supergravity. The tensor integrals in the pentagon numerators give rise to vector
contractions between left- and right-moving superfields $\langle V_1 T^m_{2,3,4,5} \delta_{mn} \tilde
V_1 \tilde T^n_{2,3,4,5}\rangle$. Components where both sides contribute through an $\epsilon_{10}$
tensor change signs between type IIA and type IIB
and only the integrated type IIB amplitude can be
written in terms of bilinears of SYM trees, see \oneloopMichael\ for details.

\ifig\figcounter{Counterexample for kinematic Jacobi relations at six-points
in the particular representation of section~\secfourfour.}
{\epsfxsize=0.65\hsize\epsfbox{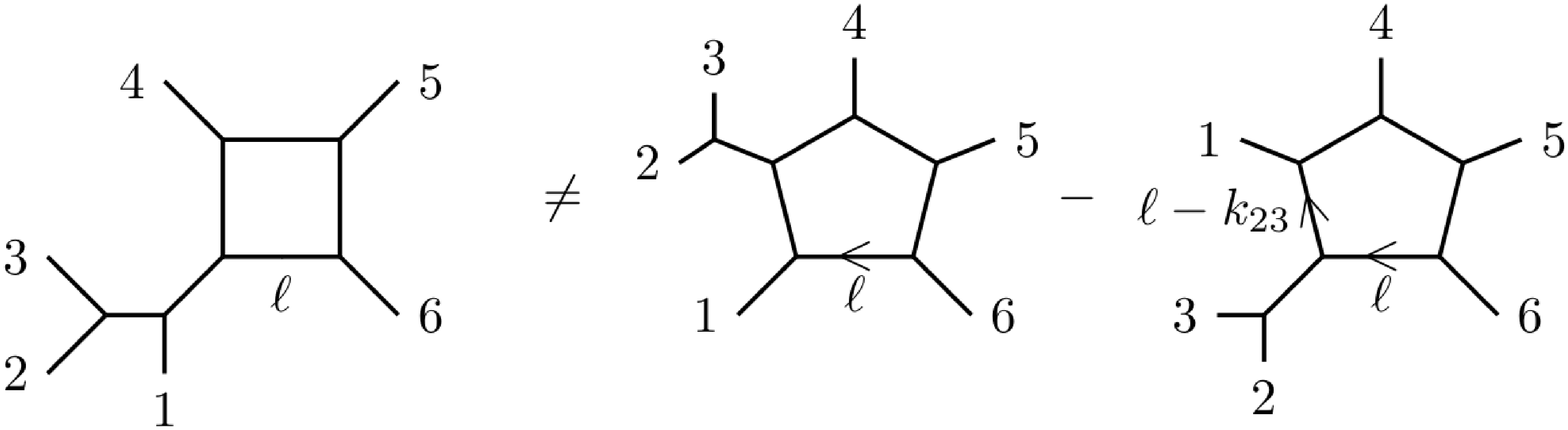}}

\subsec The six-point amplitude and BCJ duality

In spite of many encouraging antisymmetrization properties of the six-point numerators, the BCJ duality
is not satisfied by the local representation given in section~\secfourfour. That is why we restrict the
discussion to a counterexample. As depicted in \figcounter, the antisymmetrization of a massive
pentagon numerator in external trees $1$ and $23$ is related to a one-mass box by the duality,
\eqn\sixptBCJ{
\langle N_{231|4,5,6}^{(4)}\rangle \leftrightarrow \langle  N^{(5)}_{1|4,5,6,23}(\ell-k_{23})-N^{(5)}_{1|23,4,5,6}(\ell)\rangle \ .
}
The left-hand side is given by $\langle V_{231} T_{4,5,6} \rangle$, and the right-hand side can be
evaluated using the general pentagon numerators \genPent. The failure of \sixptBCJ\ is then described
by
\eqnn\sixFail
$$\eqalignno{
&N^{(5)}_{1|23,4,5,6}(\ell) - N^{(5)}_{1|4,5,6,23}(\ell-k_{23}) + N_{231|4,5,6}^{(4)} \cr
&\ \ \ = k_m^{23}V_1 T^m_{23,4,5,6} + V_{231}T_{4,5,6}
 + \big[ V_1 T_{234,5,6} + (4\leftrightarrow 5,6)\big] \ .
&\sixFail
}$$
However, since the right-hand side of \sixFail\ is not BRST closed it cannot be zero in the cohomology,
therefore \sixptBCJ\
cannot be an equality.
So the particular representation of the six-point amplitude obtained with the method of this paper does
not satisfy the color-kinematics duality\foot{See \refs{\YuanRG, \MonteiroOx} for four-dimensional BCJ
representations of one-loop amplitudes.}.

In the non-local but manifestly pseudo-invariant representations \BRSTSix, the failure of the BCJ
relation in \figcounter\ is captured by
\eqn\sixFailu{
\langle k_m^{23}C^m_{1|23,4,5,6} 
 + \big[ s_{34} C_{1|234,5,6} -s_{24} C_{1|324,5,6}   + (4\leftrightarrow 5,6)\big] \rangle
 = \langle P_{1|2|3,4,5,6}-P_{1|3|2,4,5,6} \rangle \,,
}
see equation (9.8) of \partI. The pseudo-invariant nature of the $P_{1|i|j,\ldots}$ on the right-hand side gives
rise to speculate that the failure in \sixFailu\ signals a subtle relation with the hexagon anomaly.

\newsec Conclusion

In this work, we have presented local representations for the one-loop integrand of five- and six-point
SYM amplitudes in ten dimensions. Pure spinor superspace allows to express each numerator as a compact
combination of superfields which were determined by BRST symmetry. Moreover, the multiparticle
superfields $[A_\alpha^B,A^m_B,W_B^\alpha,F^{mn}_B]$ of \eombbs\ gave rise to a universal structure for
box and pentagon numerators
\eqnn\genngon
$$\eqalignno{
N^{(4)}_{A|B,C,D}&\equiv
 V_A T_{B,C,D}
\cr
N^{(5)}_{A|B,C,D,E}(\ell) &\equiv
\ell_m  V_A T^m_{B,C,D,E} + {1\over 2} \big[V_{[A,B]} T_{C,D,E} + (B\leftrightarrow C,D,E) \big] &\genngon\cr
&\quad{} + {1\over 2} \big[V_{A} T_{[B,C],D,E} + (B,C|B,C,D,E) \big]\,,
}$$
largely independent on the external tree subdiagrams $A,B,\ldots,E$\foot{Minor redefinitions are required for massive corners subtending the first and last leg $n$ and $1$. The six-point
example has been analyzed in section \secfourfour\ and identified to play an essential role for the
hexagon anomaly.}. As explained in \eombbs, the bracketing
notation merges two multiparticle slots such as to connect the associated cubic diagrams through a
cubic vertex. Iterated bracketing is then expected to capture the structure of higher $n$-gon
numerators, e.g. the hexagon numerator in section \secfourfour\ should mostly generalize as
\eqnn\genhex
$$\eqalignno{
&N^{(6)}_{A|B,C,D,E,F}(\ell) \equiv \half \ell_m \ell_n V_A T^{mn}_{B,C,D,E,F}
+ \half \ell_m  V_{[A,B]} T^m_{C,D,E,F}  + \half \ell_m V_A T_{[B,C],D,E,F}^m   \cr
&\ \ \ \ + {1\over 6} (V_{[[A,B],C]}+ V_{[[C,B],A]}) T_{D,E,F} + {1\over 6} V_A  (T_{[[B,C],D],E,F} +  T_{[[D,C],B],E,F}) \cr
& \ \ \ \ + {1\over 4} V_{[A,B]} T_{[C,D],E,F} + {1\over 4} V_{A} T_{[B,C],[D,E],F} - {1\over 12} (k^A_m - k^B_m) V_{[A,B]} T^m_{C,D,E,F}&\genhex \cr
&\ \ \ \ - {1\over 12}(k^B_m - k^C_m) V_A T^m_{[B,C],D,E,F} - {1\over 24}V_A T^{mn}_{B,C,D,E,F} k^A_m k^A_n + {\rm permutations}  \ ,
}$$
where the sum of permutations follows the patterns of \hexNum\ and \scalarHex.

The $\ell$-dependent parts of the above numerators exhibit a recursive structure and mimic the pattern of
scalar lower-gon representatives upon adjusting the building blocks $T_{A,B,C} \rightarrow
T^{m\ldots}_{A,B,C,D,\ldots}$ to higher rank. Hence, the main leftover challenge in $n$-point
amplitudes at higher multiplicity $n \geq 7$ is posed by the scalar part of the irreducible $n$-gon
numerator which can be systematically addressed by demanding BRST invariance \wipNpt.

A manifestly BRST pseudo-invariant presentation of the amplitudes is given in section~\secfive. Most of
the integrals with leg one in a massive corner are eliminated which in turn assembles
pseudo-invariant kinematic factors $C^{m\ldots}_{1|\ldots}$ and $P_{1|6|2,3,4,5}$ classified in \partI. Their gluonic
components can be downloaded from \WWW, and the scalars
$C_{1|A,B,C}$ along with boxes can be expressed in terms of SYM trees \eombbs.

The BCJ duality between color and kinematics is imprinted into the symmetries of multiparticle
superfields \eombbs\ underlying $V_A$ and $T^{m\ldots}_{B,C,\ldots}$ and therefore respected by any tree
subdiagram. However, this does not necessarily imply the kinematic Jacobi relations between numerators of $p$-gons and $(p-1)$-gons. At five points, the bracket structure of the pentagon numerator
\genngon\ and the identity \BCJcheck\ relating $k^2_m V_1 T^m_{2,3,4,5}$ to box numerators guarantee that the representations
of the integrand in both section \secfourtwo\ and \secfivetwo\ satisfy the duality. At six points, on
the other hand, obstructions of the form $P_{1|i|j,4,5,6}$ in certain dual
Jacobi relations might signal a subtle connection to the hexagon anomaly, see \sixFailu. It would be interesting to
explore the physical meaning of this observation.

A valuable and complementary viewpoint on the SYM integrands of this work stems from the field-theory
limit of the open superstring. The five-point case is discussed in appendix A based on the worldsheet
integrand in \oneloopbb, and string amplitudes at higher multiplicity will provide a rich laboratory to
study the interplay between the hexagon anomaly cancellation in string theory \GSanomaly\ 
and its appearance in field-theory \FramptonAnomaly.

Furthermore, it would be interesting to connect the BRST structures with other approaches to loop
amplitudes such as ambitwistor strings \MasonSVA. Their pure spinor implementation \InfiniteTension\ is
know from \GomezWZA\ to reproduce the tree-level amplitudes following from BRST methods of \nptMethod.

Finally, the dictionary between cubic one-loop diagrams and superfields such as $V_AT_{B,C,D}$ suggests
a generalization to higher loops. The two-loop analysis of \twoloop\ in the minimal pure spinor
formalism and the advances at three loops in \GomezSLA\ using its non-minimal version \NMPS\ furnish an
encouraging starting point.


\bigskip \noindent{\bf Acknowledgements:} We thank Henrik Johansson for useful comments on the draft, Yu-tin
Huang and Piotr Tourkine for helpful discussions and Michael Green for collaboration on related topics.
OS is indebted to Song He for insightful discussions and collaboration on related topics. CRM and OS
cordially thank the Institute of Advanced Study at Princeton for hospitality during early stages of
this work. OS is grateful to the Department of Applied Mathematics and Theoretical Physics of the
University of Cambridge for hospitality during completion of this work. CRM and OS acknowledge
financial support by the European Research Council Advanced Grant No. 247252 of Michael Green. This
work was supported in part by National Science Foundation Grant No. PHYS-1066293 and the hospitality of
the Aspen Center for Physics.


\appendix{A}{The field-theory limit of the five-point superstring amplitude}

In this appendix, we show that the five-point one-loop SYM amplitude presented in subsection
\secfourtwo\ is reproduced by the field-theory limit of the open pure spinor superstring. This
endeavour requires a matching of the Schwinger parametrization of Feynman integrals with the $\alpha'
\rightarrow 0$ limit of the one-loop string amplitude prescription \looppresc.

\subsec Schwinger parametrization of the five-point SYM amplitude

The string-based formalism for field-theory amplitudes \worldline\ provides
a convenient worldline representation of a scalar $n$-gon integral in $D$ dimensions.
The scalar box and pentagon integrals in the five-point amplitude can be written as the worldline integral
\eqnn\worldlinebox
\eqnn\worldlinepent
$$\eqalignno{
\int d^D\ell \, I^{(4)}_{12,3,4,5}&= \pi^{4}\int^{\infty}_0{dt\over t} \, t^{4-D/2}  \! \! \! \! \! \! \! \! \! \!  \int \limits _{0 \leq \nu_{i} \leq \nu_{i+1} \leq 1} 
\! \! \! \! \! \! \! \! \! \!  d \nu_2\, d \nu_3 \,  d \nu_4 \, e^{-\pi t Q_4[k_{12},k_3,k_4,k_5]} \, \Big|_{\nu_1=0} &\worldlinebox\cr
\int d^D\ell \, I^{(5)}_{1,2,3,4,5}&= \pi^{5}\int^{\infty}_0{dt\over t} \, t^{5-D/2}  \! \! \! \! \! \! \! \! \! \!  \int \limits _{0 \leq \nu_{i} \leq \nu_{i+1} \leq 1} 
\! \! \! \! \! \! \! \! \! \!  d \nu_2\, d \nu_3 \,  d \nu_4 \,  d \nu_5 \, e^{-\pi t Q_5[k_{1},k_{2},k_3,k_4,k_5]} \, \Big|_{\nu_1=0}  &\worldlinepent
}$$
with exponents
\eqn\worldlineKN{
Q_n[k_{A_1},k_{A_2},\ldots,k_{A_n}] \equiv \sum_{i<j}^n (k_{A_i}\cdot k_{A_j}) (\nu_{ij}^2 - |\nu_{ij}|)
}
and shorthand notation $\nu_{ij} \equiv \nu_i-\nu_j$. The transformation from momentum space to the
worldline picture involves a Gaussian integration over the shifted loop momentum $\hat \ell^m \equiv
\ell^m + \sum_{i=1}^n k_{A_i}^m \nu_i$. Hence, vector integrals follow from substituting $\ell^m
\rightarrow -\sum_{i=1}^n k_{A_i}^m \nu_i$, e.g.
\eqn\vecpent{
\int d^D\ell \,I^{(5)}_{1,2,3,4,5}  \, \ell^m =  - \pi^{5}\int^{\infty}_0{dt\over t} \, t^{5-D/2}  \! \! \! \! \! \! \! \! \! \!  \int \limits _{0 \leq \nu_{i} \leq \nu_{i+1} \leq 1} 
\! \! \! \! \! \! \! \! \! \!  d \nu_2\, \ldots \,  d \nu_5 \, \sum_{i=1}^5 k_{i}^m \nu_i \, e^{-\pi t Q_5[k_{1},\ldots,k_5]}  \ .
}
\noindent In the pentagon contribution \Pentfive\ and \specPent\ to the five-point SYM amplitude,
scalar and vector parts can be cast into a unified form using the cohomology manipulations \refs{\oneloopMichael,\partI}
\eqnn\Tmone
$$\eqalignno{
\langle V_1 k^1_m T^m_{2,3,4,5} \rangle &= \langle -V_{12} T_{3,4,5} + (2\leftrightarrow 3,4,5) \rangle &\Tmone\cr
\langle V_1 k^2_m T^m_{2,3,4,5} \rangle&= \langle V_{12} T_{3,4,5} +  \big[ - V_1 T_{23,4,5}+(3\leftrightarrow 4,5) \big] \rangle
}$$
to express $ \sum_{i=1}^5 \nu_i k^i_m \langle V_1 T^m_{2,3,4,5} \rangle$ in terms of box numerators.
The pentagon integrals \worldlinepent\ and \vecpent\ then conspire to
\eqnn\schwingerpent
$$\eqalignno{
\int &d^D\ell \, \langle A_{\rm pent}(1,2,3,4,5|\ell) \rangle = \pi^{4}\int^{\infty}_0{dt\over t} \, t^{5-D/2}  \! \! \! \! \! \! \! \! \! \!  \int \limits _{0 \leq \nu_{i} \leq \nu_{i+1} \leq 1} 
\! \! \! \! \! \! \! \! \! \!  d \nu_2\, \ldots \,  d \nu_5 \, e^{-\pi t Q_5[k_{1},k_2,\ldots,k_5]}  \cr
&\ \  \times \Big \langle \big[ \partial_\nu G_{12} V_{12} T_{3,4,5} + (2\leftrightarrow 3,4,5)\big] + V_1\big[ \partial_\nu G_{23} T_{23,4,5} + (2,3|2,3,4,5) \big] \Big \rangle \  &\schwingerpent
}$$
with the derivative of the worldline Green function
\eqn\worldlineGF{
G_{ij} \equiv {\pi \over 2} \big( \nu_{ij}^2- |\nu_{ij}| \big) \ ,  \ \ \ \
\partial_\nu G_{ij} \equiv \pi \Big( \nu_{ij}- \half {\rm sgn}(\nu_{ij}) \Big) \ .
}
This is a convenient starting point to make contact with the corresponding superstring amplitude. The
sign function in \worldlineGF\ is defined to be $+1$ $(-1)$ when $\nu_i\ge\nu_j$ ($\nu_i<\nu_j$).

For the box contribution \Boxfive\ to the five-point amplitude, the Schwinger parametrization directly
follows from \worldlinebox\ and minor cyclic modifications
\eqnn\schwingerbox
$$\eqalignno{
\int d^D\ell \, \langle A_{\rm box}&(1,2,3,4,5) \rangle = \pi^{4}\int^{\infty}_0{dt\over t} \, t^{4-D/2} \! \! \! \! \! \! \! \! \! \!  \int \limits _{0 \leq \nu_{i} \leq \nu_{i+1} \leq 1} 
\! \! \! \! \! \! \! \! \! \!  d \nu_2\, d \nu_3 \,  d \nu_4 \,  \Big\{ \langle V_{12} T_{3,4,5} \rangle e^{-\pi t Q_4[k_{12},k_3,k_4,k_5]} 
\cr
& + \langle V_{1} T_{23,4,5} \rangle e^{-\pi t Q_4[k_{1},k_{23},k_4,k_5]} + \langle V_{1} T_{2,34,5} \rangle e^{-\pi t Q_4[k_{1},k_2,k_{34},k_5]}  \cr
&+ \langle V_{1} T_{2,3,45} \rangle e^{-\pi t Q_4[k_{1},k_2,k_3,k_{45}]} + \langle V_{51} T_{2,3,4} \rangle e^{-\pi t Q_4[k_{51},k_2,k_3,k_4]} \Big\} \Big|_{\nu_1=0}  \ . &\schwingerbox 
}$$
Note that the exponents $Q_4$ as in \worldlineKN\ associated with boxes arise from degenerations of the
pentagon, e.g. $Q_4[k_{1},k_{23},k_4,k_5] = Q_5[k_1,k_2,\ldots, k_5] \big|_{\nu_2=\nu_3}$.

\subsec Worldline limit of open string one-loop amplitudes

As described in \GreenFT, the field-theory limit of the superstring is obtained by setting $\ap\to 0$
and by degenerating the genus-one surface with modular parameter $\tau$ as $\Im(\tau) \to\infty$. These are the
limits in which strings shrink to point-particles, and the worldsheet surface reduces to point-particle
worldline diagrams. Moreover, these limits must be performed such that the proper time $t \equiv \ap
\Im(\tau)$ and the worldsheet positions $z_j\equiv {\rm Re}(z_j) + i \Im(\tau) \nu_j$ stay finite. In this
limit, the worldsheet Green function 
\eqn\sheetGF{
{\cal G}_{ij} \equiv  -{ \alpha' \over 2} \Big( \ln \big| \theta_1(z_{ij}|\tau) \big| - \pi { \Im(z_{ij})^2 \over\Im(\tau)} \Big) 
}
loses its dependence on the real part of $z_{ij}  \equiv  z_i-z_j$ and reproduces the worldline Green function \worldlineGF\ upon identifying $\Im(z_j)  = t \nu_j$ with the worldline insertion points $\nu_j$,
\eqn\GFlim{
{\cal G}_{ij}  \rightarrow t\, G_{ij} \ , \ \ \ \ \ \partial_z{\cal G}_{ij}  \rightarrow \partial_\nu G_{ij}  \ .
}
A key ingredient of the worldsheet correlator in the string amplitude \looppresc\ is the ubiquitous
Koba--Nielsen factor which reduces as follows under the field-theory limit \GFlim:
\eqn\KNlim{
{\cal I} \equiv
\left \langle \prod_{j=1}^n e^{i k_j \cdot x(z_j,\bar z_j) } \right \rangle = {1\over (\Im\,\tau)^5}
\prod_{j<k}e^{-2 s_{jk}{\cal G}_{jk}}
  \rightarrow {1\over t^5} e^{ -\pi t Q_n[k_1,k_2,\ldots, k_n]  } \ .
}
The plane waves from the vertex operators reproduce the exponential \worldlineKN\ of the Feynman integrals' worldline parametrization in \worldlinepent\ and \wlone. Similarly, the worldline integration over $t$ and $\nu_i$ descends from the worldsheet integration in \looppresc\ over the cylinder boundary,
\eqn\intlim{
\int^{\infty}_0{dt\over t^5} \! \! \! \! \! \! \! \! \! \!  \int \limits _{0 \leq \Im z_{i} \leq \Im z_{i+1} \leq t} 
\! \! \! \! \! \! \! \! \! \!  \! \! \! \!   d z_2\, d z_3 \,  \ldots \, d z_n
\rightarrow
\int^{\infty}_0{dt\over t} \, t^{n-D/2} \! \! \! \! \! \! \! \! \! \!  \int \limits _{0 \leq \nu_{i} \leq \nu_{i+1} \leq 1} 
\! \! \! \! \! \! \! \! \! \!  d \nu_2\, d \nu_3 \, \ldots \, d \nu_n \ .
}
However, before the combined limit of $\ap\to 0$ and $\Im(\tau) \to\infty$ can be performed in \KNlim,
an additional feature of the Koba--Nielsen factor has to be taken into account which is completely
absent in its worldline counterpart $e^{-\pi t Q_n}$: It is the source of kinematic poles when the vertex operator
positions $z_i \rightarrow z_j$ approach each other.

More precisely, the short-distance behaviour of \KNlim\ as $z_i \rightarrow z_j$ is governed by
${\cal I} \sim |z_{ij}|^{\alpha'k_i\cdot k_j}$. Additional factors of worldsheet propagators $\partial_z {\cal
G}_{ij} \rightarrow z_{ij}^{-1} + {\cal O}(z_{ij})$ modify the leading singular behavior to
${\cal I}\partial_z {\cal G}_{ij} \sim |z_{ij}|^{\alpha'k_i\cdot k_j-1}$. In this case, the integration domain where $|z_{ij}|
\ll 1$ gives rise to a pole in $\alpha'k_i\cdot k_j$ with $z_i=z_j$ on its residue\foot{This can be understood from the delta function
representation $\delta(x) = \lim_{s\rightarrow 0} s x^{s-1}$.}. In other words, if
$i$ and $j$ are adjacent on the worldsheet boundary, the following kinematic pole emerges:
\eqn\reducible{
{\cal I} \, \partial_z {\cal G}_{i,i+1}  \rightarrow
{\cal I} \,  { \delta(z_{i}-z_{i+1}) \over s_{i,i+1}} \ + \ {\cal O}(s_{i,i+1}^0) \ .
} 
This mechanism is the origin of box integrals in the five-point open string amplitude.

\subsec Worldline limit of the five-point open string amplitude

The above procedure to perform the point particle limit is now applied to the pure spinor prescription
for the five-point open superstring amplitude in \looppresc. The constraints from zero-mode saturation \multiloop\ admit one OPE among the vertex operators. Hence, the correlator\foot{See for instance \TsuchiyaVA\ for the analogous RNS computation which reproduces the bosonic components of \fiveptcorr\ up to total derivatives in $z_j$.} follows from summing over the ten
BRST blocks capturing the OPE contractions \oneloopbb,
\eqnn\fiveptcorr
$$\eqalignno{
\langle b \,&{\cal Z}\, V_1(z_1)\,  U_2(z_2) \, \ldots \, U_{5}(z_{5})   \rangle = {\cal I}\cdot {\cal K}_5
\cr
 {\cal K}_5 &\equiv \Big \langle \big[ \partial_z {\cal G}_{12} V_{12} T_{3,4,5} + (2\leftrightarrow 3,4,5)\big] + \big[ \partial_z {\cal G}_{23} V_1 T_{23,4,5} + (2,3|2,3,4,5) \big] \Big \rangle  \ . &\fiveptcorr
}$$
The kinematic factor ${\cal K}_5 $ is the generating function for the numerators of both the boxes and the pentagon. According to \KNlim\ and \intlim, the worldline limit reduces the plane wave correlator and the worldsheet integrations in the string amplitude to the Schwinger parametrization of the pentagon:
\eqnn\pentdeg
$$\eqalignno{
{\cal A}_{5}^{{\rm pent}}& \rightarrow \int^{\infty}_0{dt\over t} \, t^{5-D/2}  \! \! \! \! \! \! \! \! \! \!  \int \limits _{0 \leq \nu_{i} \leq \nu_{i+1} \leq 1} 
\! \! \! \! \! \! \! \! \! \!  d \nu_2\, \ldots \,  d \nu_5 \, e^{-\pi t Q_5[k_{1},k_2,\ldots,k_5]} \, \big( {\cal K}_5 \,\big|_{\partial_z {\cal G}_{ij} \rightarrow \partial_z G_{ij}} \big) \ . &\pentdeg
}$$
The worldline limit of the Green functions \GFlim\ directly maps the open string kinematic factor
${\cal K}_5$ to the Schwinger integrand \schwingerpent\ of the pentagon numerator
$N^{(5)}_{1|2,3,4,5}$, hence
\eqnn\pentcheck
$$\eqalignno{
{\cal A}_{5}^{{\rm pent}}& \rightarrow 
\int d^D\ell \, \langle A_{\rm pent}(1,2,3,4,5|\ell) \rangle
 \ . &\pentcheck
}$$
In addition, boxes arise from the singular limit \reducible\ when the positions $z_i, z_{i+1}$ of neighboring vertex operators collide. This applies to the cyclic orbit of the propagator $\partial_z {\cal
G}_{12}$ which leaves $\delta(z_1-z_2)$ at the residue of the pole in $s_{12}$:
\eqnn\boxdeg
$$\eqalignno{
{\cal A}_{5}^{{\rm box}} &= \int^{\infty}_0 {dt \over t}  \! \! \!  \! \! \!   \int \limits _{0 \leq \Im z_{i} \leq \Im z_{i+1} \leq t}  \! \! \!  \! \! \!   dz_2 \, \ldots \, dz_5 \,  {\cal I} \, \Big \langle { \delta(z_{12}) \over s_{12}} V_{12} T_{3,4,5} + { \delta(z_{23}) \over s_{23}} V_{1} T_{23,4,5} \cr
& \ \ \ \ \ \ \ \ \ \ \ \ \ \ \ \ \ + { \delta(z_{34}) \over s_{34}} V_{1} T_{2,34,5} + { \delta(z_{45}) \over s_{45}} V_{1} T_{2,3,45} + { \delta(z_{51}) \over s_{15}} V_{51} T_{2,3,4}\Big \rangle  \ . &\boxdeg
}$$
The same worldline limits \KNlim\ and \intlim\ applied to the irreducible part of the correlator reduce
\boxdeg\ to the Schwinger parametrization of the boxes in \schwingerbox,
\eqnn\boxcheck
$$\eqalignno{
{\cal A}_{5}^{{\rm box}}& \rightarrow 
\int d^D\ell \, \langle A_{\rm box}(1,2,3,4,5) \rangle
 \ , &\boxcheck
}$$
using $\delta(z_1-z_2)\sim t^{-1} \delta(\nu_1-\nu_2)$ and $Q_4[k_{12},k_{3},k_4,k_5] = Q_5[k_1,k_2,\ldots, k_5] \big|_{\nu_1=\nu_2}$. 

To summarize,
the point-particle limit of the five-point open string amplitude \looppresc\ contains a reducible part
caused by the singular behavior \reducible\ which reproduces the boxes of our superspace proposal, see \boxcheck. For the
irreducible part, on the other hand, the degeneration limit of the Green function \GFlim\ is
applied directly to ${\cal K}_5$ and reproduces the pentagon numerator \specPent\ found by the BRST
analysis, see \boxcheck.

An analogous analysis can be carried out for the manifestly BRST invariant form of the string amplitude \oneloopbb,
\eqnn\stringBRST
$$\eqalignno{
{\cal A}_{5} &= 
 \int^{\infty}_0 {dt\over t}  \! \! \!  \! \! \!  \! \! \int \limits _{0 \leq \Im z_{i} \leq \Im z_{i+1} \leq t}  \! \! \!  \! \! \!  \! \!  dz_2 \, \ldots \, dz_5 \, {\cal I} \,  \langle s_{23} \partial_z {\cal G}_{23} \, C_{1|23,4,5} + (2,3|2,3,4,5) \rangle \ ,
 &\stringBRST
 }$$
related to ${\cal K}_5 $ in \fiveptcorr\ by integration by parts with respect to $z_2,\ldots,z_5$.
Using the same distinction between box and pentagon contributions as above, \stringBRST\ reduces to
the manifestly BRST invariant representation \BRSTFive\ of the field-theory limit. The Schwinger parametrization of the scalar and vector
pentagons can be coherently described by $\partial G_{ij}$ once the following cohomology manipulations
are taken into account \partI
\eqnn\Cmtwo
$$\eqalignno{
\langle  k^1_m C^m_{1|2,3,4,5} \rangle &= 0  \ , \ \ \ \
\langle k^2_m C^m_{1|2,3,4,5} \rangle= - \langle  s_{23} C_{1|23,4,5}+(3\leftrightarrow 4,5)  \rangle \ . &\Cmtwo
}$$
Note that they follow from \Tmone\ under the prescription \InvMap\ to manifest BRST invariance.

\listrefs

\bye